%% file: Extended_A+.tex
\documentclass[fleqn,usenatbib]{mnras}

\usepackage{newtxtext,newtxmath}

\usepackage[T1]{fontenc}

\DeclareRobustCommand{\VAN}[3]{#2}
\let\VANthebibliography\thebibliography
\def\thebibliography{\DeclareRobustCommand{\VAN}[3]{##3}\VANthebibliography}

\usepackage{graphicx}	
\usepackage{amsmath}	
\usepackage{siunitx}
\usepackage{subcaption}
\usepackage{threeparttable}
\usepackage{float}
\usepackage{anyfontsize}
\usepackage{lscape} 
\usepackage{caption}
\DeclareCaptionLabelFormat{continued}{#1~#2 (Cont.)}
\usepackage{lscape} 
\defcitealias{Rao2021}{Paper I}

\title{Determination of dynamical ages of open clusters through the A$^+$ parameter -- II}

\author[Khushboo K. Rao et al.]{Khushboo K. Rao,$^{1}$ \thanks{E-mail: p20170419@pilani.bits-pilani.ac.in} 
Kaushar Vaidya,$^1$
Manan Agarwal,$^2$
Shanmugha Balan,$^1$ and
\newauthor Souradeep Bhattacharya$^3$\\
$^{1}$ Department of physics, Birla Institute of Technology and Science-Pilani, 333031 Rajasthan, India\\
$^{2}$ Anton Pannekoek Institute for Astronomy $\&$ GRAPPA, University of Amsterdam, Science Park 904, 1098 XH Amsterdam, The Netherlands\\
$^{3}$ Inter University Centre for Astronomy and Astrophysics, Ganeshkhind, Post Bag 4, Pune 411007, India
}
\date{Accepted}

\pubyear{2023}

\begin{document}
\label{firstpage}
\pagerange{\pageref{firstpage}--\pageref{lastpage}}
\maketitle
\input{chapters/Abstract}
\input{chapters/Introduction}
\input{chapters/Data}
\input{chapters/Results}
\input{chapters/Discussion}
\input{chapters/Final_remarks}
\input{chapters/Acknowledgements}
\section*{Data Availability}
The data underlying this article are publicly available at \url{https://gea.esac.esa.int/archive}. The identified members of the open clusters will be available at the CDS \url{https://cds.u-strasbg.fr/}. 
\bibliographystyle{mnras}
\bibliography{References} 
\input{chapters/Appendix}
\bsp	
\label{lastpage}
\end{document}

%% file: chapters/Abstract.tex
\begin{abstract}
Blue straggler stars (BSS), one of the most massive members of star clusters, have been used for over a decade to investigate mass segregation and estimate the dynamical ages of globular clusters (GCs) and open clusters (OCs). This work is an extension of our previous study, in which we investigated a correlation between theoretically estimated dynamical ages and the observed $A^+_{\mathrm{rh}}$ values, which represent the sedimentation level of BSS with respect to the reference population. Here, we use the ML-MOC algorithm on \textit{Gaia} EDR3 data to extend this analysis to 23 OCs. Using cluster properties and identified members, we estimate their dynamical and physical parameters. In order to estimate the $A^+_{\mathrm{rh}}$ values, we use the main sequence and main sequence turnoff stars as the reference population. OCs are observed to exhibit a wide range of degrees of dynamical evolution, ranging from dynamically young to late stages of intermediate dynamical age. Hence, we classify OCs into three distinct dynamical stages based on their relationship to $A^+_{\mathrm{rh}}$ and $N_{\text{relax}}$. NGC 2682 and King 2 are discovered to be the most evolved OCs, like Familly III GCs, while Berkeley 18 is the least evolved OC. Melotte 66 and Berkeley 31 are peculiar OCs because none of their dynamical and physical parameters correlate with their BSS segregation levels.
\end{abstract}
\begin{keywords}
blue stragglers -- open clusters: general -- methods: statistical
\end{keywords}

%% file: chapters/Introduction.tex
\section{Introduction} \label{sec:intro}
Star clusters play a major role in our understanding of stellar evolution, stellar dynamics, and galactic evolution \citep{Portegies2010,Renaud2018}. Older clusters (age $>$1 Gyr) are in their long-term evolutionary phase \citep{Krumholz2019,Krause2020}, during which the following factors contribute to and drive the evolution of a cluster: (i) Cluster properties: depending on their initial characteristics, such as mass function, total mass, binary fraction, number of stars, etc., clusters with comparable ages and locations in the Galaxy can be at very different stages of their dynamical evolution \citep{Ferraro2012}. (ii) Internal factors: stellar interactions and encounters between cluster members have a cumulative effect of pushing the cluster toward energy equipartition, which subsequently leads to mass segregation. As a result, the configuration of cluster members changes over time \citep{Vesperini2010}. (iii) External factors: interactions with nearby giant molecular clouds (if any) and galactic potential play a significant role in determining the shape and size of a cluster \citep{Baumgardt2003}. Due to the internal dynamics and external factors, the properties and stellar content of a cluster keep changing throughout its lifetime. For many years now, a variety of techniques have been used to determine the dynamical stages of star clusters, such as the slope of mass functions \citep{BhattacharyaandMahulkar2017,Bhattacharya2017,Bhattacharya2021}, the segregation of massive populations \citep{Allison2009,Bhattacharya2022}, and the radial distributions of blue straggler stars \citep[BSS;][]{Ferraro2012,Ferraro2018,Vaidya2020,Rao2021}.

A cluster hosts a variety of simple stellar populations as well as exotic stellar populations which are descendants of stellar interactions and binary evolution, like BSS \citep{Stryker1993,Bailyn1995}, cataclysmic variables \citep{Ritter2010}, and so on, over the course of its existence. BSS are particularly intriguing among exotic populations because they are bluer and brighter than main sequence turnoff stars (MSTOs), and thus appear to defy the standard theory of stellar evolution \citep{Sandage1953}. Their unusual location on colour-magnitude diagrams (CMD) suggests that they have gained additional mass during their history, making them one of the most massive populations of star clusters \citep{Shara1997,Fiorentino2014}. Massive stars, including BSS, are known to sink into the cluster center faster than any other cluster population due to the effect of dynamical friction \citep{Chandrasekhar1943}. Thus, BSS have been used as tools to determine the dynamical ages of globular clusters \citep[GCs;][]{Ferraro2012,Ferraro2018,Ferraro2019,Ferraro2020,Cadelano2022,Dresbach2022,Beccari2023} and open clusters \citep[OCs;][]{Bhattacharya2019,Vaidya2020,Rao2021}.
\begin{figure*}
	\includegraphics[width=0.98\textwidth]{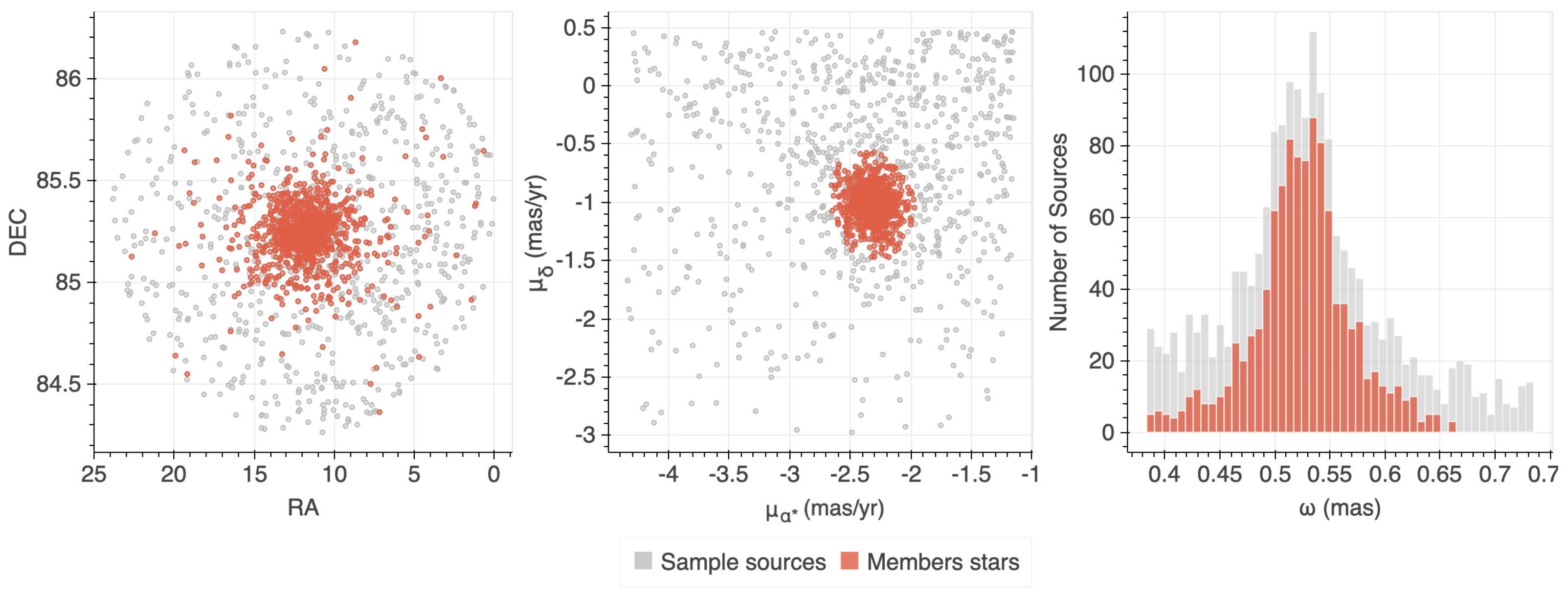}
    \caption{The spatial, proper motion, and parallax distribution of \textit{Sample} \textit{sources} (grey) and cluster members (orange) of NGC 188 OC identified by the ML-MOC algorithm.}
    \label{fig:mem_fig}
\end{figure*}
In \citet[hereafter Paper I]{Rao2021}, we estimated the dynamical ages of 11 OCs using the sedimentation level of their BSS population in the OCs' centre. There we determined the sedimentation level of BSS using the $A^+$ parameter, i.e., the area enclosed between cumulative radial distributions of BSS and reference population\footnote{In \citetalias{Rao2021}, we used MSTOs, sub giant branch stars (SGBs), red giant branch stars (RGBs), and red clump stars (RCs) as a reference population.}, given as
\begin{equation} 
A^+ = \int^x_{x_{min}} \phi_{\mathrm{BSS}}(x^{\prime})- \phi_{\mathrm{REF}}(x^{\prime})dx^{\prime} 
\end{equation} 
where $x$ = $\mathrm{\log}(r/r_{\mathrm{h}})$ and $x_{\mathrm{min}}$ are the outermost and innermost radii from the cluster center, respectively, and $r_{\mathrm{h}}$ is the half-mass radius of the cluster. $A^+$ is an observational parameter, which was first introduced by \citet{Alessandrini2016} to indicate the level of BSS segregation. They showed that $A^+$ should always rise as a cluster evolves, but the presence of neutron stars and black holes in the cluster environment delays the BSS segregation process, which in turn slows $A^+$ growth. Stellar populations of a cluster differentially experience the strength of the Galactic field. Consequently, the inner region of a cluster is least responsive to the Galactic field and thus most susceptible to mass segregation due to two-body relaxation. Therefore, we estimated $A^+$ up to $r_{\mathrm{h}}$ for OCs, as was done for GCs \citep{Lanzoni2016}, and hence named $A^+_{\mathrm{rh}}$. In order to know whether $A^+_{\mathrm{rh}}$ actually an indicator of dynamical ages for OCs as was observed for GCs \citep{Ferraro2018}, we had compared them with the theoretical estimates of dynamical ages of star clusters, namely the number of central relaxation a cluster has experienced since its formation, $N_{\text{relax}}$ (see \S\ref{sec:results} for its mathematical formula). In \citetalias{Rao2021}, we obtained a broad correlation between $A^+_{\mathrm{rh}}$ and $N_{\text{relax}}$ compared to that of GCs and showed that OCs are among the less evolved GCs. The sample size of OCs in \citetalias{Rao2021} was small compared to the 48 GCs data points. Therefore, in the present work, we increase our sample to 23 OCs (ages $>$1 Gyr), including a reanalysis of the 11 OCs studied in \citetalias{Rao2021}, to re-investigate the previously estimated relationships of $A^+_{\mathrm{rh}}$ vs dynamical and physical parameters. With these relations, we aim to explore the dynamical ages of 23 OCs.

The rest of the paper is organized as follows: In \S\ref{sec:data}, we describe the membership identification process of 23 OCs. In \S\ref{sec:results}, we present details of BSS and reference population identification, and estimation of $A^+_{\mathrm{rh}}$, physical, and dynamical parameters for 23 OCs. In \S\ref{sec:discussion}, we discuss the relationships of $A^+_{\mathrm{rh}}$ with other markers of cluster dynamical age, namely $N_{\text{relax}}$, and physical parameters of OCs, as well as how they compare to GCs. Based on those relations, we estimate their dynamical ages. In the end, in \S\ref{sec:Final_remarks}, we present the summary and conclusion of this study.

%% file: chapters/Data.tex
\section{Data and Membership identification}
\label{sec:data}
\input{Tables/Fundamental_params}
We use the ML-MOC algorithm \citep{Agarwal2021} on the \textit{Gaia} EDR3 data \citep{Gaiaedr32021} to identify members of 23 OCs. ML-MOC is the k-Nearest Neighbour \citep[kNN;][]{Cover_kNN} and Gaussian mixture model \citep[GMM;][]{mclachlan200001} based membership determination algorithm for OCs. It uses proper motions and parallax information from \textit{Gaia} data in order to identify members. ML-MOC is described in detail in \citet{Agarwal2021} on \textit{Gaia} DR2, with further applications for OCs membership determinations to \textit{Gaia} EDR3 in \citet{Bhattacharya2021,Bhattacharya2022,Rao2022,Rao2023,Panthi2022}. 
\citet{Bhattacharya2022} investigated the completeness of the Gaia EDR3 data using NGC 2248 as a representative OC. They compared its \textit{Gaia} EDR3 members with deeper Pan-STARRS1 DR2 data and demonstrated that \textit{Gaia} EDR3 data is $\sim$90 per cent complete down to G = 20 mag. \citet{Bhattacharya2022} also explored the completeness and contamination fraction of ML-MOC by comparing the identified cluster members of Berkeley 39 with its deepest available spectroscopic data \citep{Bragaglia2022}. With this investigation, they found that ML-MOC is $\sim$90$\%$ complete with $\sim$2.3 per cent contamination fraction down to G = 19.5 mag.

\input{Tables/Dynamical_params}

In Fig. \ref{fig:mem_fig}, we show the spatial, proper motion, and parallax distributions of \textit{Sample} \textit{sources} and members of NGC 188 OC as a representative. Here, the \textit{Sample} \textit{sources} are sources in the cluster region which are having a higher fraction of cluster members than field stars and members are final cluster members having membership probability $>$0.2. We have identified 1061 sources as members of NGC 188. We estimate the cluster limit beyond which probable cluster members are indistinguishable from field stars, as 30 arcmin, which we term as the cluster radius. We similarly estimated cluster radii for the rest of the 22 OCs, which are listed in column 4 of Table \ref{tab:fundamental_params}. Among 23 OCs, the CMD of Trumpler 5 is quite broad (see the middle panel of Fig. \ref{fig:DR}). In \citet{Rain2020b} and \citetalias{Rao2021}, it has been shown that the primary reason behind the broad CMD of Trumpler 5 is the effect of differential reddening and extinction along the line of sight of the cluster. Therefore, it is necessary to correct its members from differential reddening and extinction in order to identify genuine BSS candidates. Following \citet{Massari2012}, we performed differential reddening correction for members of Trumpler 5, see \S\ref{sec:appendix} for details of the method.

%% file: Tables/Fundamental_params.tex
\begin{table*}
	\centering
	\caption{The list of clusters studied in this work, their fundamental parameters, the total number of BSS, and references for ages and metallicities.}
	\label{tab:fundamental_params}
	\begin{tabular}{cccccccccl} 
		\hline
		\\
		Name & RA & DEC & R &Age & Distance & [M/H] & A$_\text{V}$ & N$_\text{BSS}$ & Reference \\
		     & (deg) & (deg) & (arcmin) & (Gyr) & (pc)      & (dex)   &    (mag)       &  & \\
		\\
		\hline
		\\
Berkeley 17   & 80.136307 & 30.573710  & 15 &9    & 3000 & $-$0.01  & 1.74   & 18 &   1, 2, 3 \\
Berkeley 18   & 80.534671        & 45.440855 &  15       & 3.7  & 5200 & $-$0.39  & 1.9   & 23 &  4, 5 \\
Berkeley 21   & 87.931125        & 21.802785 &    9    & 2.1  & 6100 & $-$0.54  & 2.5   & 10 &  6, 7  \\
Berkeley 31   & 104.406861      & 8.284359   &   5    & 3.1  & 7000 & $-$0.35  & 0.51   & 12 &   4, 8 \\
Berkeley 32   & 104.53249       & 6.436102   &   10    & 4.9  & 3250 & $-$0.3   & 0.52   & 14 &  9, 10, 11  \\
Berkeley 36   & 109.100877       & $-$13.196754 &  10      & 3.0  & 5300 & $-$0.15  & 2.05    & 18 &  12, 13, 28   \\
Berkeley 39   & 116.696526 & $-$4.668978 & 14 & 5.65    & 4200 & $-$0.15   & 0.51 & 17  &  14  \\
Collinder 261 & 189.522487 & $-$68.379857 & 20 & 6    & 2800 & $-$0.03  & 0.96   & 37 &    1, 8, 13  \\
King 2        & 12.743372        & 58.193953 &    8     & 5.8  & 5700 & $-$0.42   & 1.26   & 19 &  15, 16, 17  \\
Melotte 66    & 111.574698 & $-$47.686005 & 15 & 3.4  & 4810 & $-$0.176 & 0.43   & 14 &  14, 18  \\
NGC 188       & 11.830432      & 85.241492  & 30      & 7    & 1800 & 0.12   & 0.15   & 19 &  14  \\
NGC 1193      & 46.488114        & 44.383018  &  10       & 4.3  & 5400 & $-$0.22  & 0.6    & 15 &  4, 19   \\
NGC 2141      & 90.728093        & 10.457084  &   12      & 2.3  & 4150 & $-$0.33      & 1.25   & 16 &   13, 20, 21  \\
NGC 2158      & 91.863715 & 24.098575  & 10  & 1.9  & 4000 & $-$0.15  & 1.5    & 48 &  12, 14   \\
NGC 2506      & 120.009294     & $-$10.774234  &  20    & 2.1    & 3300 & $-$0.27  & 0.25    & 11 &   14 \\
NGC 2682      & 132.8518         & 11.8248  &    48       & 4    & 840  & 0.0      & 0.08   & 12 &  12, 22   \\
NGC 6791      & 290.223531     & 37.777353  &   15    & 8.5  & 4300 & 0.31   & 0.38   & 29 &  14  \\
NGC 6819      & 295.328552 & 40.1875522  & 18 & 2.73 & 2400 & 0.0 & 0.45   & 16 &  14, 22  \\
NGC 7789      & 359.330567 & 56.726606  & 35  & 1.6  & 1900 & $-$0.093 & 0.85   & 13 &  18, 23   \\
Pismis 2      & 124.480956       & $-$41.676602  &  10      & 1.5  & 4000 & $-$0.07  & 4.1   & 10 &  13, 24, 25\\
Tombaugh 2    & 105.772938       & $-$20.817078  &    6    & 2.05  & 8500 & $-$0.31  & 1.1    & 25 &  26, 27 \\
Trumpler 5    & 99.132815 & 9.475601  & 20 & 3    & 3050 & $-$0.403 & 1.4    & 53 &  22  \\
Trumpler 19   & 168.62611        & $-$57.565928  &   15     & 3.7  & 2420 & 0.14   & 0.5    & 12 &  13, 15 \\ 
		\\
		\hline
    \end{tabular}
    \begin{tablenotes}
        \item $^1$\citet{Bragaglia2006}, $^2$\citet{Friel2005}, $^3$\citet{Bhattacharya2019}, $^4$\citet{Overbeek2016}, $^5$\citet{Netopil2016}, $^6$\citet{Tosi1998}, $^7$\citet{Yong2005}, $^8$\citet{Bragaglia2022}, $^9$\citet{Sariya2021}, $^{10}$\citet{Vande2010}, $^{11}$\citet{Carrera2011}, $^{12}$\citet{Viscasillas2022}, $^{13}$\citet{Cantat2018}, $^{14}$\citet{Vaidya2020}, $^{15}$\citet{Dias2002}, $^{16}$\citet{Jadhav2021}, $^{17}$\citet{Kaluzny1989}, $^{18}$\citet{Rao2022}, $^{19}$\citet{Friel2010}, $^{20}$\citet{Magrini2021}, $^{21}$\citet{Rosvick1995}, $^{22}$\citet{Rao2021}, $^{23}$\citet{Vaidya2022}, $^{24}$\citet{Friel2002}, $^{25}$\citet{Janes1994}, $^{26}$\citet{Gloria2011}, $^{27}$\citet{Villanova2010}, $^{28}$\citet{Ortolani2005}
    \end{tablenotes}
\end{table*}


%% file: Tables/Dynamical_params.tex
\begin{table*}
	\caption{The dynamical and physical parameters, and the estimated values of $A^+_{\mathrm{rh}}$ and errors in $A^+_{\mathrm{rh}}$ of the OCs. Here, Column 1: Cluster name; Columns 2, 3, 4, and 5: fitted King parameters; Column 6: average stellar mass; Column 7: integrated absolute magnitudes; Columns 8 and 9: central luminosity and mass density, respectively;  Column 11: Central relaxation time.}
	\label{tab:dynamical_params}

	\begin{tabular}{cccccccccc}
		\hline
		\\
	    Cluster  & $r_{\mathrm{c}}  $ & $r_{\mathrm{h}}$ &  $r_{\mathrm{t}}$  &   c   &  \textless{}m*\textgreater{} & $I_{\mathrm{M_V}}$   &   $\rho_{_{\scriptscriptstyle  L,O}}$ &   $\rho_{_{\scriptscriptstyle  M,O}}$ &      $t_{\mathrm{rc}}$ \\
	      &   (arcmin) &   (arcmin) &   (arcmin) &    &  ($M_{\sun}$) &   (mag) &    ($L_{\sun}$/$pc^3$) &    ($M_{\sun}$/$pc^3$) &      (Myr) \\
		 \\
		\hline
		\\
        Berkeley 17 & 2.99±0.14 & 5.8$\pm$0.4 & 30.75$\pm$4.82 & 1.01$\pm$0.07 & 0.8 & $-$3.485 & 10$\pm$1.6 & 7.6$\pm$1.2 & 83$\pm$9 \\ 
        Berkeley 18 & 6.3$\pm$1 & 8.1$\pm$0.6 & 30.71$\pm$5.62 & 0.69$\pm$0.11 & 0.43 & $-$5.305 & 2.3$\pm$0.8 & 1.8$\pm$0.6 & 2700$\pm$1000 \\ 
        Berkeley 21 & 1.51$\pm$0.34 & 3.3$\pm$0.7 & 18.81$\pm$8.8 & 1.1$\pm$0.23 & 1.08 & $-$4.995 & 32$\pm$20 & 24$\pm$15 & 140$\pm$60 \\ 
        Berkeley 31 & 0.65$\pm$0.04 & 1.99$\pm$0.33 & 15$\pm$5.39 & 1.36$\pm$0.16 & 0.94 & $-$3.411 & 42$\pm$11 & 32$\pm$8 & 24$\pm$4 \\ 
        Berkeley 32 & 1.83$\pm$0.11 & 5.89$\pm$0.27 & 45.76$\pm$3.57 & 1.4$\pm$0.04 & 0.94 & $-$3.775 & 25$\pm$4 & 19.1$\pm$2.9 & 38$\pm$4 \\ 
        Berkeley 36 & 1.58$\pm$0.08 & 5$\pm$1 & 38.01$\pm$12.98 & 1.38$\pm$0.15 & 0.94 & $-$4.801 & 24$\pm$6 & 18$\pm$4 & 89$\pm$12 \\ 
        Berkeley 39 & 2.34$\pm$0.16 & 4.9$\pm$0.5 & 27.86$\pm$6.96 & 1.08$\pm$0.11 & 0.85 & $-$4.056 & 11.6$\pm$2.7 & 8.8$\pm$2.1 & 109$\pm$17 \\ 
        Collinder 261 & 3.28$\pm$0.09 & 7$\pm$0.4 & 39.66$\pm$4.75 & 1.08$\pm$0.05 & 0.85 & $-$5.01 & 34$\pm$4 & 25.5$\pm$2.7 & 130$\pm$9 \\ 
        King 2 & 0.72$\pm$0.02 & 3.1$\pm$0.13 & 27.58$\pm$1.4 & 1.583$\pm$0.025 & 1.03 & $-$3.779 & 60$\pm$4 & 44.9$\pm$3.3 & 20.4$\pm$1.1 \\ 
        Melotte 66 & 3.68$\pm$0.29 & 6.2$\pm$0.4 & 28.32$\pm$4.59 & 0.89$\pm$0.08 & 1.02 & $-$4.853 & 5.9$\pm$1.3 & 4.5$\pm$1 & 370$\pm$60 \\ 
        NGC 188 & 4$\pm$0.13 & 10.8$\pm$1 & 74.64$\pm$12.89 & 1.27$\pm$0.08 & 0.83 & $-$3.551 & 13.8$\pm$1.8 & 10.4$\pm$1.4 & 48$\pm$4 \\ 
        NGC 1193 & 0.67$\pm$0.03 & 2$\pm$0.27 & 14.79$\pm$3.71 & 1.34$\pm$0.11 & 1.06 & $-$3.283 & 77$\pm$14 & 58$\pm$11 & 13.8$\pm$1.5 \\ 
        NGC 2141 & 2.71$\pm$0.19 & 5.3$\pm$0.5 & 27.88$\pm$6.66 & 1.01$\pm$0.11 & 0.85 & $-$5.24 & 26$\pm$6 & 19$\pm$5 & 220$\pm$40 \\ 
        NGC 2158 & 1.59$\pm$0.04 & 4.05$\pm$0.26 & 26.79$\pm$3.45 & 1.23$\pm$0.06 & 0.89 & $-$6.152 & 234$\pm$23 & 177$\pm$18 & 106$\pm$6 \\ 
        NGC 2506 & 2.77$\pm$0.04 & 7.22$\pm$0.29 & 48.47$\pm$4.01 & 1.24$\pm$0.04 & 0.55 & $-$5.453 & 40.5$\pm$2.5 & 30.6$\pm$1.9 & 201$\pm$7 \\ 
        NGC 2682 & 5.38$\pm$0.23 & 21.3$\pm$1.2 & 183.73$\pm$12.82 & 1.53$\pm$0.04 & 0.66 & $-$3.942 & 56$\pm$6 & 42$\pm$5 & 28.1$\pm$2.3 \\ 
        NGC 6791 & 2.94$\pm$0.15 & 5.07$\pm$0.28 & 23.94$\pm$3.43 & 0.91$\pm$0.07 & 0.87 & $-$4.814 & 14.8$\pm$2.5 & 11.2$\pm$1.9 & 211$\pm$25 \\ 
        NGC 6819 & 2.5$\pm$0.1 & 7.1$\pm$0.8 & 50.84$\pm$10.08 & 1.31$\pm$0.09 & 0.81 & $-$4.915 & 90$\pm$14 & 68$\pm$10 & 56$\pm$5 \\ 
        NGC 7789 & 6.57$\pm$0.11 & 13.7$\pm$0.4 & 76.47$\pm$4.77 & 1.066$\pm$0.028 & 0.68 & $-$6.029 & 35.4$\pm$2.1 & 26.7$\pm$1.6 & 378$\pm$15 \\ 
        Pismis 2 & 1.25$\pm$0.09 & 2.58$\pm$0.14 & 14.25$\pm$2.1 & 1.06$\pm$0.07 & 1.27 & $-$5.29 & 4.2$\pm$0.8 & 3.2$\pm$0.6 & 41$\pm$6 \\ 
        Tombaugh 2 & 0.7$\pm$0.03 & 2.18$\pm$0.31 & 16.72$\pm$4.85 & 1.38$\pm$0.13 & 0.87 & $-$5.221 & 98$\pm$20 & 74$\pm$15 & 67$\pm$8 \\ 
        Trumpler 5 & 5.08$\pm$0.17 & 11.3$\pm$0.8 & 66.46$\pm$10.03 & 1.12$\pm$0.07 & 0.97 & $-$6.209 & 20.1$\pm$2.6 & 15.2$\pm$2 & 408$\pm$33 \\ 
        Trumpler 19 & 2.63$\pm$0.22 & 7.2$\pm$0.7 & 49.75$\pm$8.12 & 1.28$\pm$0.08 & 0.96 & $-$3.791 & 25$\pm$6 & 19$\pm$4 & 41$\pm$6 \\ 
 		\\
		\hline
	\end{tabular}
\end{table*}

%% file: chapters/Results.tex
\section{Analysis}
\label{sec:results}
\subsection{Identification of BSS and Reference population}\label{population_selections}
We identify the BSS and reference population (REF) of 23 OCs using the method established in \citetalias{Rao2021}. We describe the method briefly here. First, we plot the CMDs of the 23 OCs and fit PARSEC isochrones. In order to fit isochrones, we choose known ages and metallicities estimated in the literature. We estimate the distance to an OC as the mean distance of their bright members (G$\le$15 mag or G$\le$16 mag when there are very few members for G$\le$15 mag) taken from \citet{Bailer2021}. We have estimated the lower bound and upper bound of the distance to each OC as the mean of the lower bound and the mean of the upper bound of their bright members, respectively. The \textit{Gaia} DR3 data provides A$_\text{G}$ and E(BP$-$RP) of a maximum fraction of cluster members. We estimate the range of A$_\text{G}$ for each OC as the range of A$_\text{G}$ of their bright members. In order to fit the isochrones to 23 OCs, we choose the values of A$_\text{G}$ from the estimated range which fits best along with the estimated range of distances and known ages and metallicities. In order to visually best fit the isochrones, we also tuned ages within literature ranges. The fundamental parameters of the 23 OCs corresponding to the fitted isochrones are listed in Table \ref{tab:fundamental_params}. The last column of Table \ref{tab:fundamental_params} shows the references of metallicities and ages for 23 OCs. 

We plotted ZAMS of age 40 -- 120 Myr in order to avoid the inclusion of hot subdwarfs and gap stars located just below the BSS population. We also plotted equal mass binary isochrone in order to exclude unresolved binary located immediately above MSTO points of OCs. We then normalized OCs CMDs, isochrones, ZAMS, and equal mass binary isochrones in order to locate turnoff magnitudes of OCs at (0,0). Therefore, we denote the normalized color and magnitude of CMD by (BP$-$RP)$^*$ and G$^*$, respectively. All the OCs are divided into two categories based on the shape of their fitted isochrones, those having a blue hook near their MSTO points or those with a smooth rightward-going isochrone from their MSTO points. We then followed the steps given in \citetalias{Rao2021} to identify BSS, MSTO stars, SGBs, and RGBs based on their respective categories. In order to estimate $A^+_{\mathrm{rh}}$, we need to consider REF as an average massive population of a cluster that can represent the whole cluster \citep{Ferraro2018}. As discussed in \S \ref{sec:data}, \textit{Gaia} EDR3 and ML-MOC are almost complete down to G = 18.5 mag and OCs selected for the current work have G$_{\text{TO}} <$ 18.5 mag. Therefore, we use MSTO stars and MS stars brighter than G = 18.5 mag as REF for the current work. Fig. \ref{fig:CMD1} shows CMDs of 23 OCs with identified BSS candidates (blue-filled circles), MSTO region (black box), and SGBs and RGBs (black triangles). 
\subsection{Structural parameters} \label{sc_param}
For the current work, we estimate the structural parameters such as core radius ($r_{\text{c}}$), half-mass radius ($r_{\text{h}}$), tidal radius ($r_{\text{t}}$), and concentration parameter ($c$ = log($r_{\text{t}}$/$r_{\text{c}}$)) for 23 OCs including 11 OCs of \citetalias{Rao2021}. the Gaia EDR3 data for open clusters is 90 per cent complete down to G $\sim$19.5 mag where the contamination fraction of ML-MOC member detection is $\sim$2.3 per cent \citep{Bhattacharya2022}, therefore, to not be affected by completeness and contamination effects, we conservatively consider only those cluster members brighter than G = 18.5 mag to estimate the structural parameters.

We start by dividing a cluster radius into equal radius bins.  We then compute the number density of each radius bin and plot the logarithmic number densities against the logarithmic radii of the annular regions from the cluster center.  We have plotted  surface density profiles of various bin sizes, where the number of bins are varied from the cluster radius to twice the cluster radius. The observed number density profiles are then fitted with isotropic single-mass King models \citep{King1966} using the available software package \texttt{LIMEPY}\footnote{\url{https://github.com/mgieles/limepy}} \citep{Gieles2015,Gieles2017} In order to estimate the values of structural parameters and explore their associated uncertainties, we use \texttt{emcee}, affine-invariant Markov chain Monte Carlo (MCMC) sampler \citep{Foreman2013}, available through the \texttt{LMFIT} Python package \citep{Newville2014}. In order to generate a list of initial guesses for the input parameters, we use Differential Evolution Global Optimization Algorithm \citep{Storn1997} that returns the maximum likelihood solution. The best-fitted observed density profiles are then chosen based on the minimum value of reduced chi-square ($\chi^2_r$) and a sufficient number of stars in radial bins of the cluster's periphery to ensure statistical reliability. The value of the $\chi_r^2$ is calculated using the following formula:

\begin{equation} 
\chi^2_r = \frac{1}{N-n_p}\sum_{i=0}^{N} \frac{(D_{o,i} - D_{m,i})}{\sigma_{o,i}^2}
\end{equation} 
    
where N is the number bins, $n_{\mathrm{p}}$ is the total number of fitted model parameters which are 3 in our case, $D_{o,i}$ is the observed number density, $D_{m,i}$ is the theoretical number density predicted by the model, and $\sigma_{o,i}$ is the Poisson error in the observed number density. We then feed \texttt{emcee} with the maximum likelihood parameter values and run it for 100 walkers, 100 burn-in steps, and 4000 -- 6000 iterations. To fit the King profile, \texttt{LIMEPY} takes two parameters as input, the dimensionless central potential ($\hat{\phi_{\text{0}}}$) and any one of the radial parameters, like King radius ($r_{\text{0}}$), $r_{\text{h}}$, $r_{\text{t}}$, and virial radius ($r_{\text{v}}$) where these radii are related to each other for a constant value of $\hat{\phi_{\text{0}}}$. In addition to these two parameters, we also provide  \texttt{emcee} with the total mass of a cluster as an additional parameter because its default value of it in \texttt{LIMEPY} is 10$^5$ M$_{\odot}$ which is excessively large for OCs.

The fitted King profiles for 23 OCs are shown in Fig. \ref{fig:kings_profiles}. The best-fitted values (the median over all the post-burn-in iterations) and 1$\sigma$ uncertainties of $r_{\text{c}}$, $r_{\text{h}}$, $r_{\text{t}}$, and $c$ are listed in columns 2, 3, 4, and 5 of Table \ref{tab:dynamical_params}, respectively. 
\subsection{Mass functions}\label{PDMFs}
In order to estimate the average stellar masses and total number of stars of OCs, we derive their present-day mass functions (PDMFs). The mass function is denoted by $\zeta(M)$ and is given as
\begin{equation}
    \zeta(M) = \frac{dN}{dM}
\end{equation}
where dN is the number of stars between mass interval M to M$+$dM.

In order to plot the PDMF of a cluster, we choose MS stars located within the range of MSTO magnitude and G = 18.5 mag on the cluster CMD. We then obtain the masses of these selected MS stars using the fitted PARSEC isochrones. We use interpolation to get masses of MS stars as a function of their G mag from the fitted isochrones. As we impose G = 18.5 mag cut, thus the effect of incompleteness on PDMFs has been eliminated. The PDMFs of 23 OCs are shown in Fig. \ref{fig:PDMFs}. The slope of the derived PDMF is calculated using the relation $\mathrm{log}(\frac{dN}{dM}) \, = \, -(1+\chi) \times \mathrm{log}(m) \, + \, constant $. The obtained  $\chi$ values of all the 23 OCs are smaller than the $\chi$ = 1.37 derived by \citet{Salpeter1955} for solar neighbourhood conditions. The flatter slopes of PDMFs are indicative of dissolving OCs which have lost their low mass members due to various disruptive effects, like two-body relaxation, interactions with nearby GMCs, and tidal shocks.

In order to get the average stellar masses of an OC, the total number of stars and total mass are estimated using the following equations
\begin{equation}\label{eq:N_tot}
    N_{\mathrm{tot}} = \int_{M_{\mathrm{min}}}^{M_{\mathrm{max}}} \, \zeta(M) dM,
\end{equation}

\begin{equation}\label{eq:M_tot}
    M_{\mathrm{tot}} = \int_{M_{\mathrm{min}}}^{M_{\mathrm{max}}} \, M\zeta(M) dM
\end{equation}
where $M_{\mathrm{min}}$ is 0.08 M$_{\odot}$ which is the hydrogen burning limit of stars and $M_{\mathrm{max}}$ is the mass of the stars at MSTO point. The masses and numbers of MSTOs, SGBs, RGBs, and RCs are added manually after integrating the equations (\ref{eq:N_tot}) and (\ref{eq:M_tot}). From the slopes of the PDMF, it is evident that all 23 OCs lack low-mass members even in the visible mass range. This implies that a significant number of cluster members fainter than the mass corresponding to G = 18.5 mag are no longer bound to their host cluster. In addition, the purpose of estimating the average mass in this work is to determine the central relaxation times of OCs, which must be calculated for the average conditions in the cluster's central region. Given that the OCs examined in this study are older than 1 Gyr, their central regions must be dominated by massive stars. In Balan et al. (in preparation), it is observed that even in less evolved clusters with $A^+_{\mathrm{rh}}$ values close to zero, low MS stars are significantly less concentrated than high-mass stars. Therefore, despite yielding a slightly larger value for the average stellar mass, it is still worth estimating the central relaxation times of OCs.  The average stellar masses of 23 OCs are listed in Table \ref{tab:dynamical_params}.

\subsection{Dynamical parameters}
We estimate N$_{\text{relax}}$ for 23 OCs that is given as C$_{\text{age}}$/$t_{\text{rc}}$ as in \citetalias{Rao2021}, where $t_{\text{rc}}$ is central relaxation time of a cluster which is given by equation 10 of \citet{Djorgovski1993}
\begin{equation}
    t_{\mathrm{rc}} = 1.491 \times 10^7 yr \times \frac{k}{ln(0.4N_{*})} <m_{*}>^{-1} \rho_{_{\scriptscriptstyle  M,O}}^{1/2} r^3_{c}
\end{equation}
where $k \sim 0.5592$, $r_{\mathrm{c}}$ is the core radius, $<m_{*}>$ is the average mass of the cluster members, $\rho_{_{\scriptscriptstyle M,O}}$ is the central mass density of the cluster, and $N_*$ is the total number of the cluster members. 

\begin{figure}
	\includegraphics[width=0.48\textwidth]{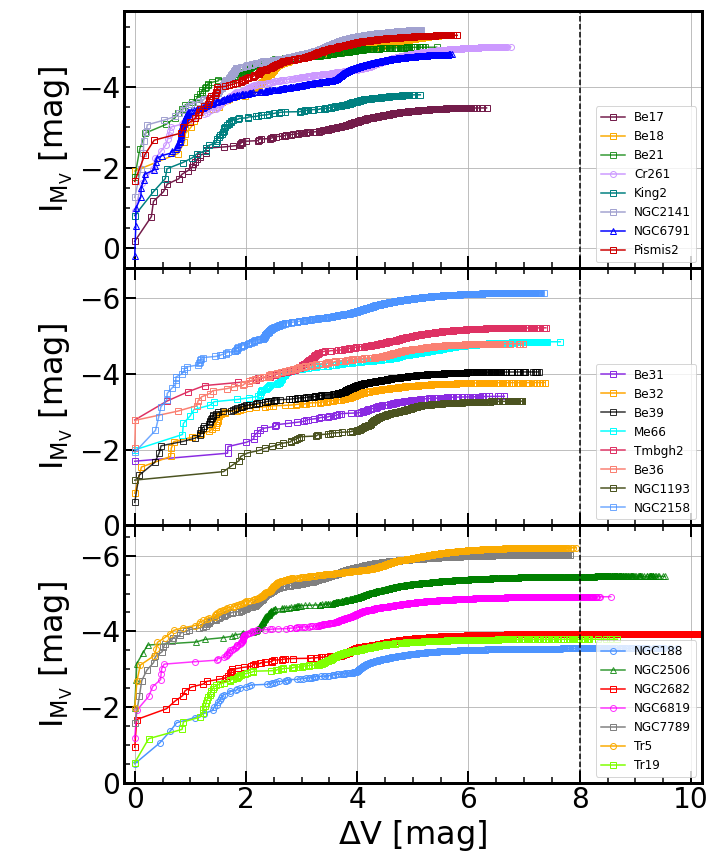}
    \caption{Integrated absolute magnitude profiles for the 23 OCs. $\Delta$V is the magnitude difference between each cluster member and the brightest cluster member. The black dashed line shows the saturation level of the integrated absolute magnitude profiles.}
    \label{fig:unseen_stars}
\end{figure}

In order to calculate $\rho_{_{\scriptscriptstyle M,O}}$, we first calculate the integrated magnitude of the cluster, and then using it we calculate the required quantities. For this, the G magnitudes of cluster members are converted into V magnitudes using the conversion formula given on the Gaia ESO website\footnote{\url{https://gea.esac.esa.int/archive/documentation/GDR3/Data_processing/chap_cu5pho/cu5pho_sec_photSystem/cu5pho_ssec_photRelations.html}}. We then estimate apparent integrated V magnitudes of OCs using the following equation given by \citet{Piskunov2008}.
\begin{equation}
I_{\mathrm{V}} = -2.5 \mathrm{log} \left( \sum_i^{N_i} 10^{-0.4 V_i} + 10^{-0.4 \Delta I_{\mathrm{V}}} \right)
\end{equation}
where $N_i$ and $V_i$ are the numbers and the apparent V magnitude of the cluster members. $\Delta I_{\mathrm{V}}$ is the term proposed to perform unseen stars correction, i.e., to make $I_{\mathrm{V}}$ and $I_{M_{\mathrm{V}}}$ independent of the extent of the stellar magnitudes observed in a cluster. We then convert apparent integrated V magnitudes to absolute integrated V magnitudes using distances and A$_{\text{V}}$ listed in Table \ref{tab:fundamental_params}. Fig. \ref{fig:unseen_stars} shows the absolute integrated magnitude profiles of 23 OCs. We have divided 23 OCs into three categories based on maximum $\Delta V$, i.e., the magnitude difference between their brightest member to the faintest member. From Fig. \ref{fig:unseen_stars}, we see that clusters in the middle and lower panels have a maximum $\Delta V$ reaching up to 7 or more than that, and their absolute integrated magnitude profiles reach a plateau or saturation before the maximum $\Delta V$. Therefore, the faintest members of the OCs in these two categories will have a negligible contribution to their absolute integrated magnitudes, and unseen stars correction is not required. In contrast, the 8 OCs in the upper panel of Fig. \ref{fig:unseen_stars} have the lowest maximum $\Delta V$ and do not reach a plateau in their absolute integrated magnitude profiles, requiring unseen stars correction. To perform this correction, we use NGC 2506 as a template cluster. We calculate the unseen stars correction term by subtracting the absolute integrated magnitude of NGC 2506 at the brightest magnitude of the OC under consideration plus $\Delta V$ = 8 from the absolute integrated magnitude of NGC 2506 at the faintest magnitude of the same OC under consideration. The unseen stars correction term for the 8 OCs ranges between 0.001 mag and 0.01 mag.

Using equations (7), (8), and (9) of \citet{Djorgovski1993} and integrated magnitudes of OCs, we estimate luminosities ($L_\text{V}$) and central luminosity densities ($\rho_{_{\scriptscriptstyle L,O}}$) of OCs. We then use the relation between log(age) and log(M/L$_{\text{V}}$) given by \citet{Piskunov2011} to convert $\rho_{_{\scriptscriptstyle L,O}}$ values into $\rho_{_{\scriptscriptstyle M,O}}$. The estimated structural parameters (\S\ref{sc_param}), $N_*$ and $<m_{*}>$ (as derived in \S\ref{PDMFs}) and $\rho_{_{\scriptscriptstyle M,O}}$  are then plugged into equation (3) to obtain $t_{\text{rc}}$ values for the OCs. In the end, the N$_{\text{relax}}$ of OCs is estimated by dividing the OCs' ages by their $t_{\text{rc}}$ values. The values of $I_{M_{\mathrm{V}}}$, $\rho_{_{\scriptscriptstyle L,O}}$, $\rho_{_{\scriptscriptstyle M,O}}$, and $t_{\text{rc}}$ are listed in Table \ref{tab:dynamical_params} and the values of N$_{\text{relax}}$ are listed in Table \ref{tab:class}.
\input{Tables/eval_stage}
\begin{figure*}
	\includegraphics[width=0.98\textwidth]{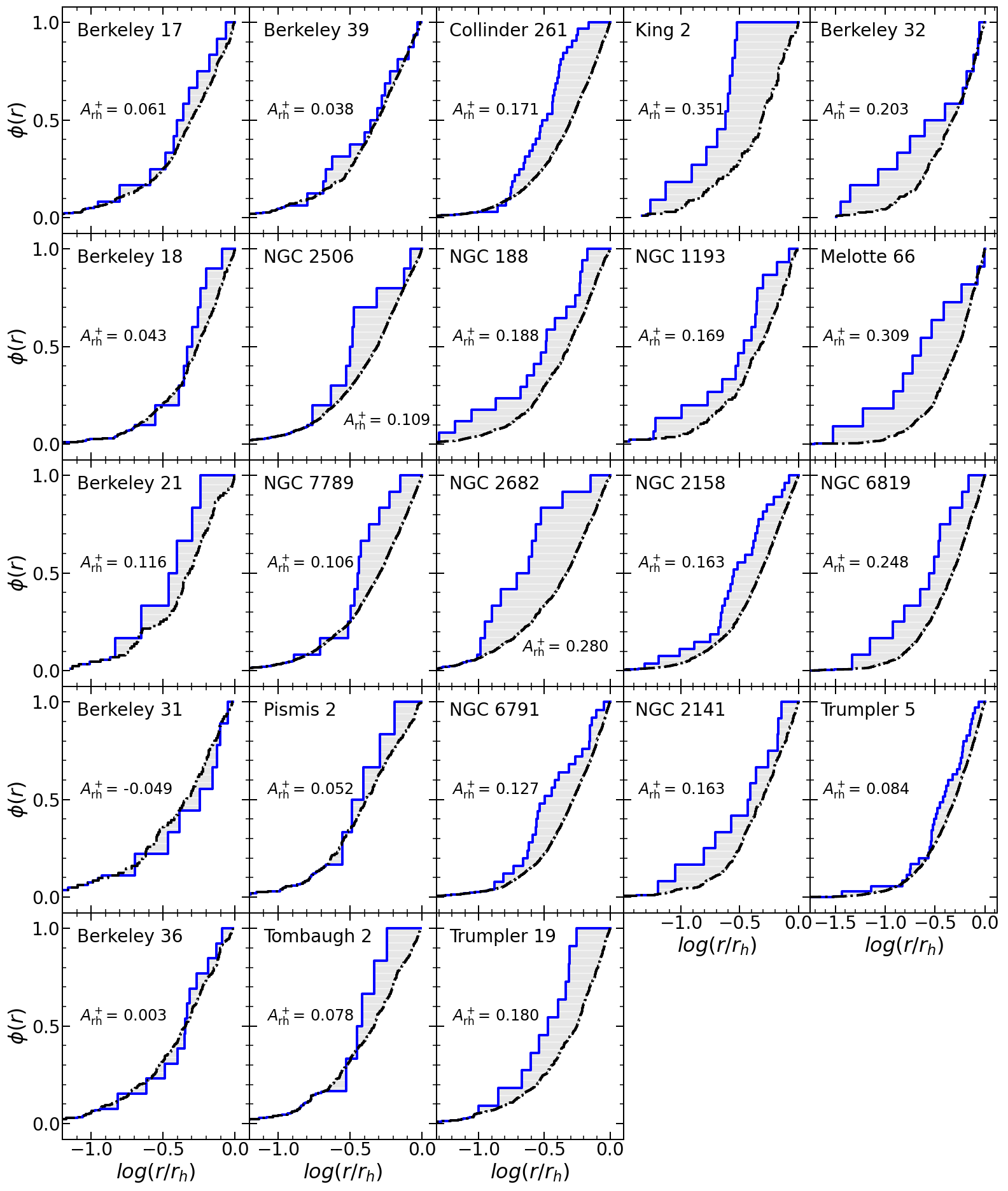}
    \caption{The cumulative radial distributions of the BSS (blue) and the REF population (black), plotted against the logarithm of the radial distance from the cluster center in the units of $r_{\mathrm{h}}$,  for 12 OCs. The values of $A^+_{\mathrm{rh}}$ shown on each plot correspond to the grey-shaded portion between the cumulative radial distributions of the BSS and REF population.}
    \label{fig:crd1}
\end{figure*}
\subsection{Estimation of \texorpdfstring{$A^+$}{A+}}
As described in the \S\ref{sec:intro}, the central area of a cluster is most suitable for exploring mass segregation and hence useful in estimating its dynamical age. Therefore, we estimate $A^+_{\mathrm{rh}}$ up to $r_{\text{h}}$ for OCs. $A^+_{\mathrm{rh}}$ is then estimated using equation (1), where MSTOs and MS stars (G $<$ 18.5 mag) are used as REF. The errors in $A^+_{\mathrm{rh}}$ values ($\epsilon_{A^+_{\text{rh}}}$) for 23 OCs are estimated using the Bootstrap method, where the process of estimating $A^+_{\mathrm{rh}}$ is iterated 1000 times by bootstrapping the BSS and REF samples and the 1$\sigma$ of the $A^+_{\mathrm{rh}}$ distribution is considered as $\epsilon_{A^+_{\text{rh}}}$. Fig. \ref{fig:crd1} shows cumulative radial distributions of BSS and REF of 23 OCs. The values of $A^+_{\mathrm{rh}}$ and $\epsilon_{A^+_{\text{rh}}}$ for 23 OCs are listed in Table \ref{tab:class}. 

%% file: Tables/eval_stage.tex
\begin{table}
	\caption{The dynamical physical parameter, estimated values of $A^+_{\mathrm{rh}}$ and errors in $A^+_{\mathrm{rh}}$  and evolutionary stage of the 23 OCs.}
	\label{tab:class}
	\begin{tabular}{ccccc}
		\hline
		\\
	    Cluster  &  $N_{\mathrm{relax}}$ &  $A^+_{\mathrm{rh}}$ &   Error & Class \\
	      &    &   &   ($\epsilon_{A^+}$) & \\
		 \\
		\hline
		\\
        Berkeley 17 & 108$\pm$12 & 0.061 & 0.064 & II \\ 
        Berkeley 18 & 1.4$\pm$0.5 & 0.043 & 0.042 & I \\ 
        Berkeley 21 & 15$\pm$7 & 0.116 & 0.078 & II \\ 
        Berkeley 31 & 129$\pm$20 & $-$0.049 & 0.085 & -- \\ 
        Berkeley 32 & 131$\pm$14 & 0.203 & 0.138 & II \\ 
        Berkeley 36 & 34$\pm$5 & 0.003 & 0.061 & II \\ 
        Berkeley 39 & 52$\pm$8 & 0.038 & 0.062 & II \\ 
        Collinder 261 & 46$\pm$3 & 0.171 & 0.034 & II \\ 
        King 2 & 285$\pm$15 & 0.351 & 0.076 & III \\ 
        Melotte 66 & 9.1$\pm$1.6 & 0.309 & 0.127 & -- \\ 
        NGC 188 & 145$\pm$11 & 0.188 & 0.076 & II \\ 
        NGC 1193 & 311$\pm$34 & 0.169 & 0.083 & II \\ 
        NGC 2141 & 10.4$\pm$1.7 & 0.163 & 0.088 & II \\ 
        NGC 2158 & 18$\pm$1.1 & 0.163 & 0.054 & II \\ 
        NGC 2506 & 10.4$\pm$0.4 & 0.109 & 0.066 & II \\ 
        NGC 2682 & 142$\pm$11 & 0.28 & 0.073 & III \\ 
        NGC 6791 & 40$\pm$5 & 0.127 & 0.049 & II \\ 
        NGC 6819 & 49$\pm$5 & 0.248 & 0.096 & II \\ 
        NGC 7789 & 4.23$\pm$0.16 & 0.106 & 0.046 & II \\ 
        Pismis 2 & 36$\pm$5 & 0.052 & 0.058 & II \\ 
        Tombaugh 2 & 30.4$\pm$3.5 & 0.078 & 0.047 & II \\ 
        Trumpler 5 & 7.4$\pm$0.6 & 0.084 & 0.046 & II \\ 
        Trumpler 19 & 91$\pm$14 & 0.18 & 0.062 & II \\  		
        \\
		\hline
	\end{tabular}
\end{table}

%% file: chapters/Discussion.tex
\section{Discussion}
\label{sec:discussion}
\subsection{\texorpdfstring{$A^+_{\mathrm{rh}}$}{A+} vs \texorpdfstring{$N_{\mathrm{relax}}$}{Nrelax}}
$A^+_{\mathrm{rh}}$ is an observational parameter that represents the level of BSS segregation in a cluster center and increases as the cluster evolves \citep{Alessandrini2016}. N$_{\text{relax}}$, on the other hand, is estimated based on cluster properties and depicts the degree of mass segregation that the cluster should have attained by this point. Despite the fact that both parameters are independent of one another, they are measuring the same phenomenon, namely the mass segregation effect. As a result, the relationship between $A^+_{\mathrm{rh}}$ and N$_{\text{relax}}$ has been investigated in the literature to see if BSS can be used as a probe to estimate the dynamical ages of clusters \citep{Lanzoni2016,Ferraro2018,Ferraro2019,Ferraro2020,Rao2021,Cadelano2022,Dresbach2022,Beccari2023}. \citet{Ferraro2018} demonstrated that $A^+_{\mathrm{rh}}$ does, in fact, represent a measure of the dynamical ages of GCs.

\begin{figure}
	\includegraphics[width=0.48\textwidth]{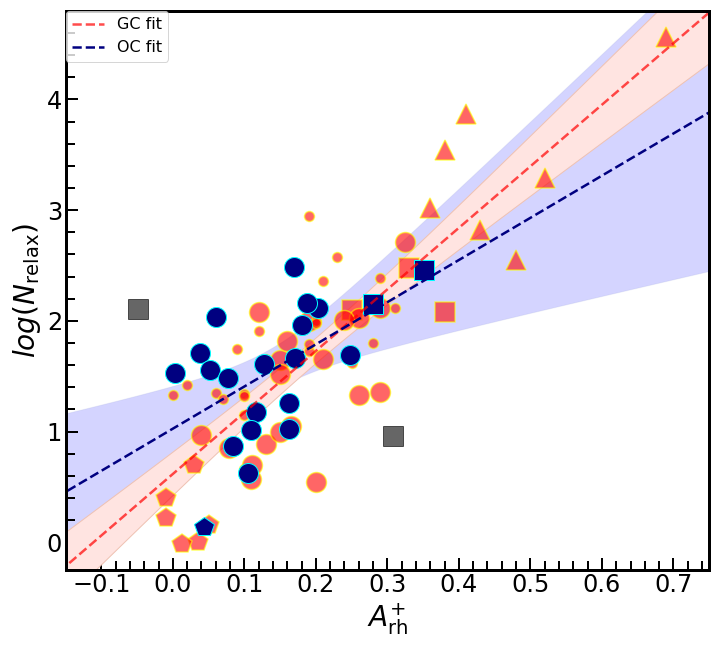}
        \caption{The correlation between the values of $A^+_{\mathrm{rh}}$ and the number of current central relaxation, $N_{\mathrm{relax}}$, for 21 OCs (blue) and 58 GCs \citep[red:][]{Ferraro2018,Ferraro2019,Cadelano2022,Dresbach2022,Beccari2023}. The black filled square shows Melotte 66 and Berkeley 31 OCs. Triangles represent core-collapsed GCs, while filled squares, circles, and pentagons represent Class III, II, and I clusters, respectively (see \ref{sec:discussion} section). The red dots represent GCs that are not classified individually in any dynamical stage. The blue dashed line represents the best-fitted line for the 21 OCs excluding Melotte 66 and Berkeley 31, while the red dashed line represents the best-fitted line for the GCs \citep{Ferraro2018}. The blue and red shaded regions represent errors in fitted correlations within the 95 per cent confidence interval for OCs and GCs, respectively.}
        \label{fig:A_plus}
\end{figure}

In the present work, we estimated $A^+_{\mathrm{rh}}$ and $N_{\text{relax}}$ for 23 OCs, including 11 OCs of \citetalias{Rao2021}. Fig. \ref{fig:A_plus} depicts the correlation of $A^+_{\mathrm{rh}}$ against log($N_{\text{relax}}$) for OCs, with GCs data points taken from \citet{Ferraro2018} are also included for comparison. Fig. \ref{fig:A_plus} shows that Melotte 66 and Berkeley 31 (black squares) are  outliers in the plot. Melotte 66 has the highest value of $A^+_{\mathrm{rh}}$, but in comparison, the value of $N_{\text{relax}}$ is quite low (see \S\ref{Me66} for details). Berkeley 31 has the lowest value of $A^+_{\mathrm{rh}}$, while its $N_{\text{relax}}$ is quite high (see \S\ref{Be31} for details). Therefore, we exclude Melotte 66 and Berkeley 31 from the fit and compute the best-fit relation for the remaining 21 OCs plotted in Fig. \ref{fig:A_plus}, which is given as
\begin{equation}\label{eq:ext_A+}
	\mathrm{log}(N_{\mathrm{relax}}) = 3.8(\pm 1.3) \times A^{+}_{\mathrm{rh}}+1.03(\pm 0.22)
\end{equation}
Whereas the best-fit relation for 58 GCs  \citep{Ferraro2018,Ferraro2019,Cadelano2022,Dresbach2022,Beccari2023}
\begin{equation}\label{eq:GC_A+}
	\mathrm{log}(N_{\mathrm{relax}}) = 5.6(\pm 0.5) \times A^{+}_{\mathrm{rh}}+0.62(\pm 0.12)
\end{equation} 
From equations (\ref{eq:ext_A+}) and (\ref{eq:GC_A+}), we observe that the slope of the best-fit relation between $A^+_{\mathrm{rh}}$ and $N_{\text{relax}}$ for OCs is smaller than GC as well as errors in the slope and intercept values are still large.  For reference, when compared with the correlation found with 11 OCs in \citetalias{Rao2021}, given as
\begin{equation}
	\mathrm{log}(N_{\mathrm{relax}}) = 4.0(\pm 2.1) \times A^{+}_{\mathrm{rh}}+1.42(\pm 0.30)
\end{equation}

\input{Tables/cocor_tool}

By almost doubling our sample size of the OCs, we still obtain a value of slope quite close to our previous result, but with a smaller error in the fitting. The slope for the OC is similar to that known for the GCs within the errors. According to Fig.  \ref{fig:A_plus}, OCs inhabit the same parameter space as dynamically young to intermediate dynamical age GCs of the advanced evolutionary phases, with the highest concentration of OCs  observed in the middle and bottom portions of the plot. Despite being significantly younger than GCs, OCs have reached the same level of central relaxation as GCs of advanced intermediate dynamical age.

In order to estimate the strength of the relationship between $A^+_{\mathrm{rh}}$ and $N_{\text{relax}}$ for OCs and GCs, we calculated Pearson and Spearman rank correlation coefficients for OCs and GCs of young to intermediate dynamical age, which is listed in Table \ref{tab:cocor_tool}. As the sample sizes of OCs and GCs are different, we compare the correlation coefficients using the COCOR tool\footnote{\url{http://comparingcorrelations.org/}}, which uses the Fisher test \citep{fisher1992} and the Zou test \citep{zou2007}. The results of the comparison listed in Table \ref{tab:cocor_tool} show that estimated correlation coefficients for GCs and OCs are not different, but the errors for OCs are quite large.

We also calculated $A^+_{\mathrm{rh}}$ using all cluster populations except BSS, such as SGBs, RGBs, RCs, MSTOs, and MS stars, as REF and plotted relation of $A^+_{\mathrm{rh}}$ against $N_{\text{relax}}$ as shown in Fig. \ref{fig:A_plus_all}.

\subsection{\texorpdfstring{$A^+_{\mathrm{rh}}$}{A+} vs physical parameters}

\begin{figure}
	\includegraphics[width=0.48\textwidth]{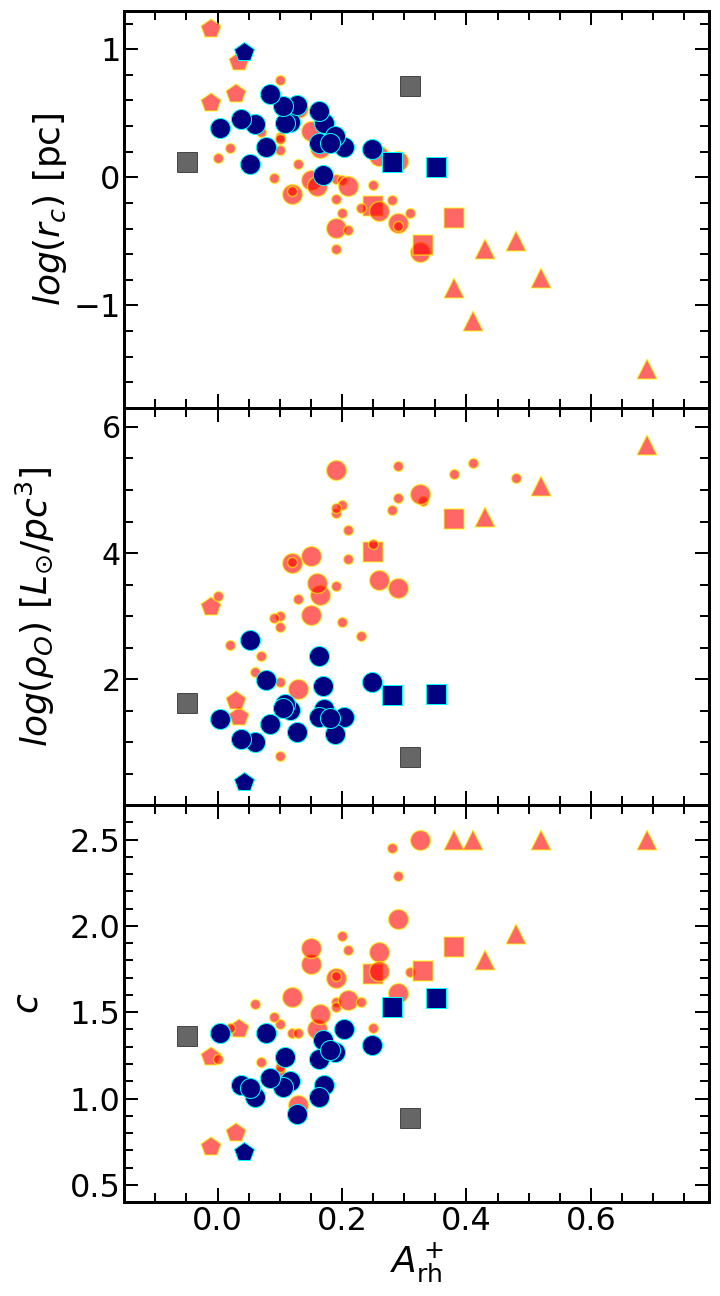}
        \caption{Relationship between the physical parameters of clusters, including their core radius (upper panel), central luminosity density (middle parameter), and concentration parameter (c), for 22 OCs (blue) and GCs (red). The different symbols have the same meanings as in Fig. \ref{fig:A_plus}. The black-filled square shows Melotte 66 OC. We utilized the values of $A^+_{\mathrm{rh}}$ and physical parameters of GCs from \citet{Lanzoni2016}, \citet{Ferraro2018}, and the references therein.}
    \label{fig:Struct_param}
\end{figure}

The clusters which are in the long-term, post-gaseous phase of evolution, as they evolve, their physical parameters like $r_{\mathrm{c}}$, central luminosity density, and $c$ are also expected to change. With the evolution of a cluster in this phase, the values of central luminosity density and $c$ increase, while $r_{\mathrm{c}}$ decreases. We, therefore, plot $A^+_{\mathrm{rh}}$ vs $r_{\mathrm{c}}$ (upper panel), central luminosity densities (middle panel), and $c$ (lower panel) of OCs in Fig. \ref{fig:Struct_param}, where we also show GCs, to see how these parameters correlate with $A^+_{\mathrm{rh}}$. Melotte 66 and Berkeley 31 are outliers in each plot of Fig. \ref{fig:Struct_param}, just as it is in the $A^+_{\mathrm{rh}}$ vs $N_{\text{relax}}$ plot. Therefore, we exclude these from the calculation of statistical parameters. Similar to Fig. \ref{fig:A_plus}, Fig. \ref{fig:Struct_param} demonstrates that their $r_{\mathrm{c}}$ and $c$ values also fall into the category of young to intermediate age GCs, whereas their central luminosity densities fall into the category of less evolved GCs. As shown in Fig. \ref{fig:Struct_param}, for a given $A^+_{\mathrm{rh}}$ value, OCs have systematically lower $c$ values than GCs (bottom panel), but slightly higher $r_{\mathrm{c}}$ values (top panel) and nearly constant luminosity density (middle). As OCs have less dense cores, this results in lower $c$ values,  and if the luminosity density remains constant, the less dense core would require slightly larger $r_{\mathrm{c}}$. The result shown in Fig. \ref{fig:Struct_param} is the consequence of OCs being sparser than GCs.

Consequently, even though OCs have achieved the same level of relaxation as GCs in their core, they are less likely to achieve the same level as GCs in $A^+_{\mathrm{rh}}$ vs physical parameter spaces. Therefore, we compare OCs to less evolved GCs here. We calculated the Pearson and Spearman rank correlation coefficients for OCs and less advanced GCs and compared them using the COCOR tool. The estimated values are noted in Table \ref{tab:cocor_tool}.

\subsection{Evolutionary stages of OCs} 
We classify OCs into three different classes of dynamical evolution by comparing their $A^+_{\mathrm{rh}}$ and $N_{\text{relax}}$ values to those of previously classified GCs into different dynamical stages by \citet{Ferraro2012} and \citet{Ferraro2018}: (i) Class I represents the least evolved OCs (filled-pentagons in Fig. \ref{fig:A_plus}) that belong to the Family I GCs (dynamically young); (ii) Class II represents the intermediate dynamical age I OCs (filled-circles in Fig. \ref{fig:A_plus}) that belong to the Family II GCs (intermediate dynamical age); (iii) Class III represents the intermediate dynamical age II OCs (filled-squares in Fig. \ref{fig:A_plus}) that are in advanced stages of their evolution and belong to the Family III GCs (evolved). Additionally, the GCs which have undergone the core-collapsed phase are shown as red triangles. Although there is some overlap between $A^+_{\mathrm{rh}}$ and $N_{\text{relax}}$ values of Family II (red filled circles) and Family III GCs (red filled squares), the Family II GCs (M92 and NGC 6752) that coincide with Family III GCs are in advanced stages of intermediate-age dynamical evolution, with the majority of the BSS segregated in the cluster center \citep{Ferraro2012}. The evolutionary classes of the 21 OCs except for Melotte 66 and Berkeley 31 are listed in Table \ref{tab:class}.

\textbf{Class I:} Only one OC, Berkeley 18, fall into this class. Berkeley 18 has the lowest $N_{\text{relax}}$ value among 23 OCs, and its $A+_{\mathrm{rh}}$ value is consistent with its $N_{\text{relax}}$ value. It consistently occupies the region of the least evolved cluster in all plots, as shown in Figs. \ref{fig:A_plus} and \ref{fig:Struct_param}.

\textbf{Class II:} Of 23 OCs, 18 OCs fall into this class of evolutionary stage. These OCs have a large spread in their values of physical and dynamical parameters. Additionally, as can be seen from Fig. \ref{fig:A_plus}, the maximum number of GCs also fall into this category. As these OCs have a large range in their physical and dynamical parameters, being classified in just one class does not necessarily imply that they all are at just one fixed dynamical age, but rather in different stages of evolution. The mass segregation is active in these OCs with some being less evolved while some of the clusters reached quite far in their evolution, like Berkeley 32, Collinder 261, NGC 188, NGC 1193, NGC 6819, and Trumpler 19.

Berkeley 17 has previously been identified as a cluster with tidal tails \citep{Bhattacharya2017}, double BSS sequences \citep{Rao2023}, and of intermediate dynamical age \citep{Bhattacharya2019,Rao2021}. Our current findings about Berkeley 17 are consistent with previous identifications in the literature. \citet{Vaidya2020} classified Berkeley 39 as a dynamically young cluster based on its flat BSS radial distribution. However, as its $N_{\text{relax}}$ was large, they speculated that it could be an evolved cluster. Here, we classify it as an intermediate dynamical age cluster, which is consistent with \citetalias{Rao2021}. Collinder 261 was previously identified as having a flat BSS radial distribution \citep{Rain2020a}, but in \citetalias{Rao2021}, we determined it as an intermediate dynamical age, which is consistent with our current findings. Trumpler 5 was previously categorized as the OC that had undergone the least amount of evolution in \citetalias{Rao2021}, however, according to our current findings, it falls in the category of intermediate dynamical age. \citet{Vaidya2020} categorized NGC 6819 as a young cluster due to its flat BSS radial distribution, but based on its cumulative radial distribution, they hypothesized that it may in fact be an evolved OC. We classified it as an intermediate dynamical age cluster in \citetalias{Rao2021}, which is consistent with the current findings. The middle panel of Fig. \ref{fig:Struct_param} shows that Pismis 2 has the highest central luminosity density among 23 OCs, however, its $A^+_{\mathrm{rh}}$ value is inconsistent with it. Additionally, it is located the same parameter space as less evolved GCs in  Fig. \ref{fig:A_plus}.

\textbf{Class III:} Of 23 OCs, 2 OCs, King 2 and NGC 2682, are classified into this category. These OCs have the highest values of $N_{\text{relax}}$ which are consistent with their $A^+_{\mathrm{rh}}$ values. According to previous works in literature, NGC 2682 is identified as a dynamically evolved cluster based on signatures of extra-tidal sources and mass segregation \citep{Fan19968,Bonatto2003,Carrera2019}. Therefore, we classify NGC 2682 as a Class III OC that is in an advanced evolutionary stage. 

    \textbf{Melotte 66:} \label{Me66} It is one of our target OCs with highly segregated BSS in the cluster core. On the contrary, its dynamical and physical parameters are comparable to those of Class II OCs. Based on the radial distribution of BSS, \citet{Vaidya2020} has shown that the cluster is of intermediate dynamical age. \citet{Carraro2014} investigated this cluster in the context of multiple stellar populations using photometric and spectroscopic data. They observed that its BSS are highly segregated in the cluster center when compared to MS stars and clump stars. In addition, they determined that it has a binary fraction of $\ge$30 per cent. On the basis of the highly segregated BSS and the binary fraction, they speculated that its BSS must have been formed through the process of binary evolution, in which the primordial binaries sank into the cluster's core prior to the BSS formation. \citet{Rao2022} investigated the formation channels of hot stellar populations of Melotte 66, including 14 BSS, by constructing spectral energy distributions from multi-wavelength photometric data. Out of the 14 BSS, they discovered two binary systems. One BSS is part of an eclipsing binary system discovered using a light curve from \textit{TESS} data \citep{Ricker2015}, while another has a low-mass white dwarf companion discovered using excess flux in ultraviolet data. However, their study was based on the near-ultraviolet data available for Melotte 66 from the \textit{Swift}/UVOT telescope \citep{Roming2005}, it is possible that only the hot companion of the recently formed BSS has been detected, while others remain undiscovered. Therefore, this cluster must be thoroughly analyzed to determine the formation mechanisms of its BSS population.

\textbf{Berkeley 31:} \label{Be31} In contrast to Melotte 66, Berkeley 31 has the lowest value of $A^+_{\mathrm{rh}}$ among all the OCs examined in this work, while its dynamical and physical parameters fall within the range of Class II OCs.  It is interesting to note that even though it has a noticeably smaller number of cluster members than the other 22 OCs, it still has a reasonable number of BSS. The reason behind its unusual nature is unclear. The cluster and its BSS must be analysed in detail to unravel this peculiar nature.

%% file: Tables/cocor_tool.tex
\begin{table*}
	\caption{The COCOR tool results of OCs and GCs for the correlation between their $A^+_{\mathrm{rh}}$ vs dynamical and physical parameters. Column 1 gives fitted correlations, column 2 gives name of the correlation coefficients where P refers to the Pearson correlation coefficient and S refers to the Spearman rank correlation coefficient, columns 3 and 4 list the calculated correlation coefficients for the OCs and GCs, column 5 and 6 give the results obtained by employing the statistical tests to compare the correlation coefficients of the OCs and GCs using the COCOR tool, the last column denotes whether the null hypothesis is or is not rejected, with a $\checkmark$ sign implying that the null hypothesis is not rejected.}
	\label{tab:cocor_tool}
	\begin{tabular}{ccccccc}
		\hline
		\\
		Correlation  & \begin{tabular}{c} Correlation \\ coefficient \end{tabular} & OCs & GCs & \begin{tabular}{c} p-value \\ (Fisher test) \end{tabular} & \begin{tabular}{c} CI$^{\star}$ \\ (Zou test) \end{tabular} & COCOR tool result \\
		 \\
		\hline
		\\
		  $A^+_{\mathrm{rh}}$ vs log($N_{\mathrm{relax}}$) & \begin{tabular}{c} P \\ S \end{tabular} & \begin{tabular}{c} $+$0.544 \\ $+$0.571 \end{tabular} & \begin{tabular}{c} $+$0.713 \\ $+$0.739 \end{tabular} & \begin{tabular}{c}   0.3014 \\   0.2776 \end{tabular} &  \begin{tabular}{c}   $-$0.5832 -- $+$0.1276 \\   $-$0.5676 -- $+$0.1123 \end{tabular} &  \begin{tabular}{c} $\checkmark$ \\ $\checkmark$ \end{tabular} \\
		\\
		$A^+_{\mathrm{rh}}$ vs log($r_{\mathrm{c}}$) & \begin{tabular}{c} P \\ S \end{tabular} & \begin{tabular}{c} $-$0.536 \\ $-$0.487  \end{tabular} & \begin{tabular}{c} $-$0.638 \\ $-$0.657 \end{tabular} &  \begin{tabular}{c}   0.5986 \\   0.3933 \end{tabular} &  \begin{tabular}{c}   $-$0.2586 -- $+$0.5361 \\   $-$0.2014 -- $+$0.6158 \end{tabular} &  \begin{tabular}{c} $\checkmark$ \\ $\checkmark$ \end{tabular} \\
		\\
		$A^+_{\mathrm{rh}}$ vs log(${\rho_{_{\scriptscriptstyle L,O}}}$) & \begin{tabular}{c} P \\ S \end{tabular} & \begin{tabular}{c} $+$0.279 \\ $+$0.317 \end{tabular} & \begin{tabular}{c} $+$0.411   \\ $+$0.270 \end{tabular} & \begin{tabular}{c}  0.6783 \\   0.8875 \end{tabular} &  \begin{tabular}{c} $-$0.7000 -- $+$0.4949 \\   $-$0.5585 -- $+$0.6781 \end{tabular} &  \begin{tabular}{c} $\checkmark$ \\ $\checkmark$ \end{tabular}\\
		\\
		$A^+_{\mathrm{rh}}$ vs $c$ & \begin{tabular}{c} P \\ S \end{tabular} & \begin{tabular}{c} $+$0.605 \\ $+$0.522 \end{tabular} & \begin{tabular}{c} $+$0.688 \\ $+$0.701 \end{tabular} & \begin{tabular}{c}   0.6315 \\   0.3332 \end{tabular} &  \begin{tabular}{c}   $-$0.4813 -- $+$0.2383 \\   $-$0.6070 -- $+$0.1663 \end{tabular} &  \begin{tabular}{c} $\checkmark$ \\ $\checkmark$ \end{tabular}\\
		\\
		\hline
	\end{tabular}
	\begin{tablenotes}
		\item {$^{\star}$Confidence interval}
	\end{tablenotes}
\end{table*}

%% file: chapters/Final_remarks.tex
\section{Summary and Conclusion}
\label{sec:Final_remarks}
We estimated the dynamical ages of 23 OCs in this study using relationships between $A^+_{\mathrm{rh}}$, an observational marker of mass segregation \citep{Alessandrini2016,Ferraro2018,Rao2021}, and $N_{\text{relax}}$, a theoretical parameter indicating the expected mass segregation a cluster should have undergone by now \citep{Djorgovski1993}. In this work, we doubled the sample size (OCs) from \citetalias{Rao2021}. We have identified members of 23 OCs using ML-MOC on \textit{Gaia} EDR3 data. These OCs have ages ranging from 1.6 to 9 Gyr and distances ranging from 0.84 to 8.5 kpc. We calculated $A^+_{\mathrm{rh}}$ of these clusters, which is a representation of the degree of BSS sedimentation in the core of clusters, using MSTO and MS stars above G = 18.5 mag as a REF. We calculated the physical and dynamical parameters of 23 OCs. There are three following differences between this paper and \citetalias{Rao2021}: (i) The members of the 11 OCs used in \citetalias{Rao2021} are based on \textit{Gaia} DR2 data, and the membership identification procedure is also different, so the number of BSS and members in \citetalias{Rao2021} differs from the current one. (ii) In \citetalias{Rao2021}, we estimated $A^+_{\mathrm{rh}}$ using MSTOs, SGBs, RGBs, and RCs as REF, whereas in the current work, we are using MSTOs and MS stars as REF. (iii) In \citetalias{Rao2021}, we limited our members of OCs to G = 17 mag due to the completeness of \textit{Gaia} DR2 data, whereas in the current work, we are going down to G = 18.5 mag. Thus, \citetalias{Rao2021} and this work estimate different physical and dynamical parameters and $A^+_{\mathrm{rh}}$ values.

We plot $A^+_{\mathrm{rh}}$ against $N_{\text{relax}}$ and physical parameters like $r_{\mathrm{c}}$, $c$, and central luminosity density for 23 OCs. We also show GC data points from \citep{Ferraro2018} to investigate the differences and similarities between OCs and GCs across these parameter spaces. In this study, we found that the relationship between $A^+_{\mathrm{rh}}$ and $N_{\text{relax}}$ is still broad, despite increasing the sample size by up to twice that of \citetalias{Rao2021}. We categorize OCs into three dynamical age groups based on the relationship between $A^+_{\mathrm{rh}}$ and $N_{\text{relax}}$. Class I OCs are the least evolved OCs. Berkeley 18 has been discovered to be the least evolved OC among 23 OCs and GCs. Class II OCs are intermediate dynamical age I OCs in which dynamical friction has begun sinking BSS in the cluster core but is only effective near the cluster core, making them among the less evolved GCs. Class III OCs are intermediate dynamical age II OCs that are in the advanced stages of evolution and are located in the middle of the $A^+_{\mathrm{rh}}$ vs $N_{\text{relax}}$ plot. NGC 2682 and King 2 are the most evolved of the 23 OCs studied in this work. The BSS of Melotte 66 are highly segregated in the cluster core, resulting in the highest value of $A^+_{\mathrm{rh}}$ among the 23 OCs. Its dynamical and physical parameters, on the other hand, are quite low, making it an outlier in all plots of $A^+_{\mathrm{rh}}$ vs physical and dynamical parameters. In contrast to this, BSS of Berkeley 31 are the least segregated in the cluster center, yielding the lowest value of $A^+_{\mathrm{rh}}$. While its physical and dynamical parameters are not consistent with its $A^+_{\mathrm{rh}}$. One must conduct a detailed analysis to investigate the possible cause of ambiguities in these clusters.

%% file: chapters/Acknowledgements.tex
\section*{Acknowledgements}
We thank the referee for their comments that helped improve the manuscript. SB is funded by the INSPIRE Faculty Award (DST/INSPIRE/04/2020/002224), Department of Science and Technology (DST), Government of India. This work has made use of the early third data release from the European Space Agency (ESA) mission {\it Gaia} (\url{https://www.cosmos.esa.int/gaia}), Gaia EDR3 \citep{Gaiaedr32021}, processed by the {\it Gaia} Data Processing and Analysis Consortium (DPAC, \url{https://www.cosmos.esa.int/web/gaia/dpac/consortium}). Funding for the DPAC has been provided by national institutions, in particular, the institutions participating in the {\it Gaia} Multilateral Agreement. This research has made use of the VizieR catalog access tool, CDS, Strasbourg, France. This research made use of {\small ASTROPY}, a {\small PYTHON} package for astronomy \citep{Astropy2013}, {\small NUMPY} \citep{Harris2020}, {\small MATPLOTLIB} \citep{Hunter4160265}, {\small SCIPY} \citep{Virtanen2020}. This research also made use of the Astrophysics Data System (ADS) governed by NASA (\url{https://ui.adsabs.harvard.edu}).

%% file: chapters/Appendix.tex
\appendix
\label{sec:appendix}
\section{Differential reddening correction of Trumpler 5}
\begin{figure*}
    \centering
	\begin{subfigure}[b]{0.48\textwidth}
    		\includegraphics[width=1.0\textwidth]{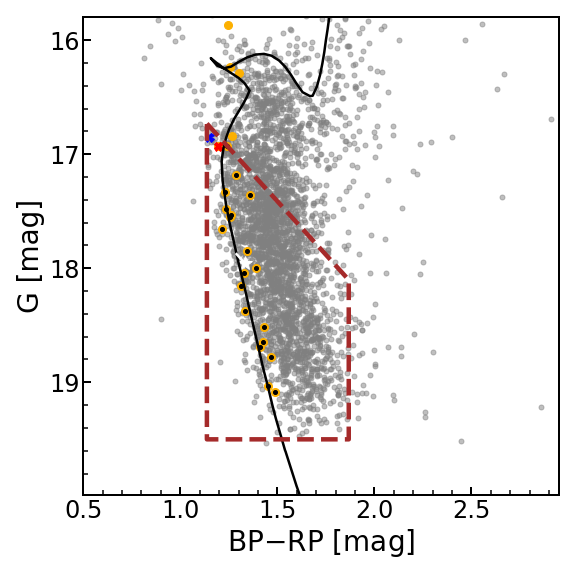}
		\caption*{}
	\end{subfigure}
	\vspace{-0.8cm}
	\begin{subfigure}[b]{0.48\textwidth}
   		\includegraphics[width=1.0\textwidth]{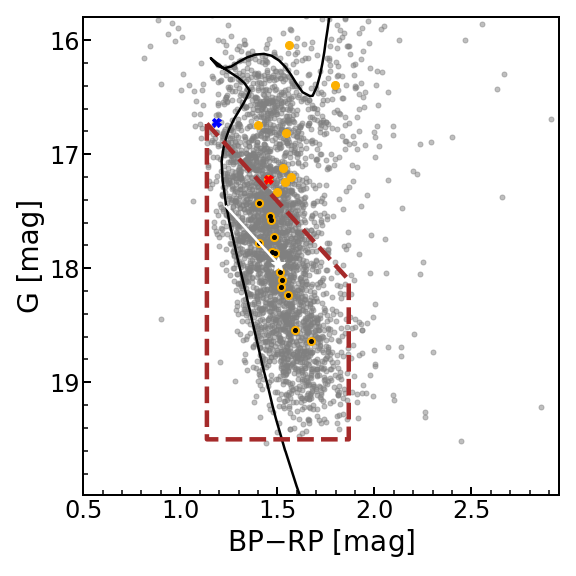}
		\caption*{}
	\end{subfigure}
	\caption{\textit{Gaia} EDR3 CMDs of Trumpler 5 zoomed in the MS region (grey dots) with plotted PARSEC isochrones (black solid lines). The selection box for the cluster members used to compute the differential reddening correction is indicated by the brown dashed lines (which are the same in both panels). The orange dots represent the 25 cluster members closest to the one we want to correct for differential reddening correction and extinction. Black dots show cluster members located inside the brown selection box which are used to compute differential reddening correction and extinction. The red and blue crosses represent the observed and reddening corrected positions of the cluster member for which reddening correction is being performed. The mean color and magnitudes of the stars in the MS grid (black dots) are denoted by a white star. The color excess of the cluster members is estimated by measuring the shift required to project this mean point onto the plotted PARSEC isochrone along the reddening vector (white solid line). The members least affected by reddening are shown on the left panel which nicely follow the plotted isochrone. The members highly affected by reddening are shown on the right panel.}
	\label{fig:DR_steps}
\end{figure*}

\begin{figure*}
    \centering
	\begin{subfigure}[b]{0.34\textwidth}
    		\includegraphics[width=1.0\textwidth]{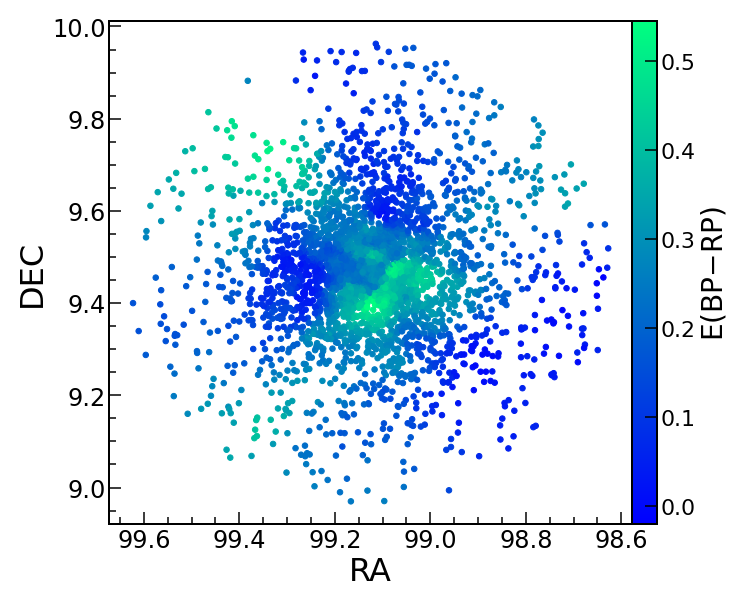}
		\caption*{}
	\end{subfigure}
		\vspace{-0.8cm}
	\begin{subfigure}[b]{0.31\textwidth}
   		\includegraphics[width=1.0\textwidth]{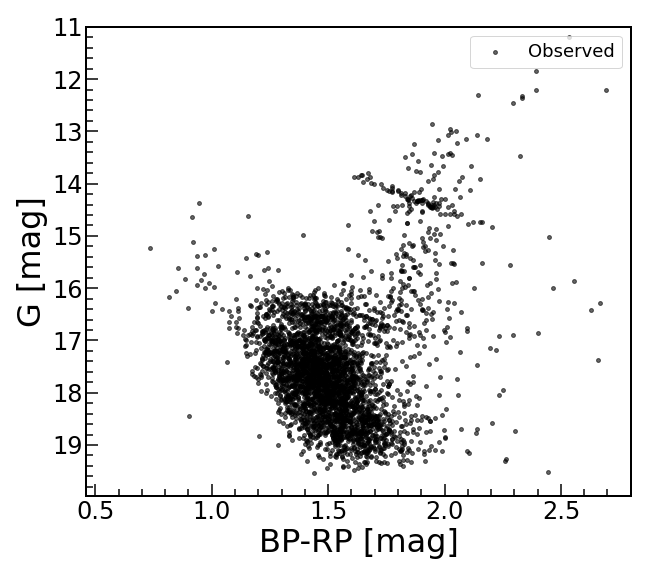}
		\caption*{}
	\end{subfigure}
	\begin{subfigure}[b]{0.31\textwidth}
   		\includegraphics[width=1.0\textwidth]{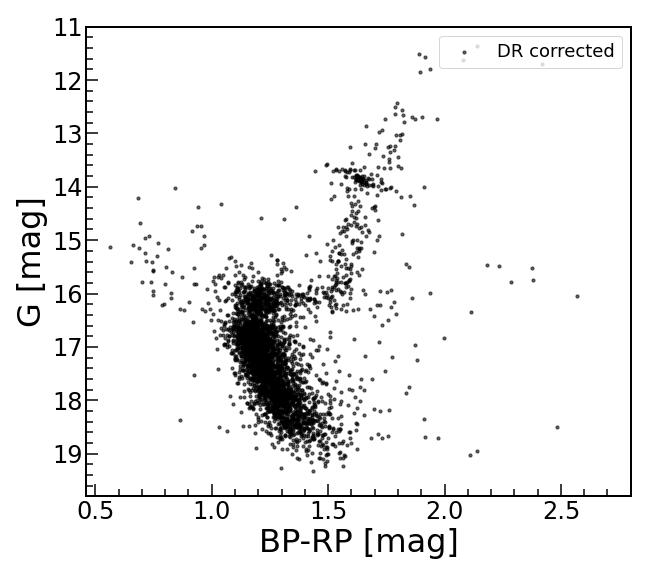}
		\caption*{}
	\end{subfigure}
	\caption{The left panel shows the reddening map of Trumpler 5, the middle panel shows observed CMD and the right panel shows reddening corrected CMD.}
	\label{fig:DR}
\end{figure*}

We follow the approach outlined in \citet{Massari2012} to carry out differential reddening correction for members of Trumpler 5. We start by plotting a CMD of the cluster members. The isochrone is then plotted to the observed cluster CMD with the following parameters: age = 2.0 Gyr, distance = 4150 pc, [M/H] = $-$0.40, and A$_{\text{V}}$ = 1.9. This isochrone serves as a reference line for reddening correction. We convert A$_{\text{V}}$ to A$_{\text{G}}$ and E(BP$-$RP) using the relation given in \citet{Wang2019}, and then we use this information to calculate the reddening vector, R$_{\text{G}}$ = A$_{\text{G}}$/E(BP$-$RP), as 1.875. The reddening vector's direction perfectly corresponds to the RCs' distortion's direction. To perform reddening correction, we only use the MS stars. This estimated reddening vector is used to create a grid on MS. The top border of the grid is made of the calculated reddening vector, while the bottom is limited to Trumpler 5's maximum G magnitude. As shown in Fig. \ref{fig:DR_steps}, the right and left borders are chosen so that they enclose the MS. We choose one Trumpler 5 member, apply the kNN algorithm to find the 25 members that are closest to it in the cluster's spatial distribution, and over-plot the CMD for those 25 members on the CMD of Trumpler 5. The number 25 was chosen so that there would be at least four sources in the MS grid (see Fig. \ref{fig:DR_steps}). Following that, we calculate the mean color ($\langle BP-RP \rangle$) and magnitude ($\langle G \rangle$) of the stars enclosed within the MS grid. In order to estimate the amount of reddening and correction, we calculate the excess in color and magnitude of the members by measuring the shift required to match the position of the mean point ($\langle BP-RP \rangle$, $\langle G \rangle$) to the plotted isochrone along the direction of the reddening vector. We use these estimated values to correct the cluster member for reddening and extinction. We similarly follow this process for each member of Trumpler 5 to get their corrected G mag and BP$-$RP mag. The reddening map of Trumpler 5 is plotted in the left panel of Fig. \ref{fig:DR}, the observed CMD is plotted in the middle panel of Fig. \ref{fig:DR}, and the corrected CMD is plotted in the right panel of Fig. \ref{fig:DR}. In comparison to the uncorrected CMD, the corrected CMD appears much thinner, with each evolutionary sequence clearly visible. One thing to note here is that the resolution of the reddening map decreases as we move away from the cluster center.
\section{Additional figures}
Here, we show CMDs with fitted isochrones and members classified into different population as described in \S\ref{population_selections}, observed surface density profiles fitted with isotropic single-mass king models \citep{King1966} as described in \S\ref{sc_param}, and derived mass function as described in \S\ref{PDMFs} for 23 OCs.

\begin{figure*}
    \centering
	\begin{subfigure}[b]{0.32\textwidth}
    		\includegraphics[width=0.99\textwidth]{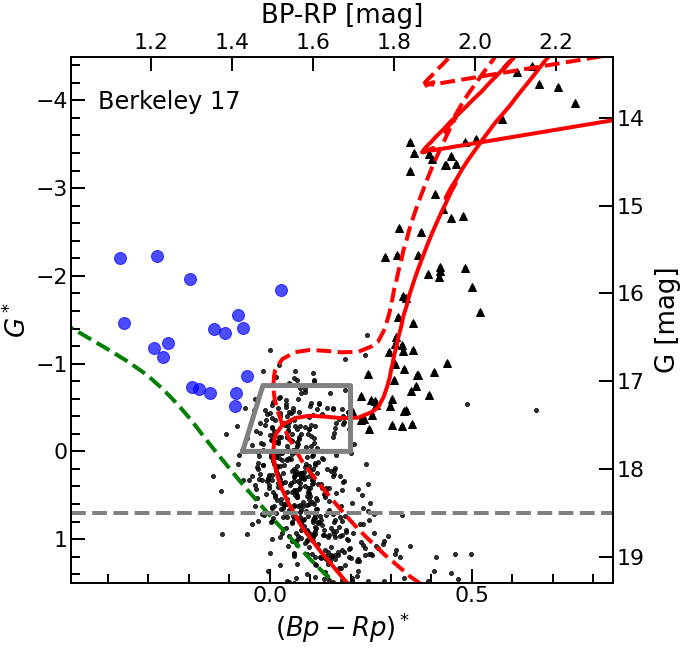}
		\caption*{}
	\end{subfigure}
	\begin{subfigure}[b]{0.32\textwidth}
   		\includegraphics[width=0.99\textwidth]{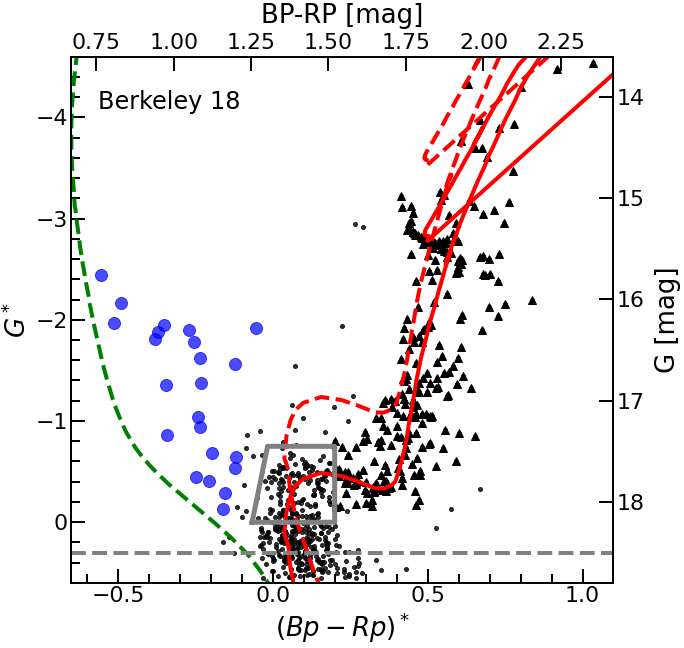}
		\caption*{}
	\end{subfigure}
	\vspace{-0.5cm}
	\begin{subfigure}[b]{0.32\textwidth}
   		\includegraphics[width=0.99\textwidth]{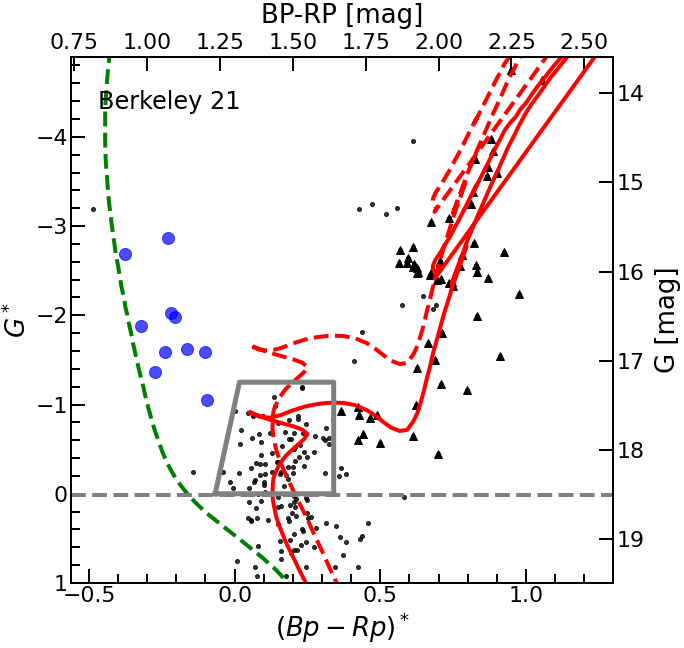}
		\caption*{}
	\end{subfigure}
	\vspace{-0.5cm}
	\begin{subfigure}[b]{0.32\textwidth}
   		\includegraphics[width=0.99\textwidth]{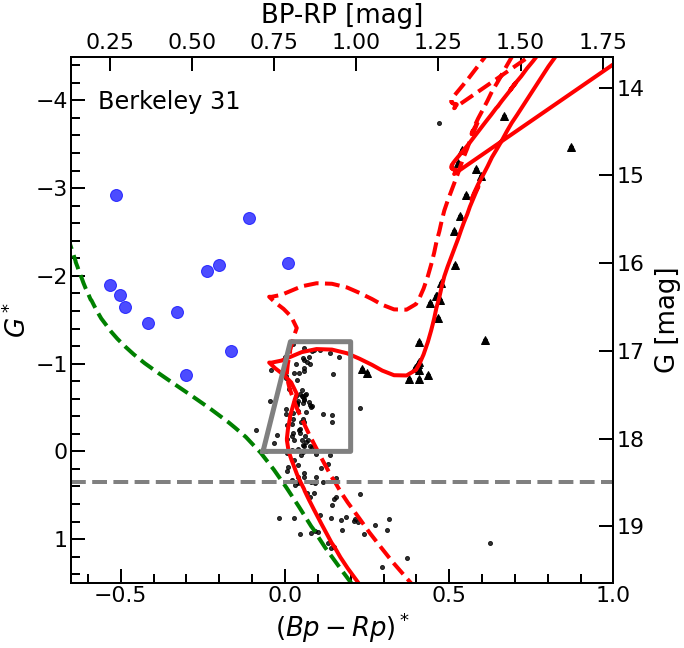}
		\caption*{}
	\end{subfigure}
	\begin{subfigure}[b]{0.32\textwidth}
   		\includegraphics[width=0.99\textwidth]{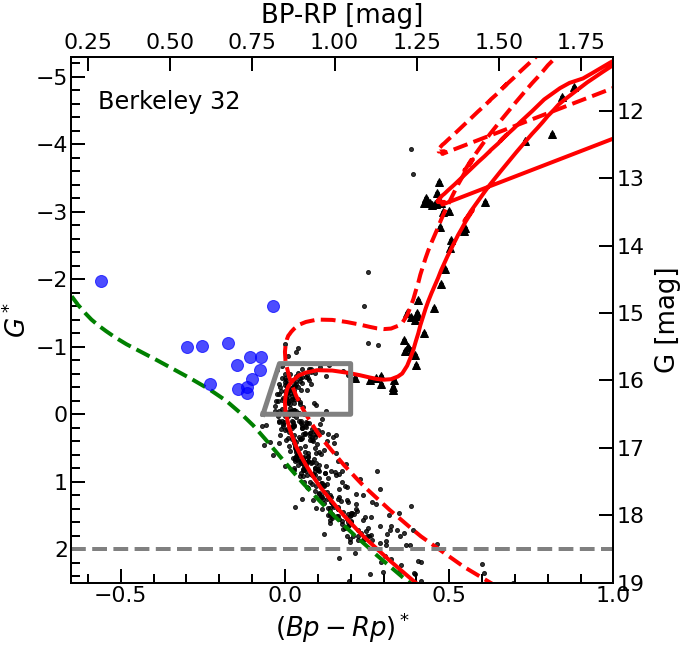}
		\caption*{}
	\end{subfigure}
	\begin{subfigure}[b]{0.32\textwidth}
   		\includegraphics[width=0.99\textwidth]{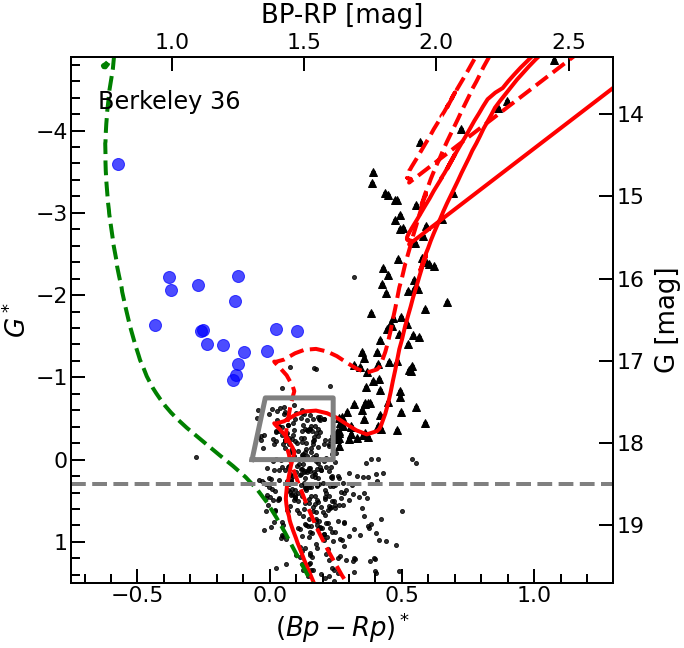}
		\caption*{}
	\end{subfigure}
	\begin{subfigure}[b]{0.32\textwidth}
   		\includegraphics[width=0.99\textwidth]{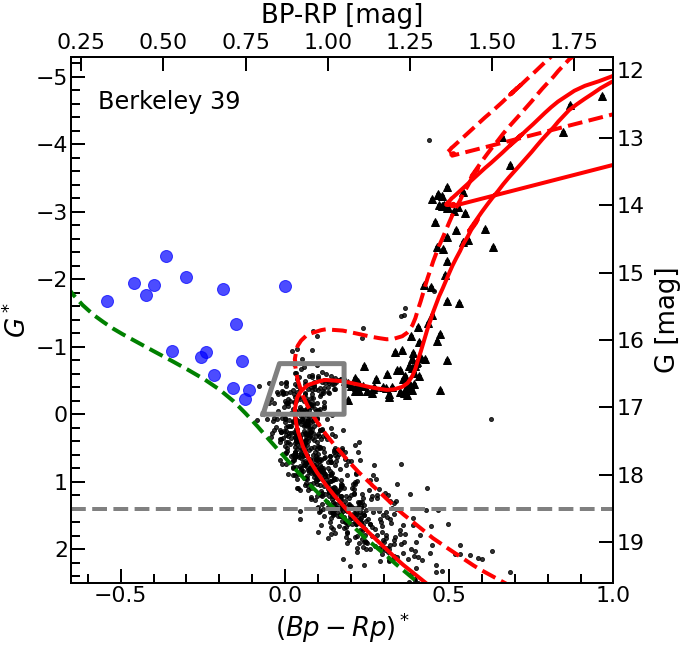}
		\caption*{}
	\end{subfigure}
	\begin{subfigure}[b]{0.32\textwidth}
   		\includegraphics[width=0.99\textwidth]{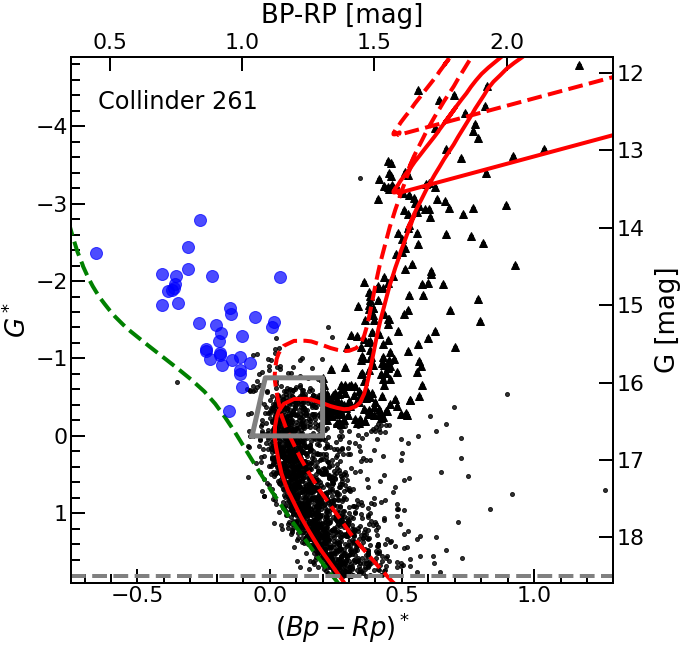}
		\caption*{}
	\end{subfigure}
	\vspace{-0.5cm}
	\begin{subfigure}[b]{0.32\textwidth}
   		\includegraphics[width=0.99\textwidth]{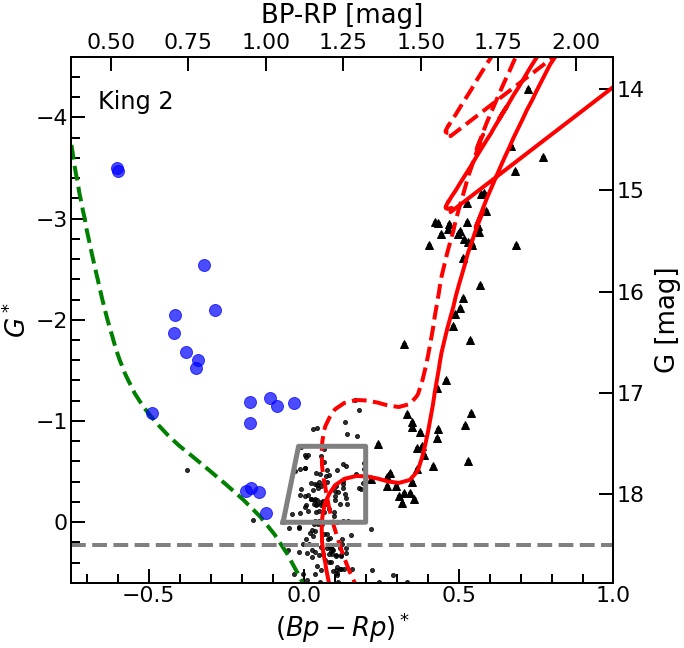}
		\caption*{}
	\end{subfigure}
	\vspace{-0.5cm}
	\begin{subfigure}[b]{0.32\textwidth}
   		\includegraphics[width=0.99\textwidth]{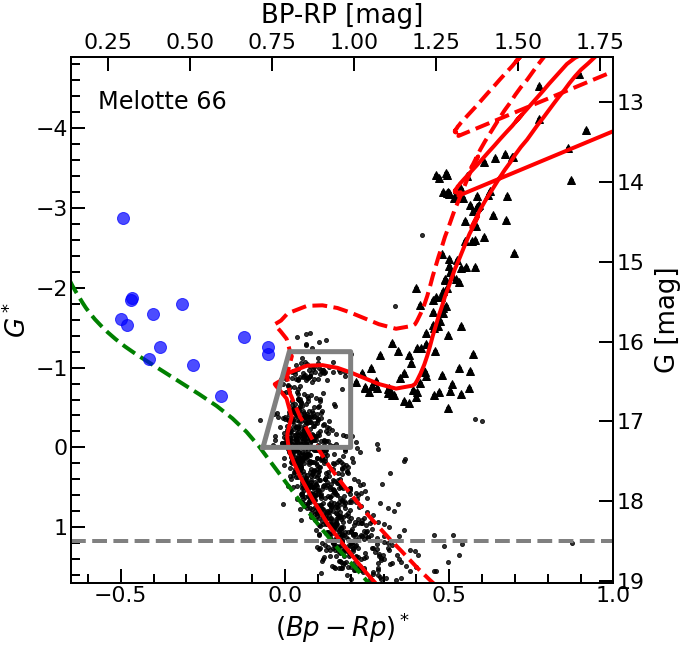}
		\caption*{}
	\end{subfigure}
	\begin{subfigure}[b]{0.32\textwidth}
   		\includegraphics[width=0.99\textwidth]{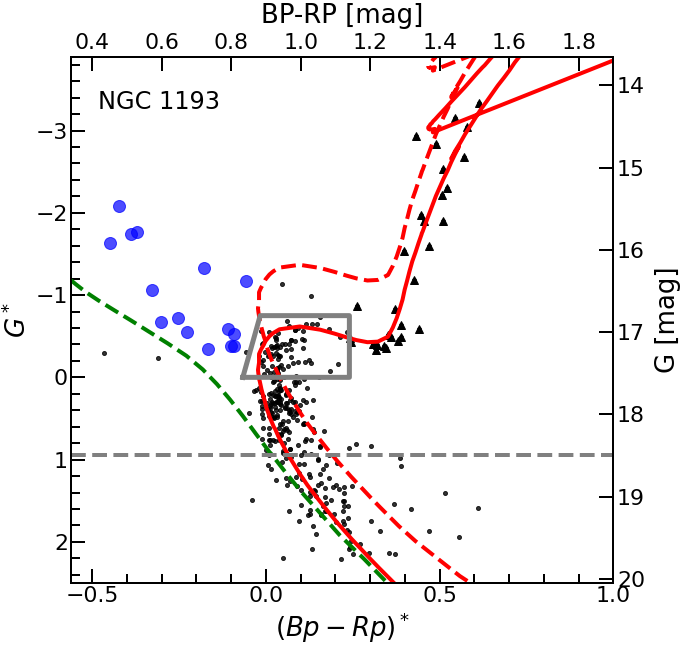}
		\caption*{}
	\end{subfigure}
	\begin{subfigure}[b]{0.32\textwidth}
   		\includegraphics[width=0.99\textwidth]{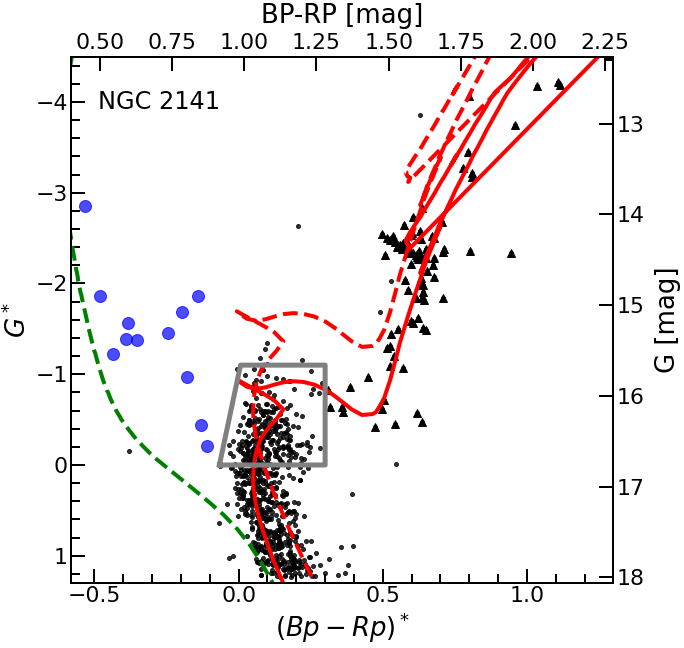}
		\caption*{}
	\end{subfigure}
    \vspace{-0.2cm}
	\caption{The \textit{Gaia} EDR3 CMDs of 12 OCs with fitted PARSEC isochrones (red solid lines). The bottom and left axes represent the normalised color and magnitude of cluster members to locate the MSTO point at (0,0). The top and right axes represent the observed color and magnitudes of OCs members. Orange dashed lines and brown dashed lines represent equal mass binary isochrone and ZAMS of ages between 90 -- 140 Myr, respectively. BSS candidates are shown as blue-filled circles. SGBs, RGBs, and RCs are shown as black triangles. Black polygons plotted on each CMD represent the MSTO region, and black dots represent the rest of the members of OCs. The grey-dashed line drawn at G = 18.5 mag represents the completeness level of \textit{Gaia} EDR3 data.}
	\label{fig:CMD1}
\end{figure*}

\begin{figure*}
    \ContinuedFloat
    \centering
	\begin{subfigure}[b]{0.32\textwidth}
   		\includegraphics[width=0.99\textwidth]{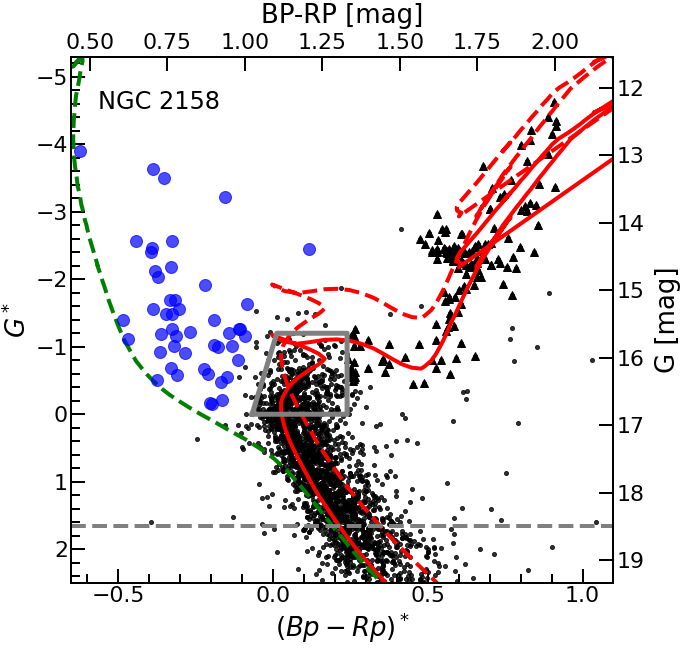}
		\caption*{}
	\end{subfigure}
	\begin{subfigure}[b]{0.32\textwidth}
   		\includegraphics[width=0.99\textwidth]{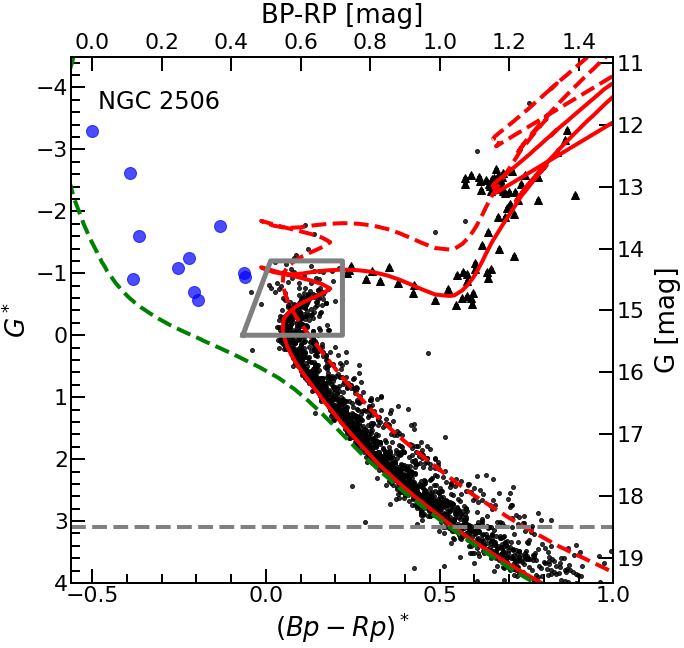}
		\caption*{}
	\end{subfigure}
	\vspace{-0.5cm}
	\begin{subfigure}[b]{0.32\textwidth}
   		\includegraphics[width=0.99\textwidth]{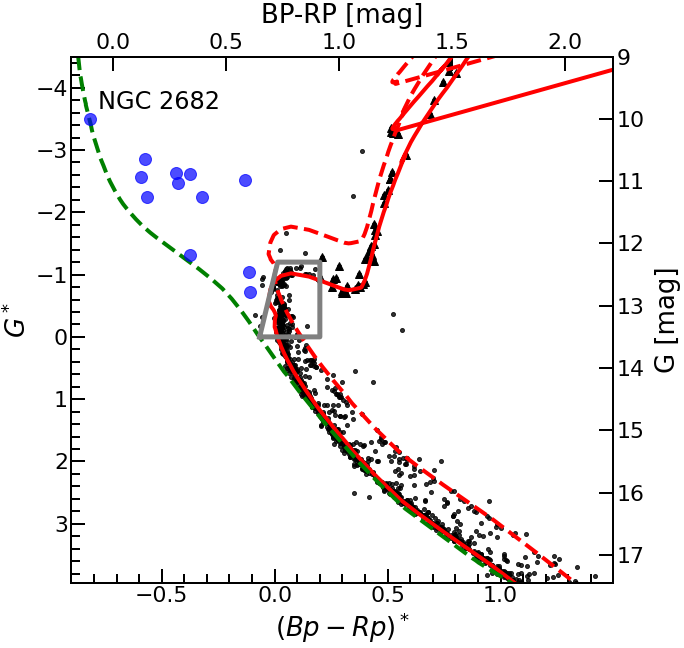}
		\caption*{}
	\end{subfigure}
	\vspace{-0.5cm}
	\begin{subfigure}[b]{0.32\textwidth}
   		\includegraphics[width=0.99\textwidth]{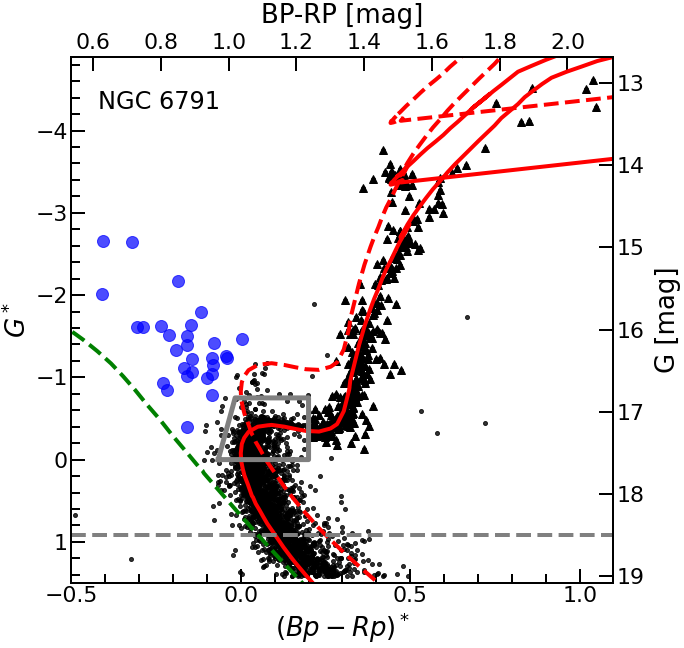}
		\caption*{}
	\end{subfigure}
	\begin{subfigure}[b]{0.32\textwidth}
   		\includegraphics[width=0.99\textwidth]{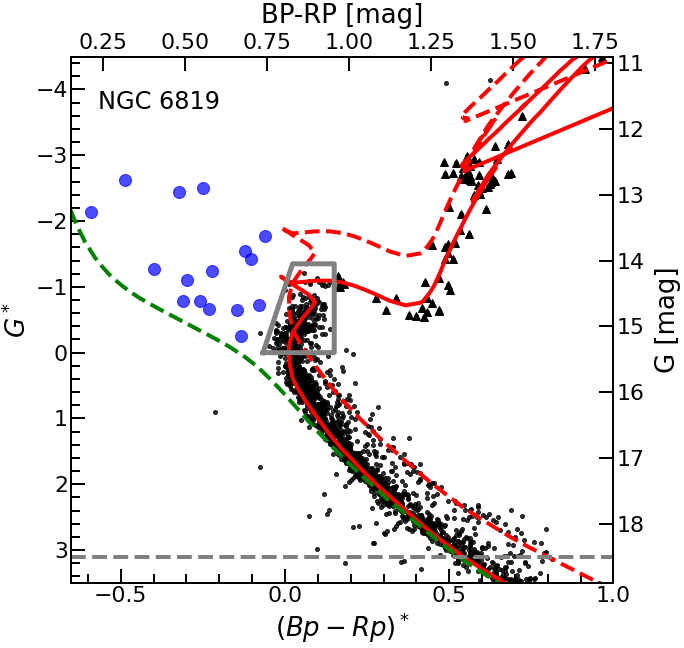}
		\caption*{}
	\end{subfigure}
	\begin{subfigure}[b]{0.32\textwidth}
   		\includegraphics[width=0.99\textwidth]{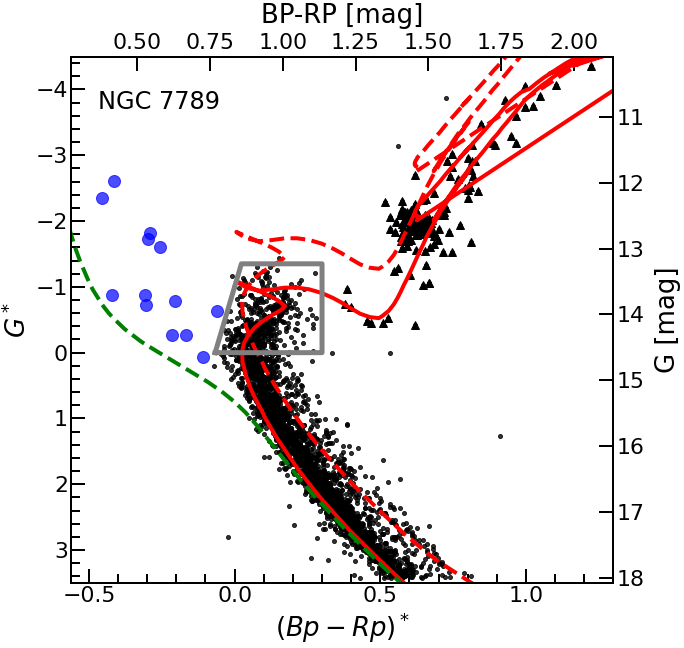}
		\caption*{}
	\end{subfigure}
	\begin{subfigure}[b]{0.32\textwidth}
   		\includegraphics[width=0.99\textwidth]{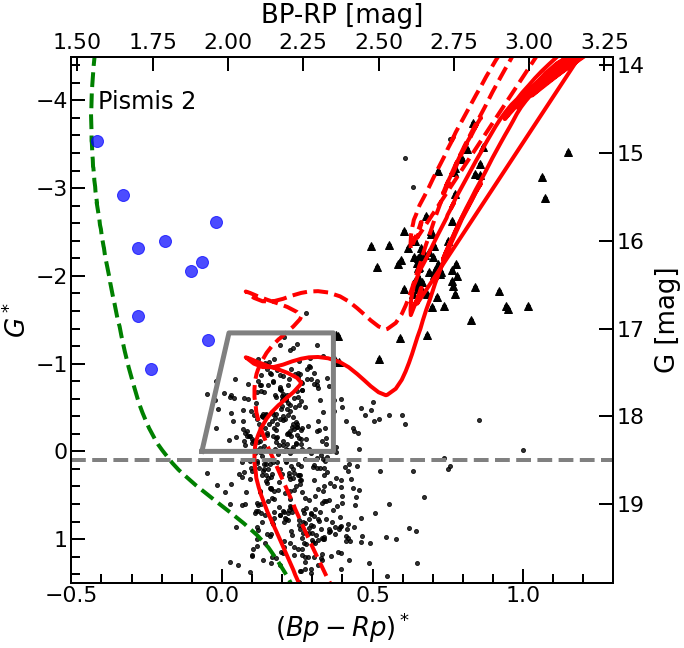}
		\caption*{}
	\end{subfigure}
    \vspace{-0.5cm}
    \begin{subfigure}[b]{0.32\textwidth}
   		\includegraphics[width=0.99\textwidth]{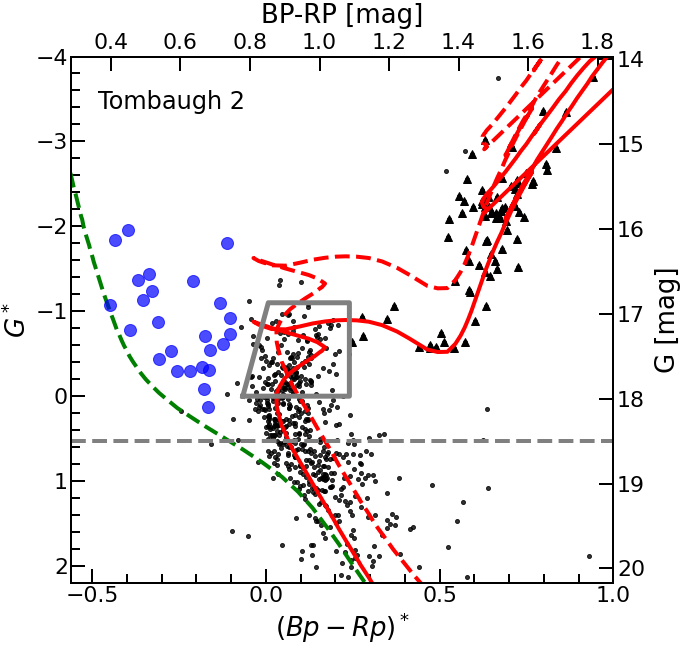}
		\caption*{}
	\end{subfigure}
	\begin{subfigure}[b]{0.32\textwidth}
   		\includegraphics[width=0.99\textwidth]{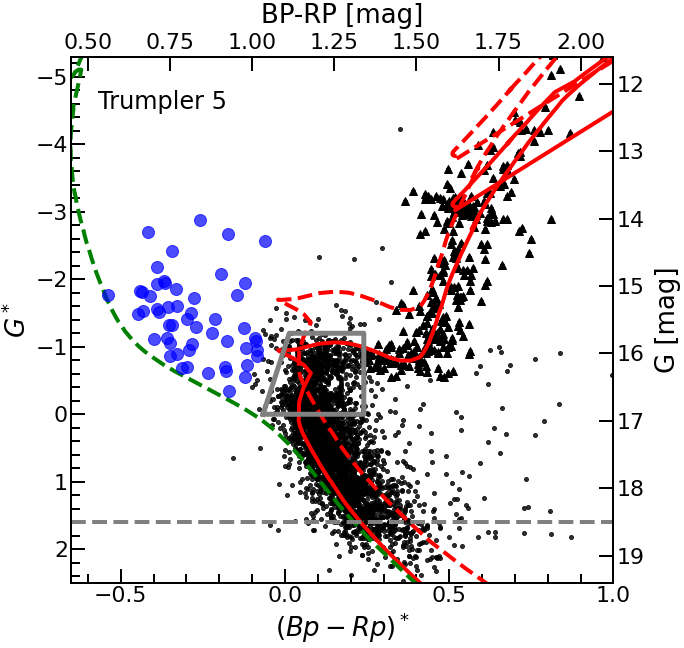}
		\caption*{}
	\end{subfigure}
	\begin{subfigure}[b]{0.32\textwidth}
   		\includegraphics[width=0.99\textwidth]{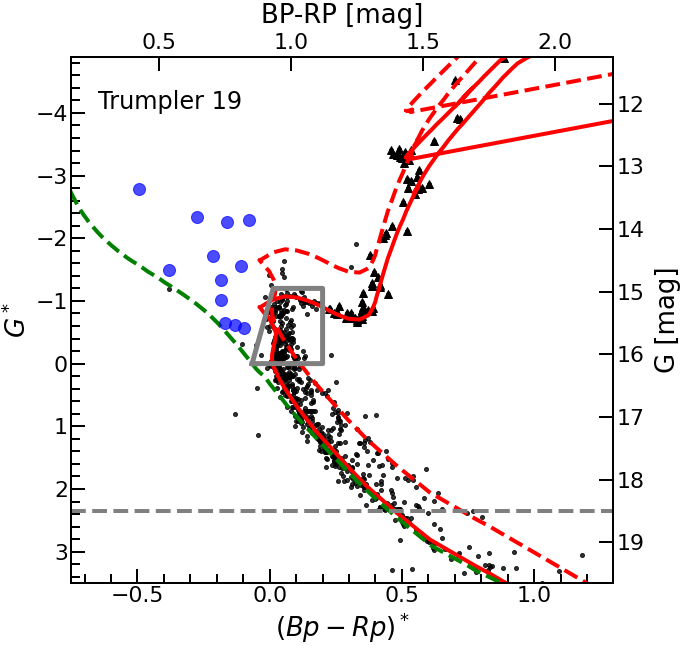}
		\caption*{}
	\end{subfigure}
	\begin{subfigure}[b]{0.32\textwidth}
   		\includegraphics[width=0.99\textwidth]{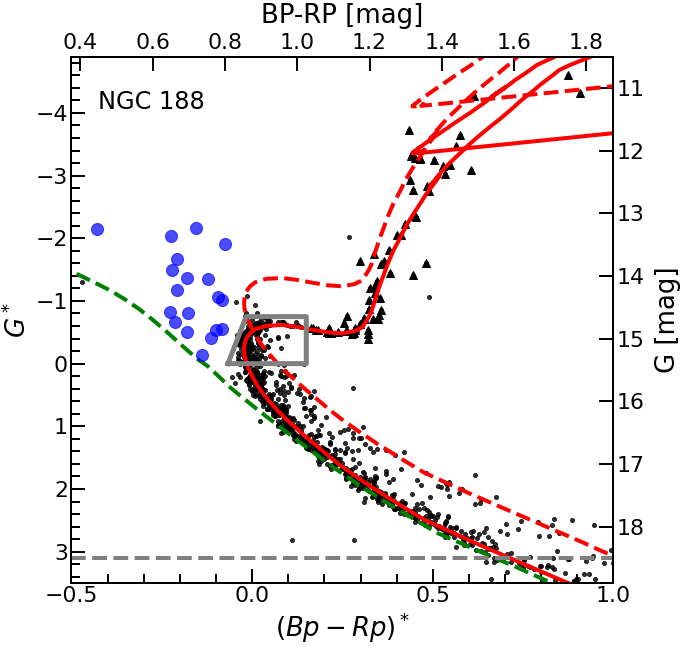}
		\caption*{}
	\end{subfigure}
	\caption{Same as Fig. \ref{fig:CMD1} for the other 11 OCs.}
\end{figure*}

\begin{figure*}
    \centering
	\begin{subfigure}[b]{0.32\textwidth}
    		\includegraphics[width=0.99\textwidth]{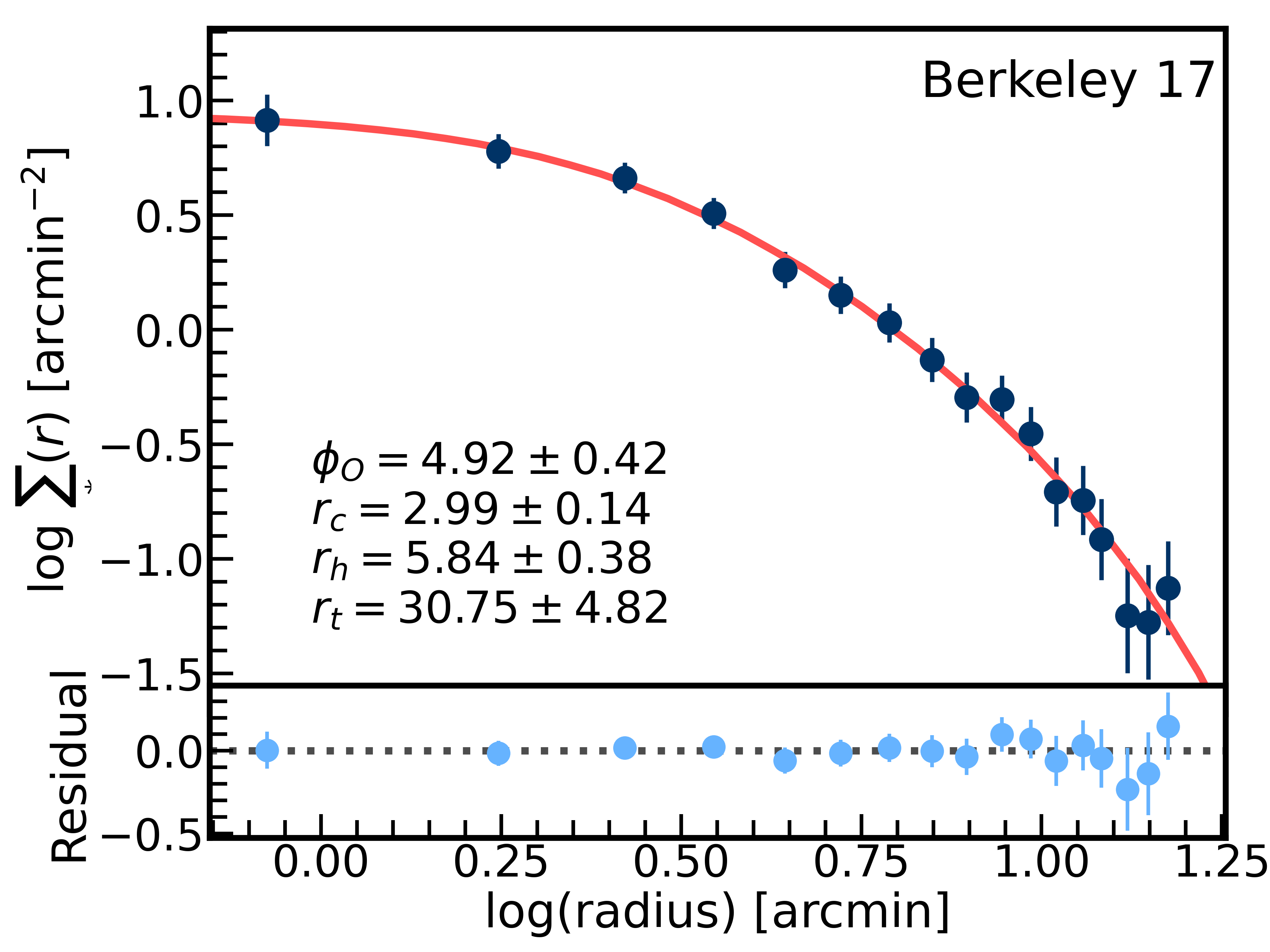}
		\caption*{}
	\end{subfigure}	 
	\begin{subfigure}[b]{0.32\textwidth}
   		\includegraphics[width=0.99\textwidth]{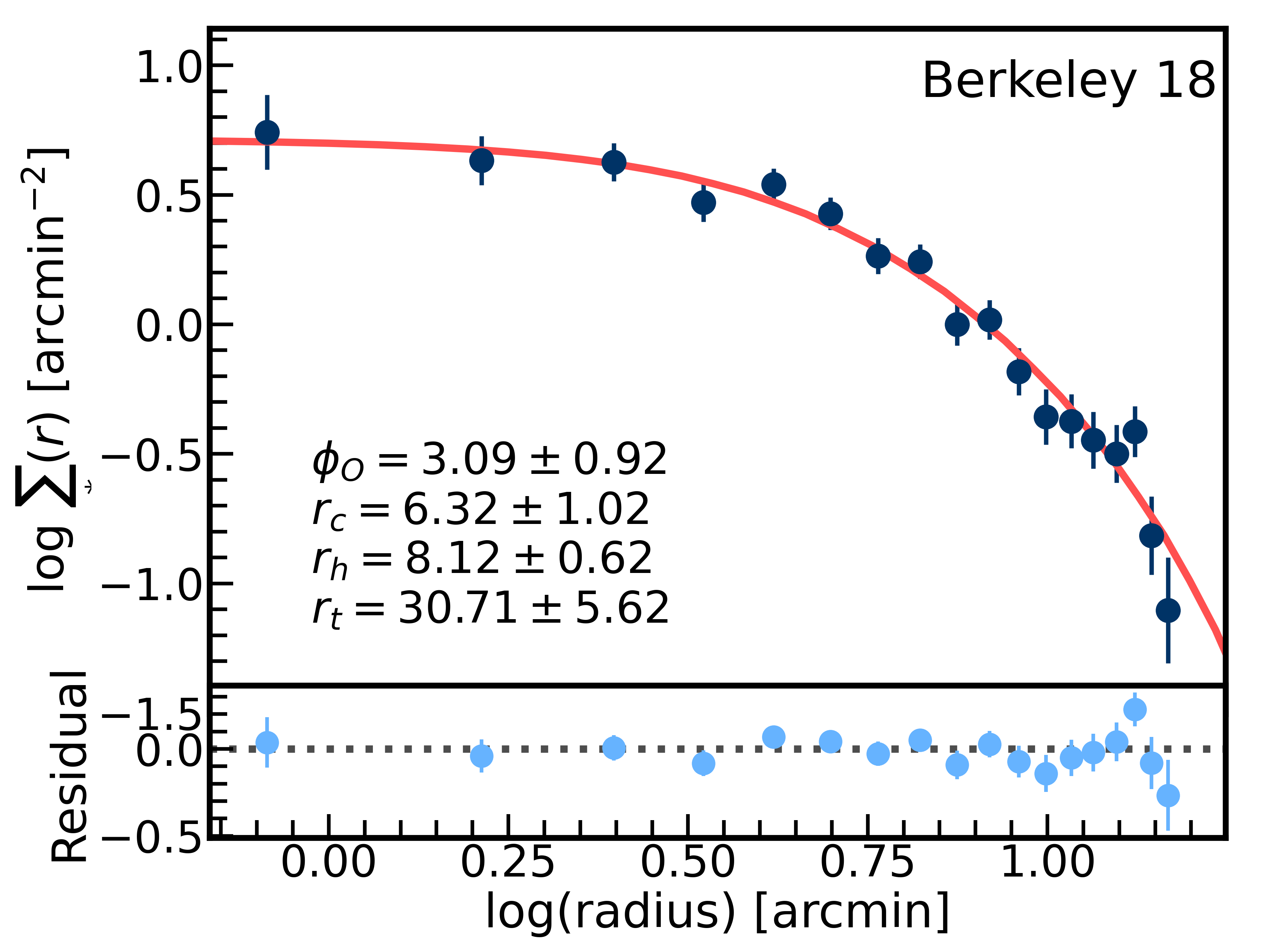}
		\caption*{}
	\end{subfigure}
	\begin{subfigure}[b]{0.32\textwidth}
   		\includegraphics[width=0.99\textwidth]{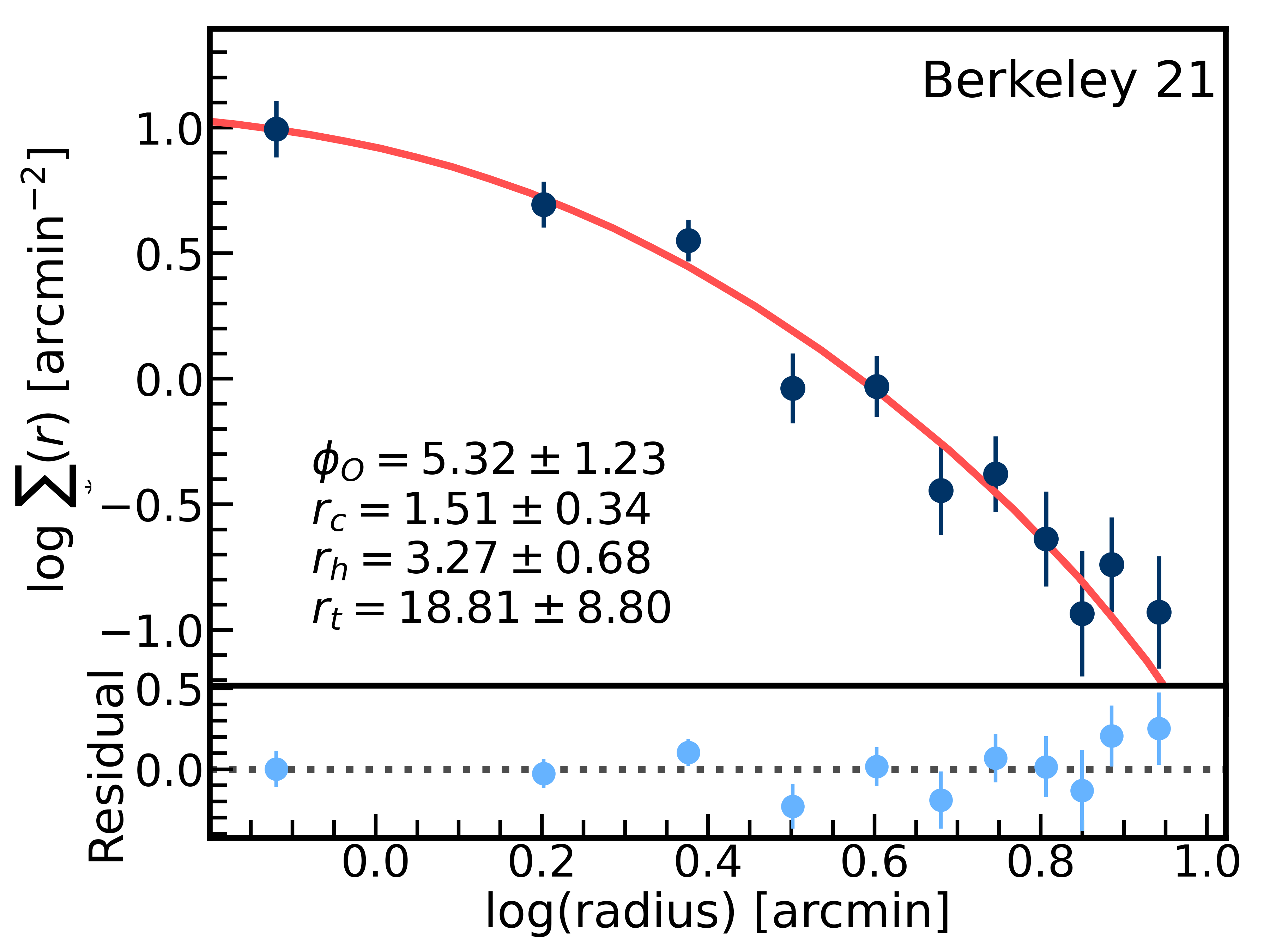}
		\caption*{}
	\end{subfigure}
	\begin{subfigure}[b]{0.32\textwidth}
   		\includegraphics[width=0.99\textwidth]{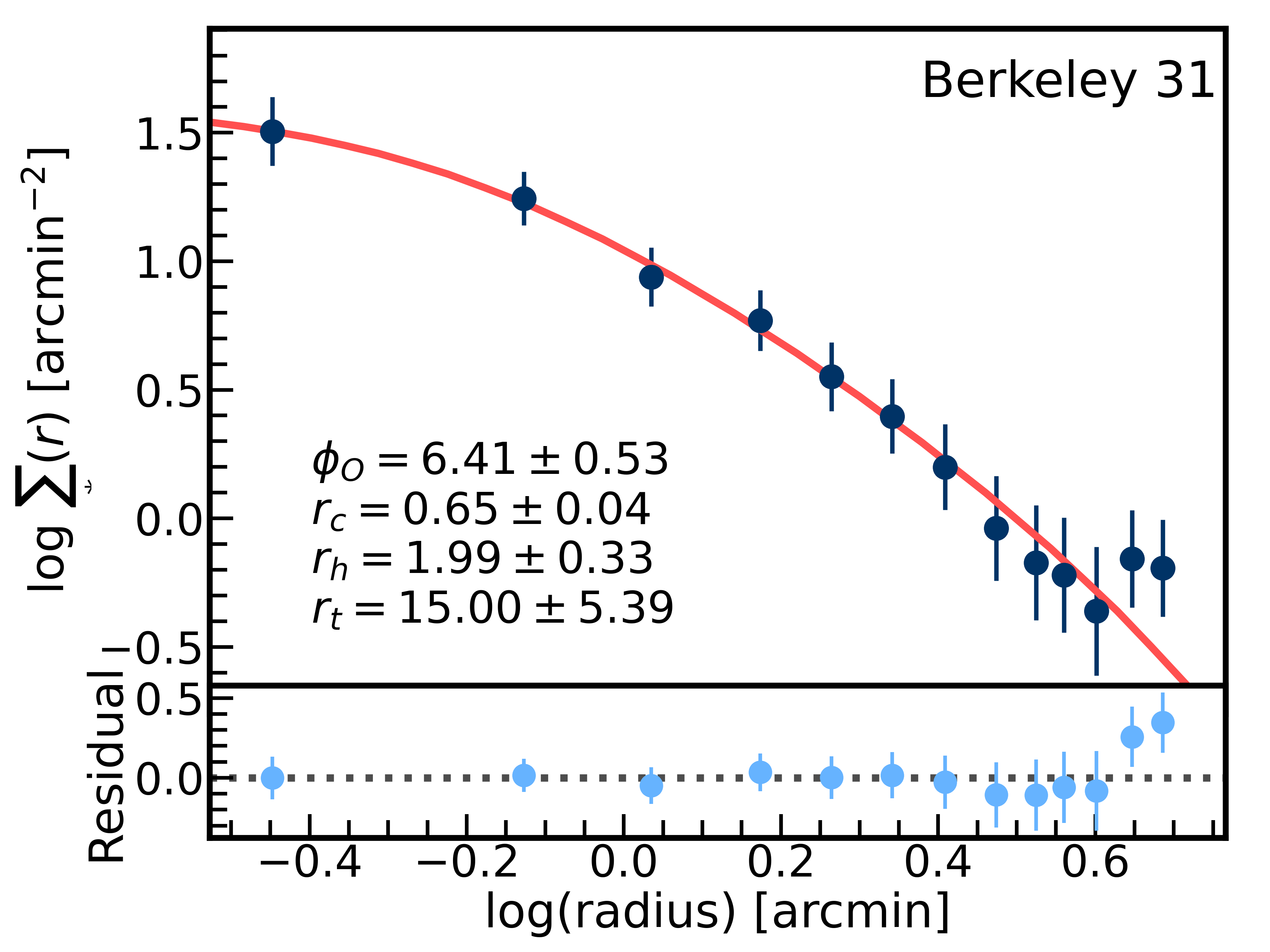}
		\caption*{}
	\end{subfigure}
	\begin{subfigure}[b]{0.32\textwidth}
   		\includegraphics[width=0.99\textwidth]{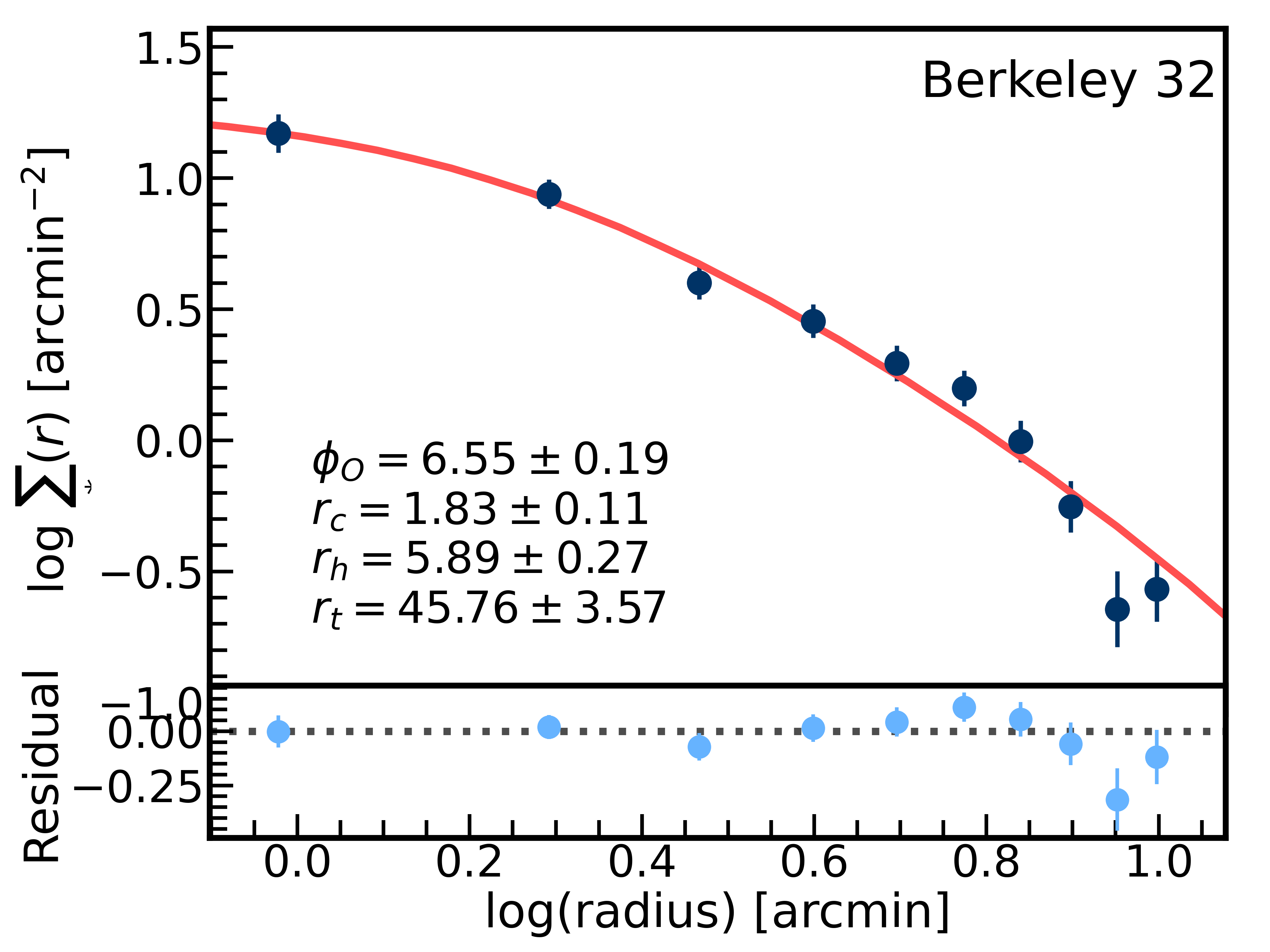}
		\caption*{}
	\end{subfigure}
	\begin{subfigure}[b]{0.32\textwidth}
   		\includegraphics[width=0.99\textwidth]{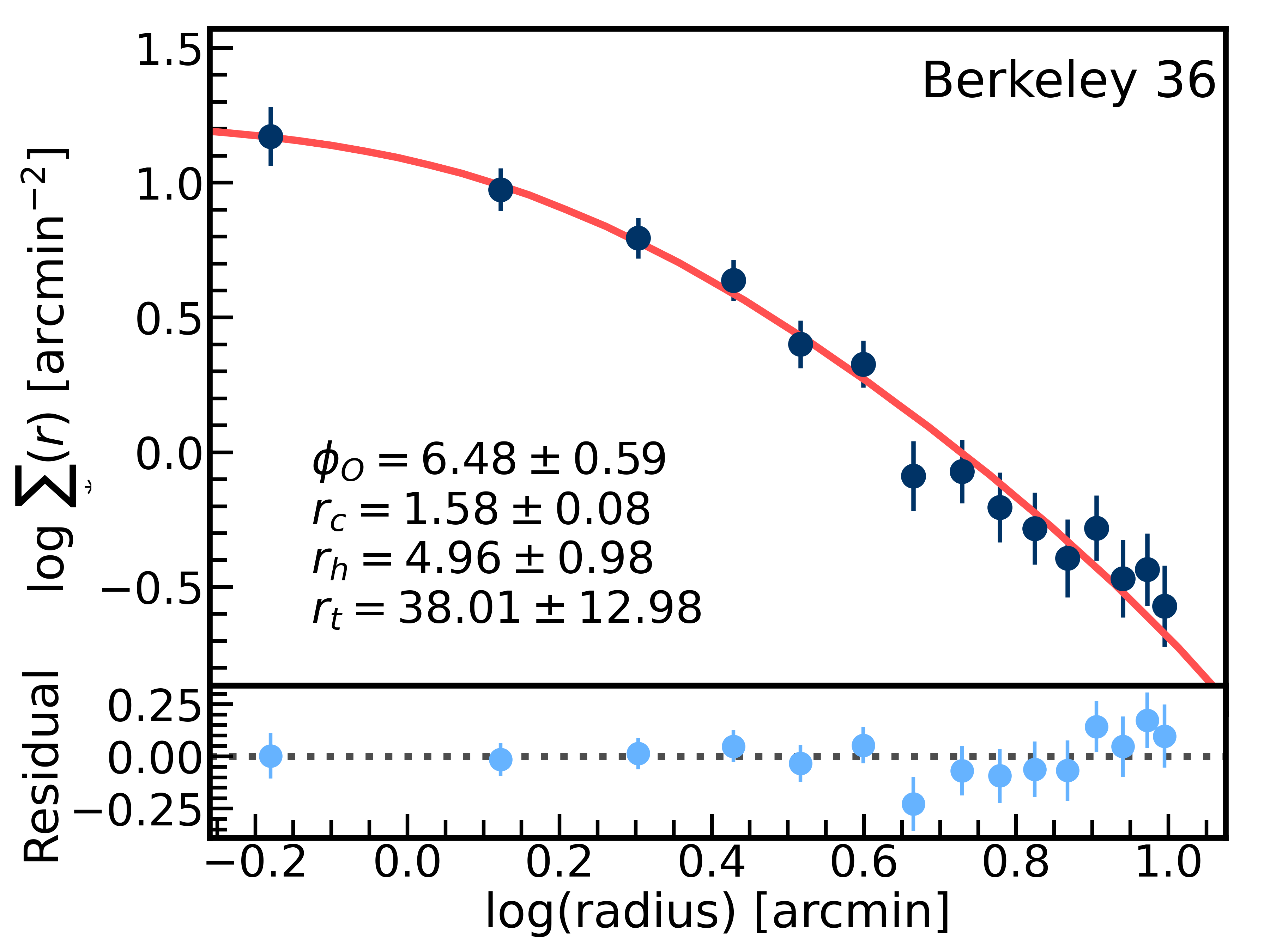}
		\caption*{}
	\end{subfigure}
	\begin{subfigure}[b]{0.32\textwidth}
   		\includegraphics[width=0.99\textwidth]{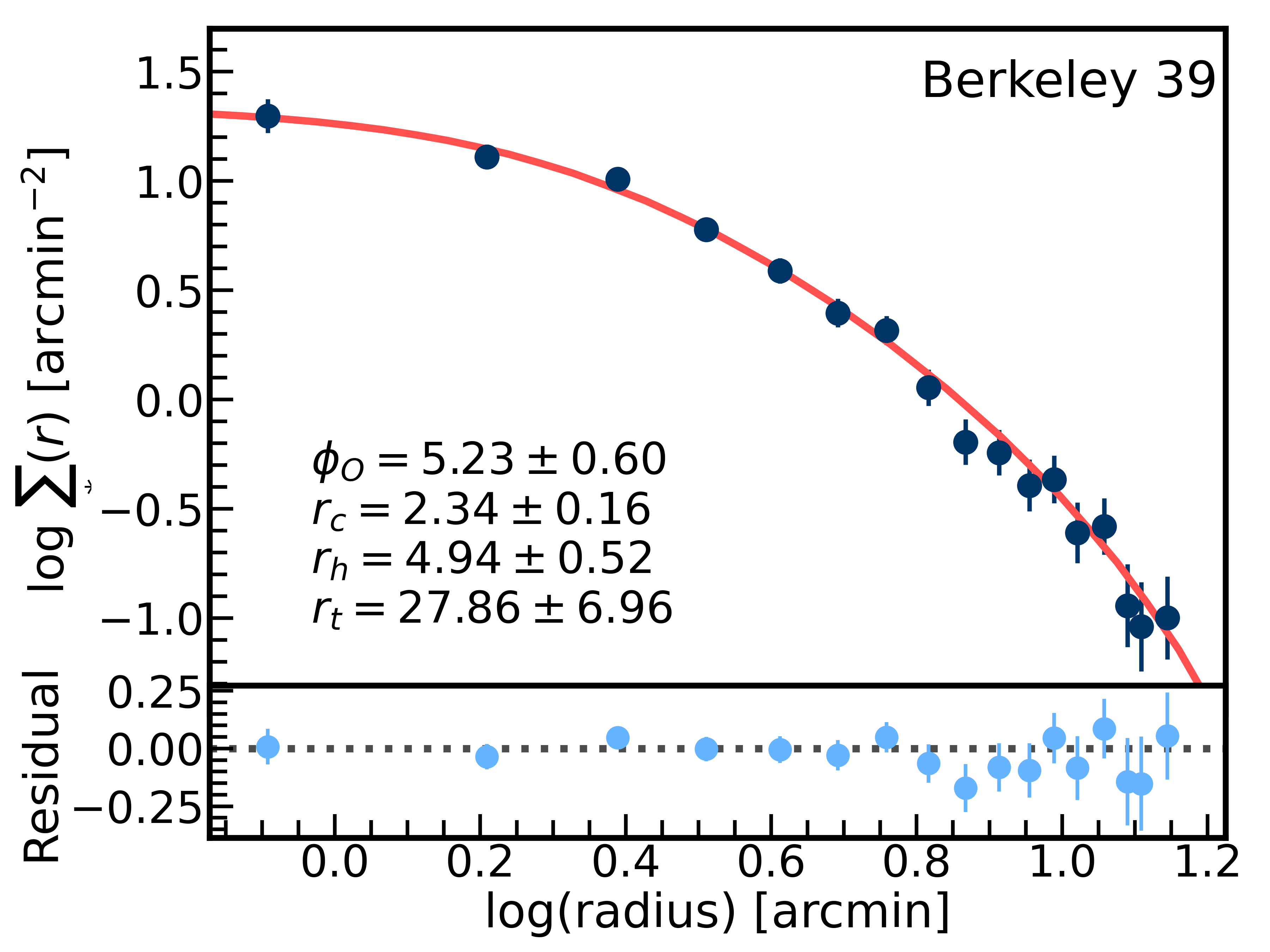}
		\caption*{}
	\end{subfigure}
	\begin{subfigure}[b]{0.32\textwidth}
   		\includegraphics[width=0.99\textwidth]{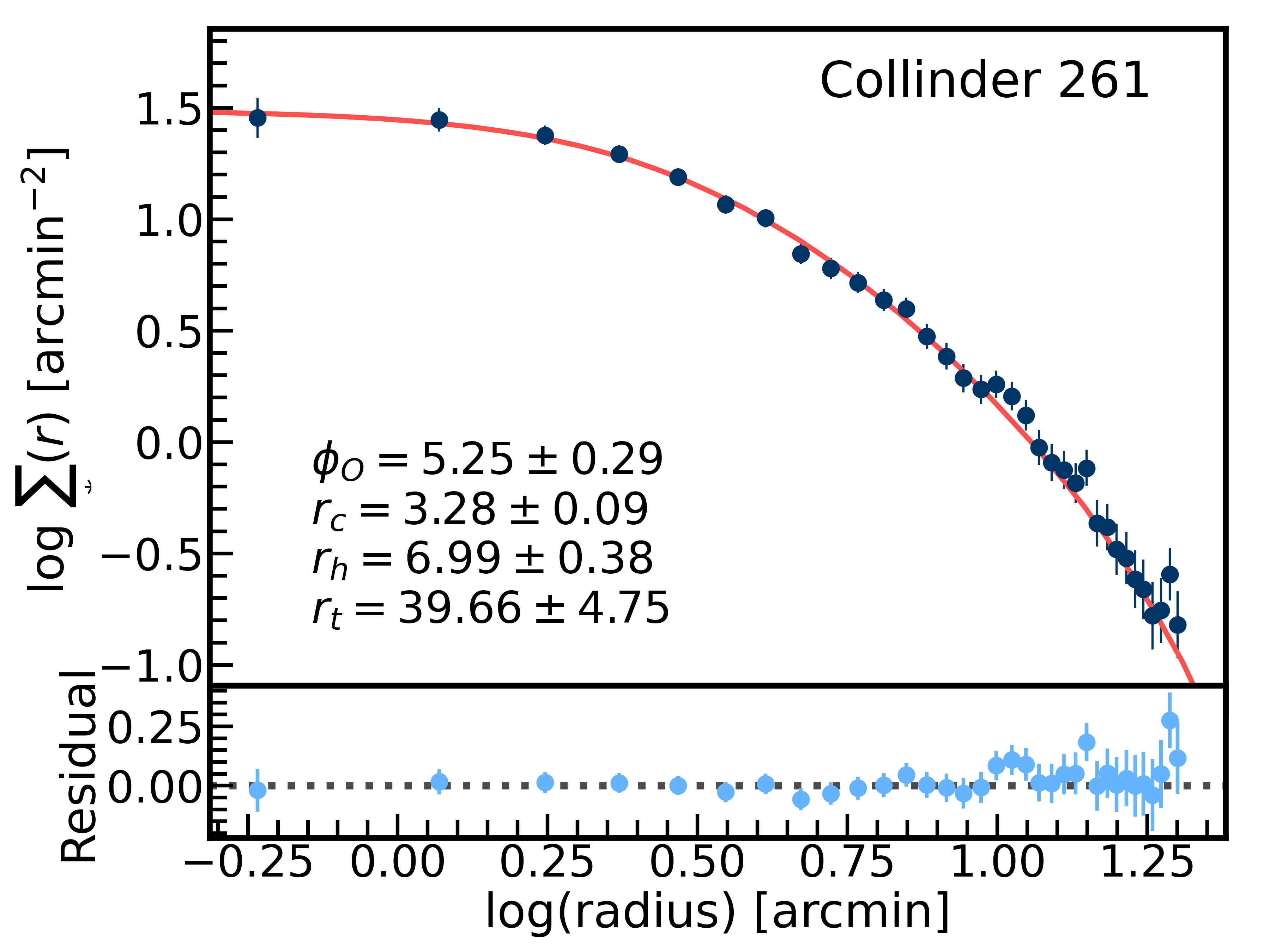}
		\caption*{}
	\end{subfigure}
	\begin{subfigure}[b]{0.32\textwidth}
   		\includegraphics[width=0.99\textwidth]{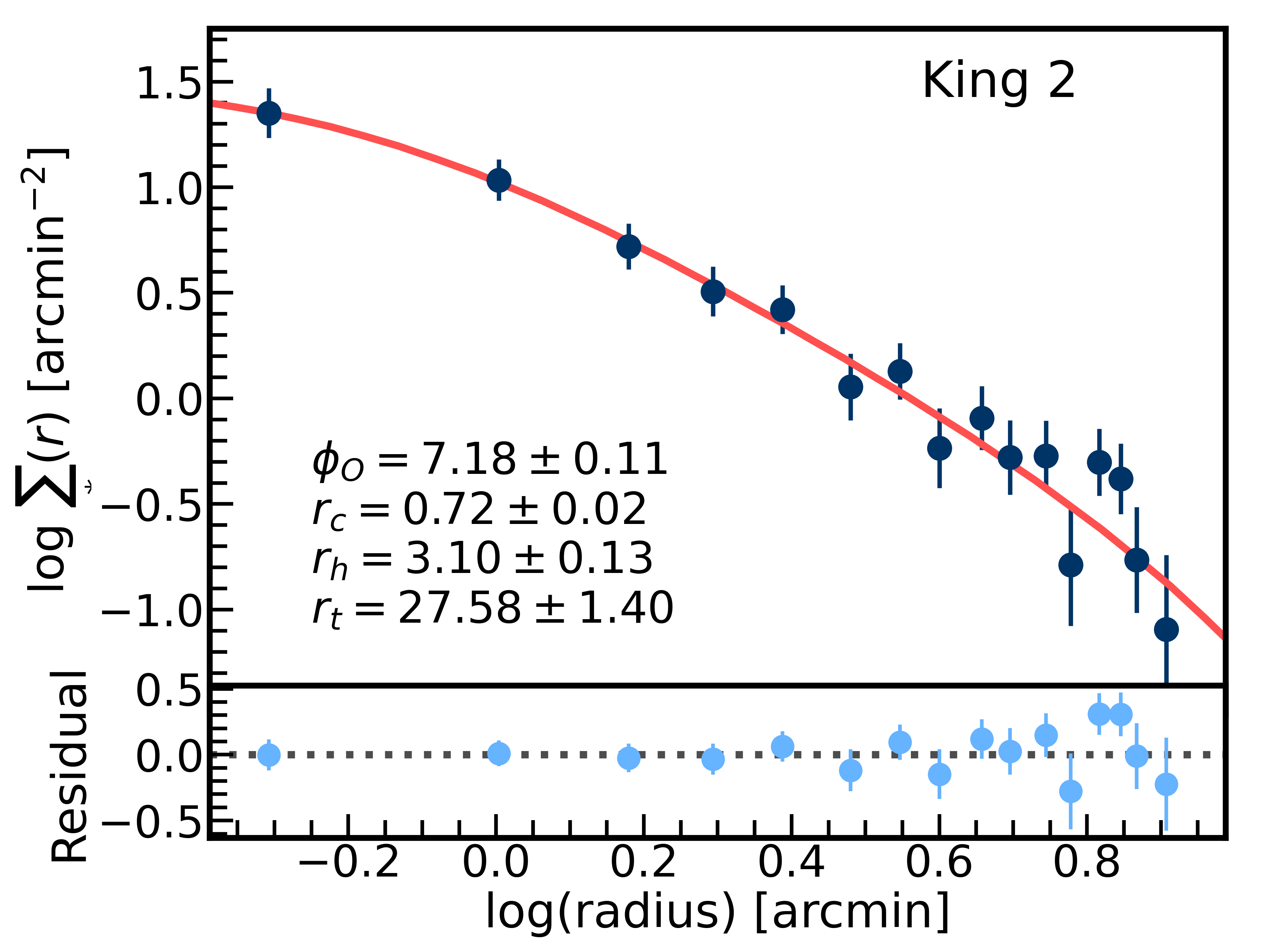}
		\caption*{}
	\end{subfigure}
	\begin{subfigure}[b]{0.32\textwidth}
   		\includegraphics[width=0.99\textwidth]{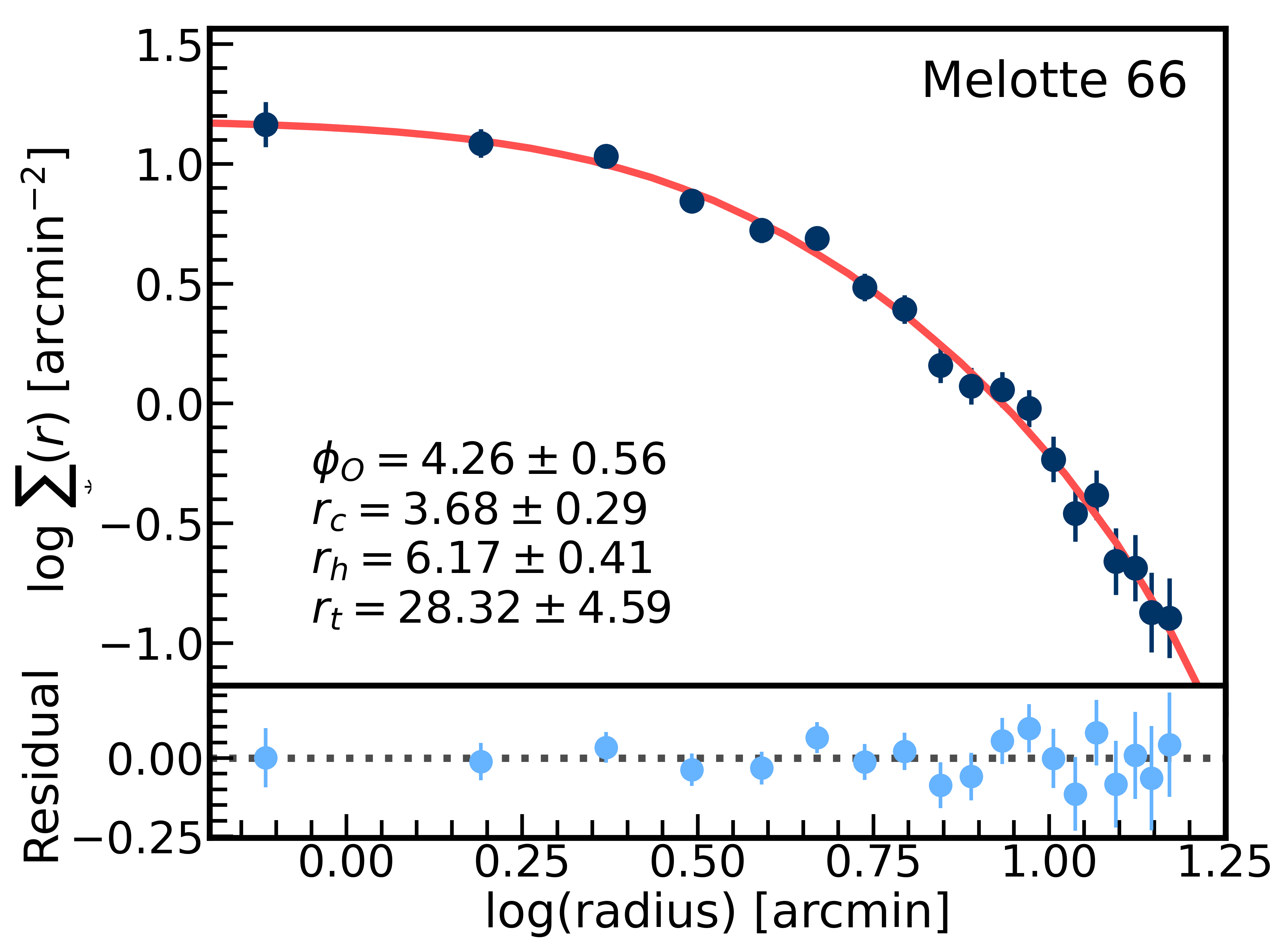}
		\caption*{}
	\end{subfigure}
	\begin{subfigure}[b]{0.32\textwidth}
   		\includegraphics[width=0.99\textwidth]{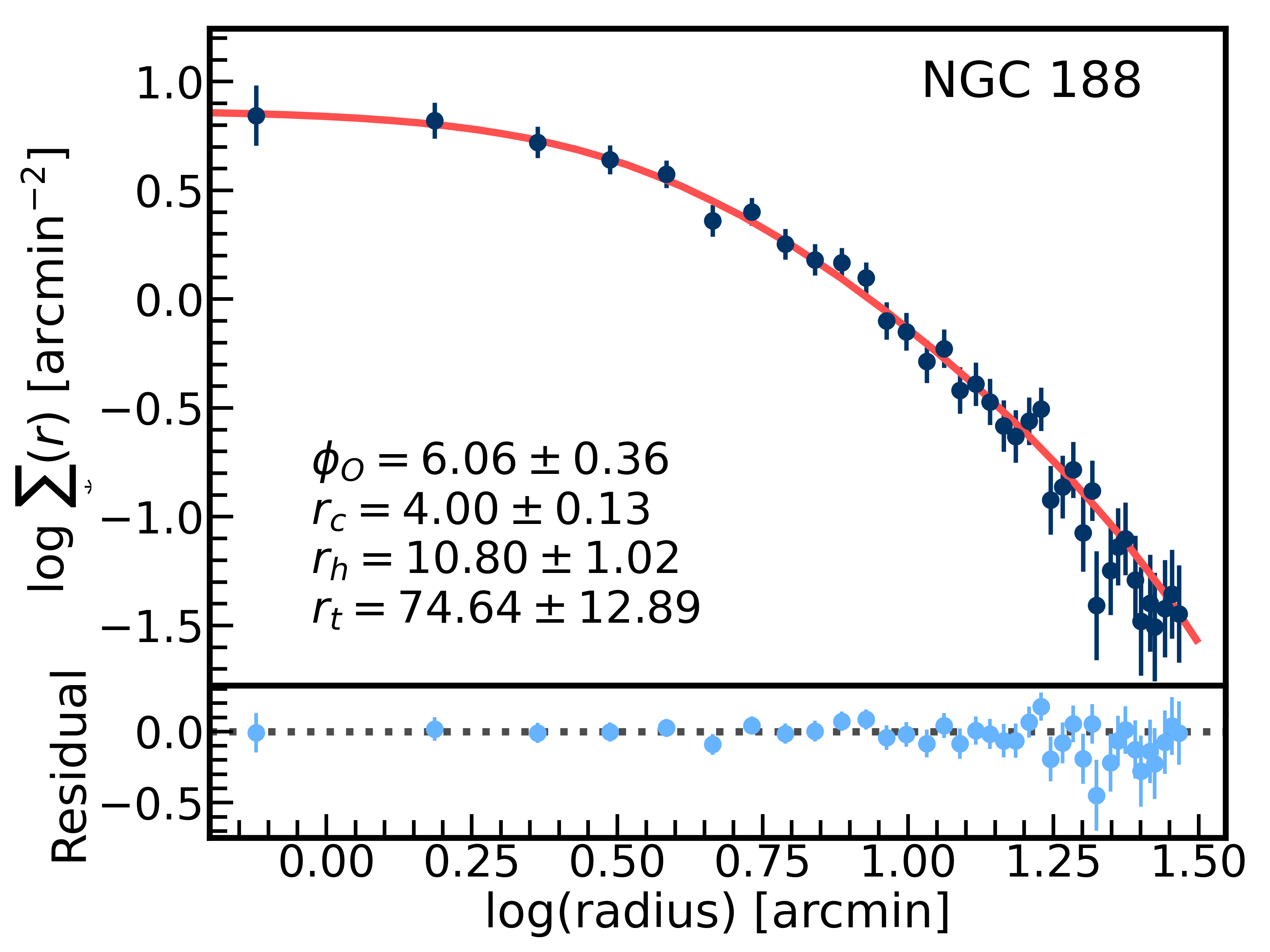}
		\caption*{}
	\end{subfigure}
	\begin{subfigure}[b]{0.32\textwidth}
   		\includegraphics[width=0.99\textwidth]{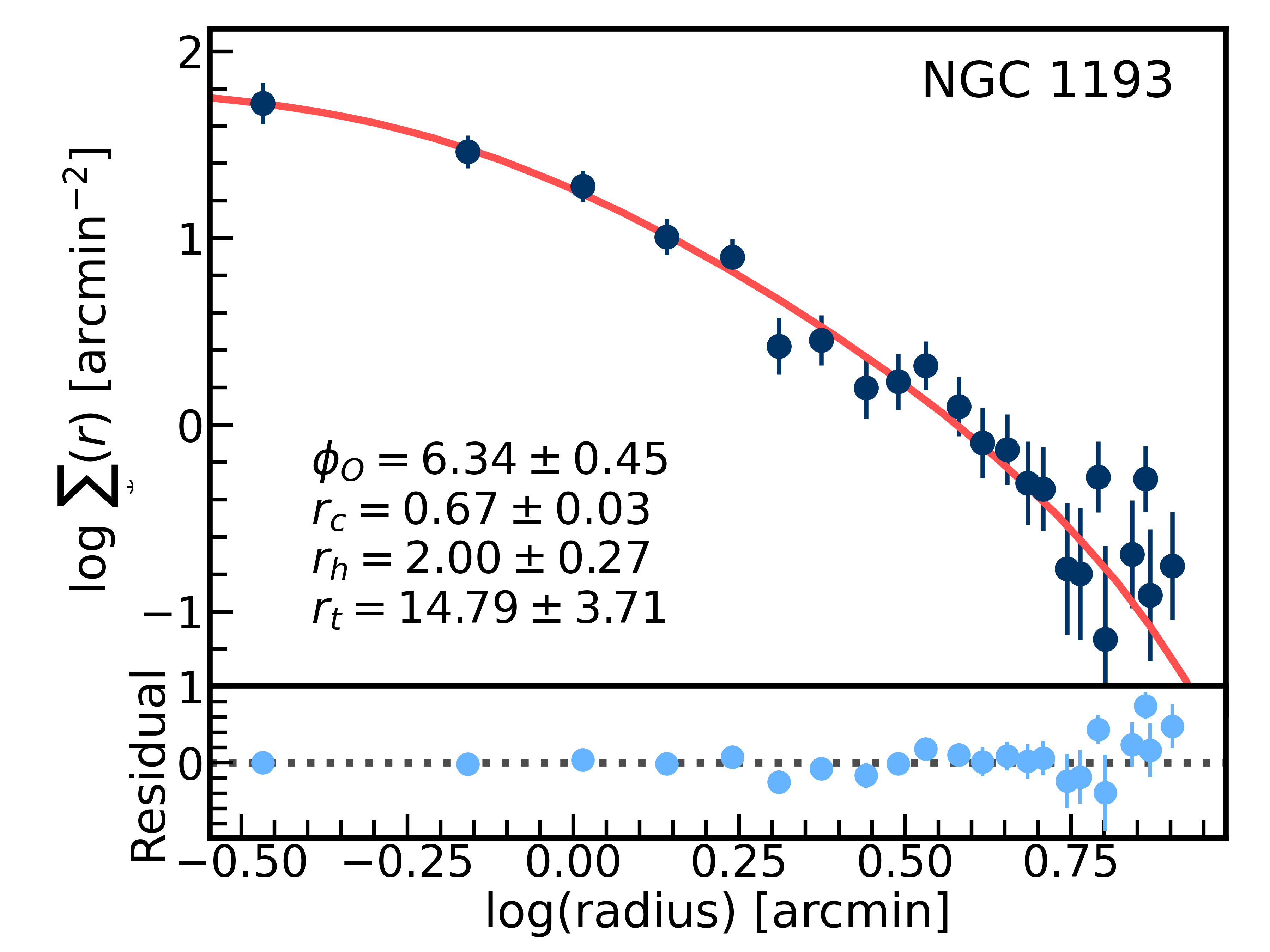}
		\caption*{}
	\end{subfigure}
        \vspace{-0.5cm}
	\caption{The observed surface density profiles fitted with an isotropic single-mass King model \citep{King1966} and the residual of the fit of the model with each observed point are shown in the upper and lower panels of each density plot of 12 OCs, respectively. The error bars are the 1$\sigma$ Poisson errors. The model producing best-fit parameters, the dimensionless central potential ($\hat{\phi_{\text{0}}}$), the core radius ($r_{\mathrm{c}}$ ), the half-mass radius ($r_{\mathrm{h}}$ ) and the tidal radius ($r_{\mathrm{t}}$) of each cluster are marked on their respective plots, where $r_{\mathrm{c}}$, $r_{\mathrm{h}}$, and $r_{\mathrm{t}}$ are in the units of arcmin.}
	\label{fig:kings_profiles}
\end{figure*}

\begin{figure*}
    \ContinuedFloat
    \centering
	\begin{subfigure}[b]{0.32\textwidth}
   		\includegraphics[width=0.99\textwidth]{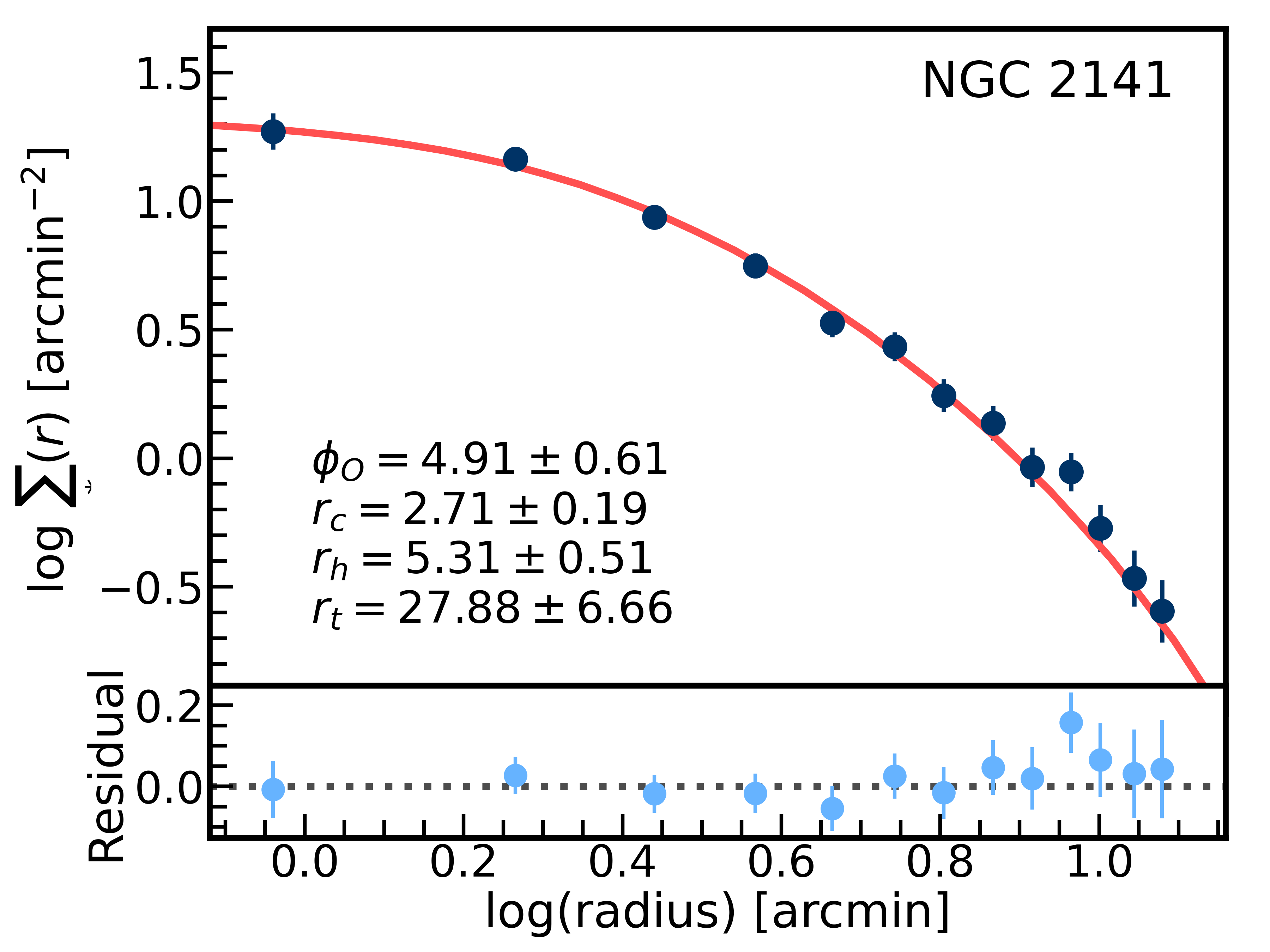}
		\caption*{}
	\end{subfigure}
	\begin{subfigure}[b]{0.32\textwidth}
   		\includegraphics[width=0.99\textwidth]{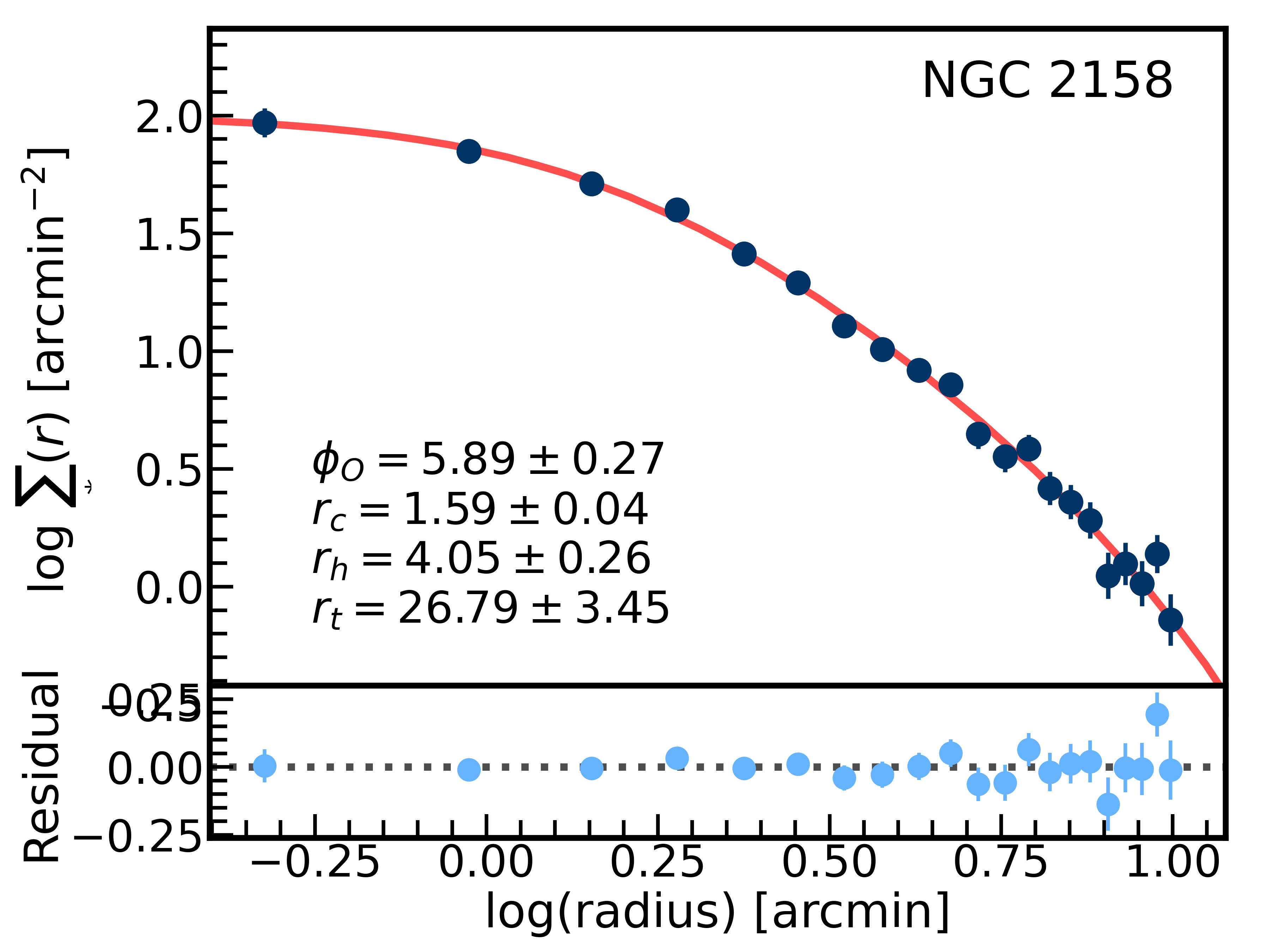}
		\caption*{}
	\end{subfigure}
	\begin{subfigure}[b]{0.32\textwidth}
   		\includegraphics[width=0.99\textwidth]{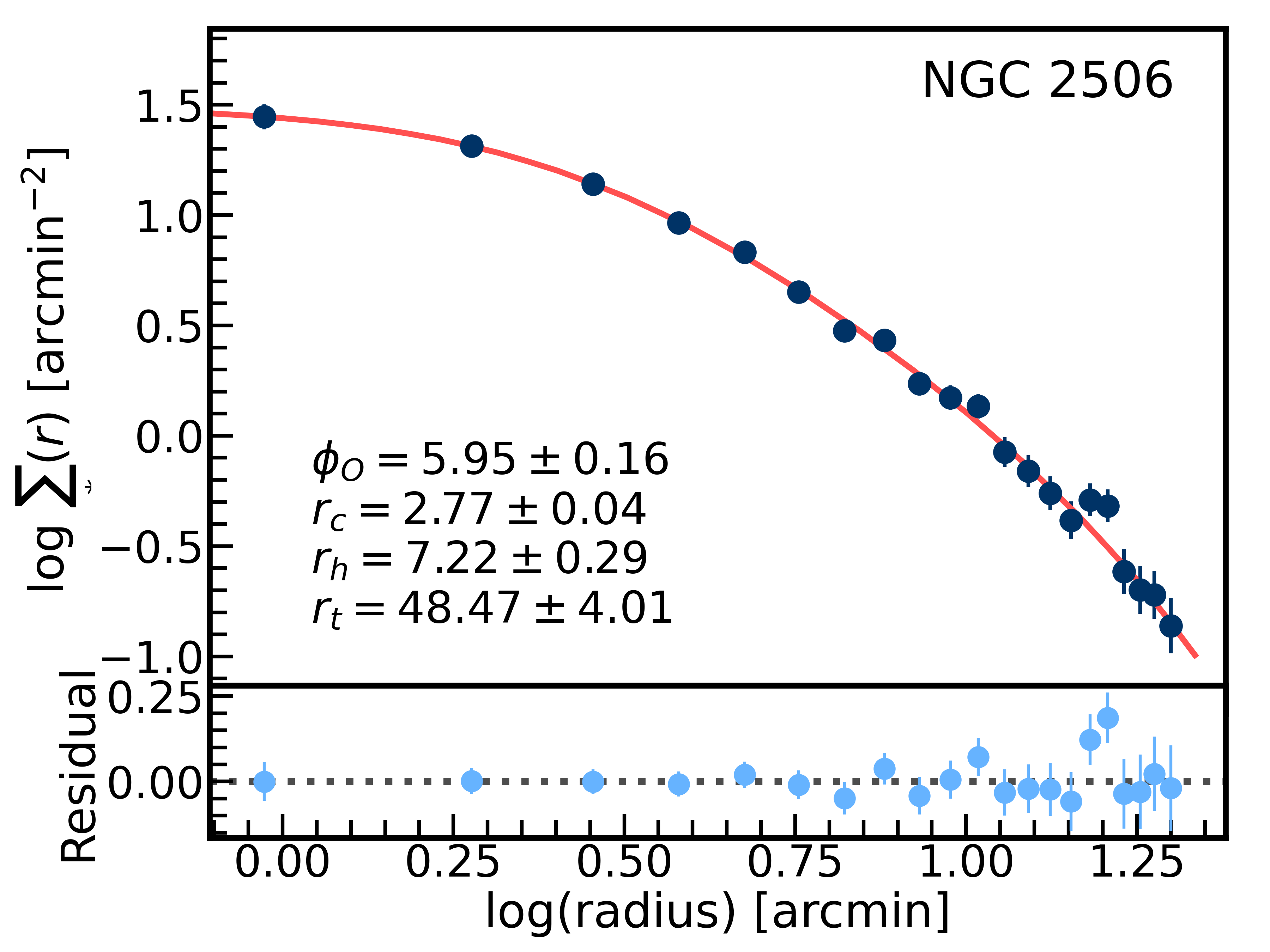}
		\caption*{}
	\end{subfigure}
	\begin{subfigure}[b]{0.32\textwidth}
   		\includegraphics[width=0.99\textwidth]{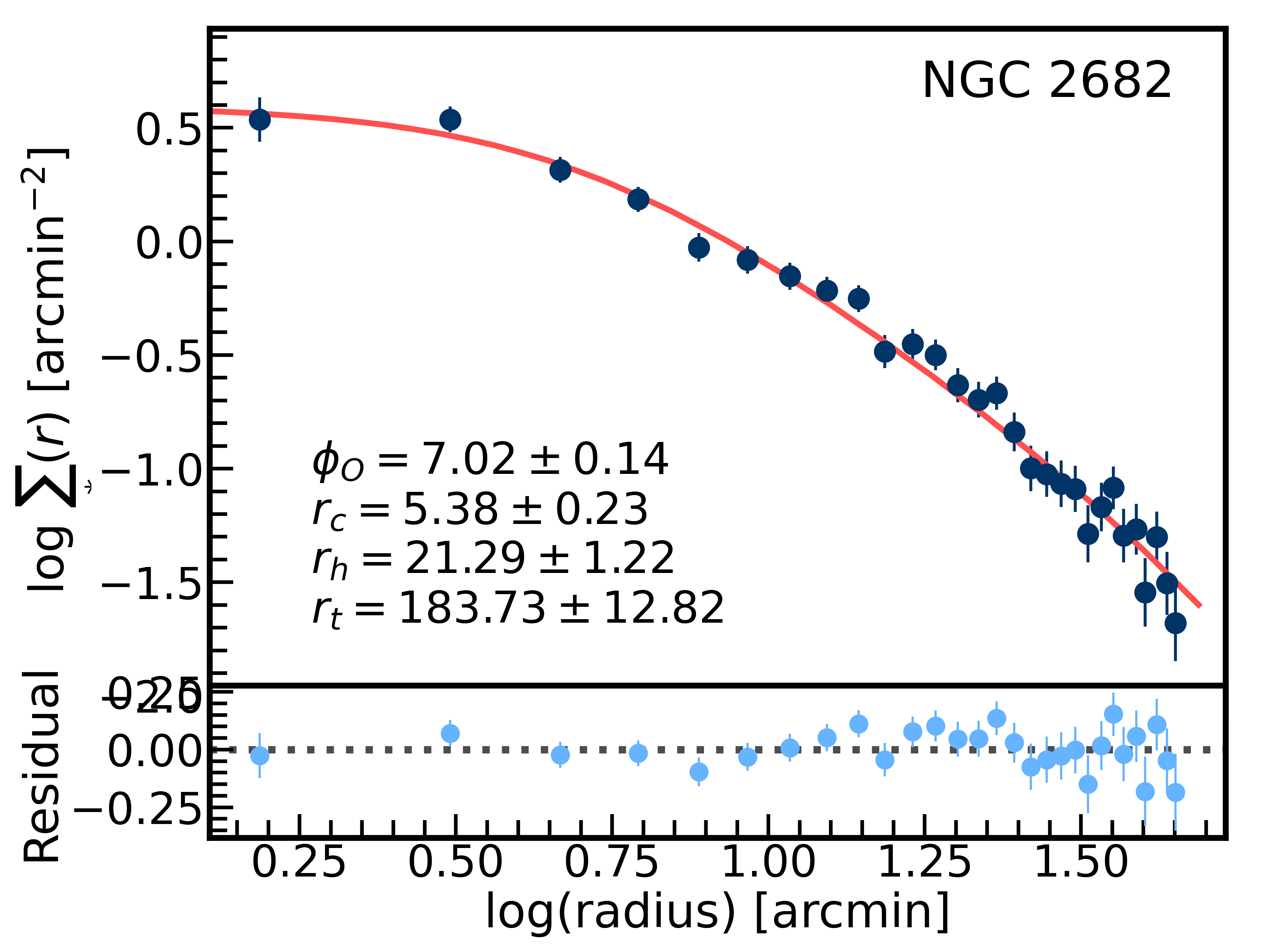}
		\caption*{}
	\end{subfigure}
	\begin{subfigure}[b]{0.32\textwidth}
   		\includegraphics[width=0.99\textwidth]{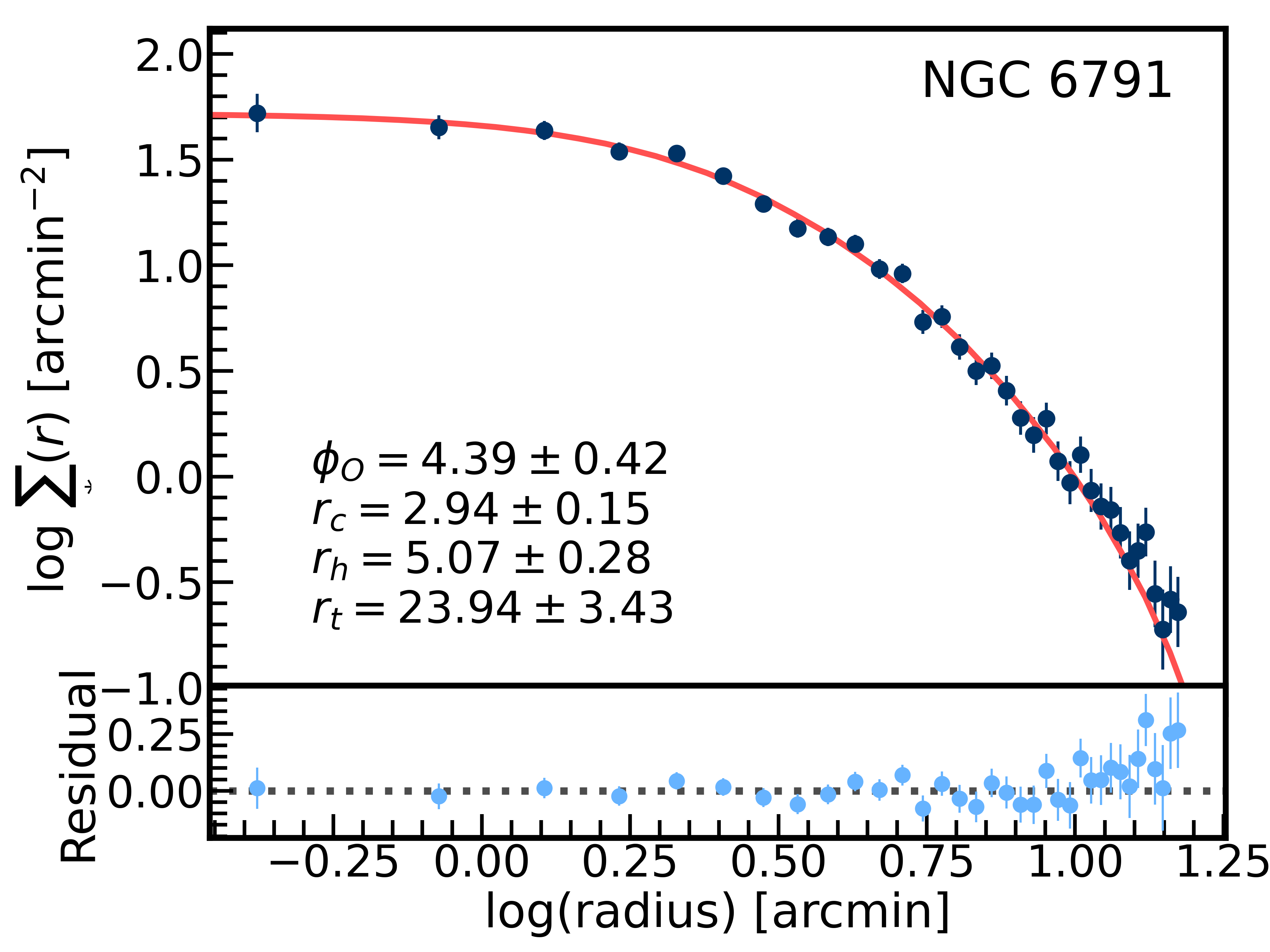}
		\caption*{}
	\end{subfigure}
	\begin{subfigure}[b]{0.32\textwidth}
   		\includegraphics[width=0.99\textwidth]{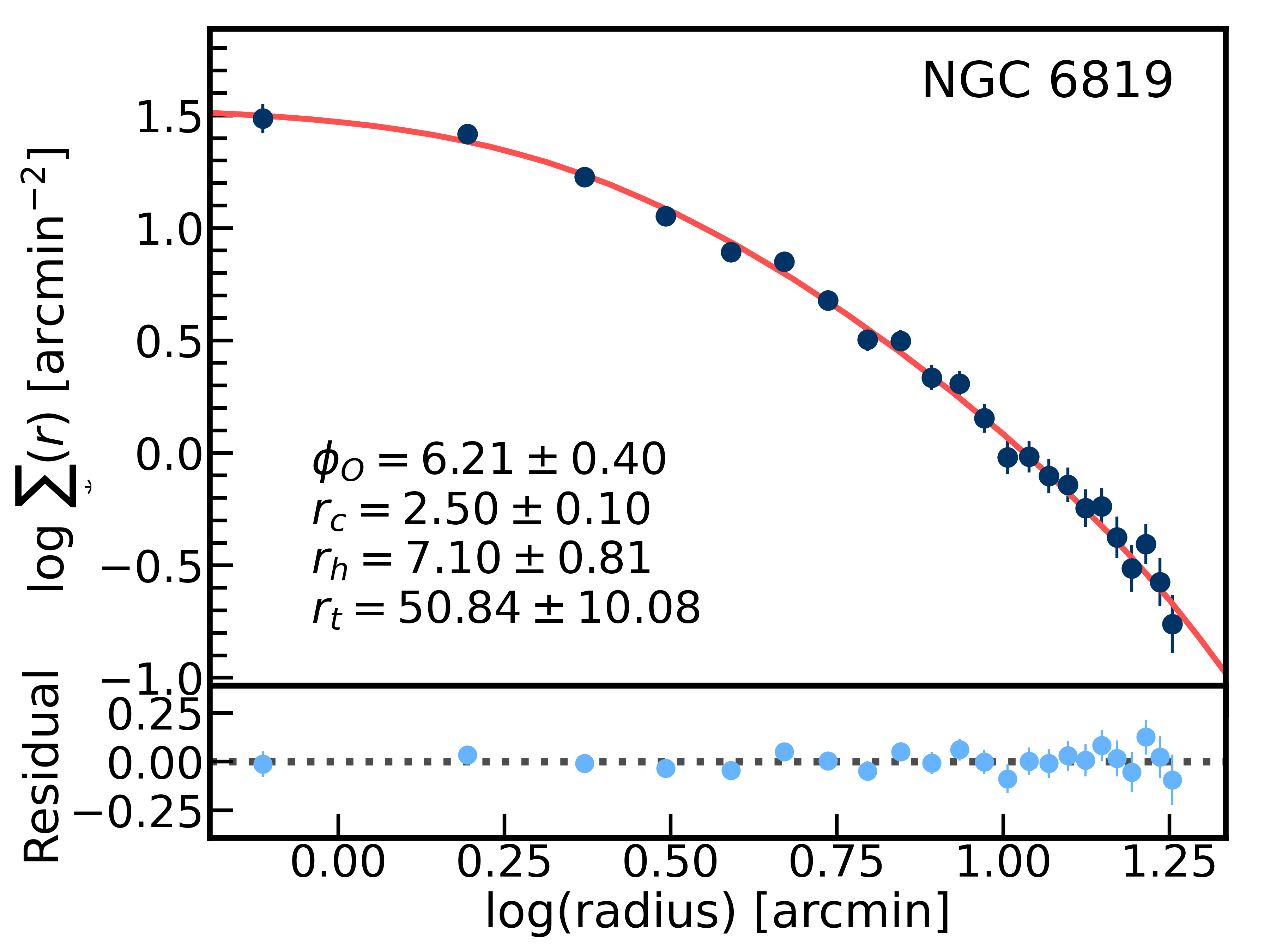}
		\caption*{}
	\end{subfigure}
		\begin{subfigure}[b]{0.32\textwidth}
   		\includegraphics[width=0.99\textwidth]{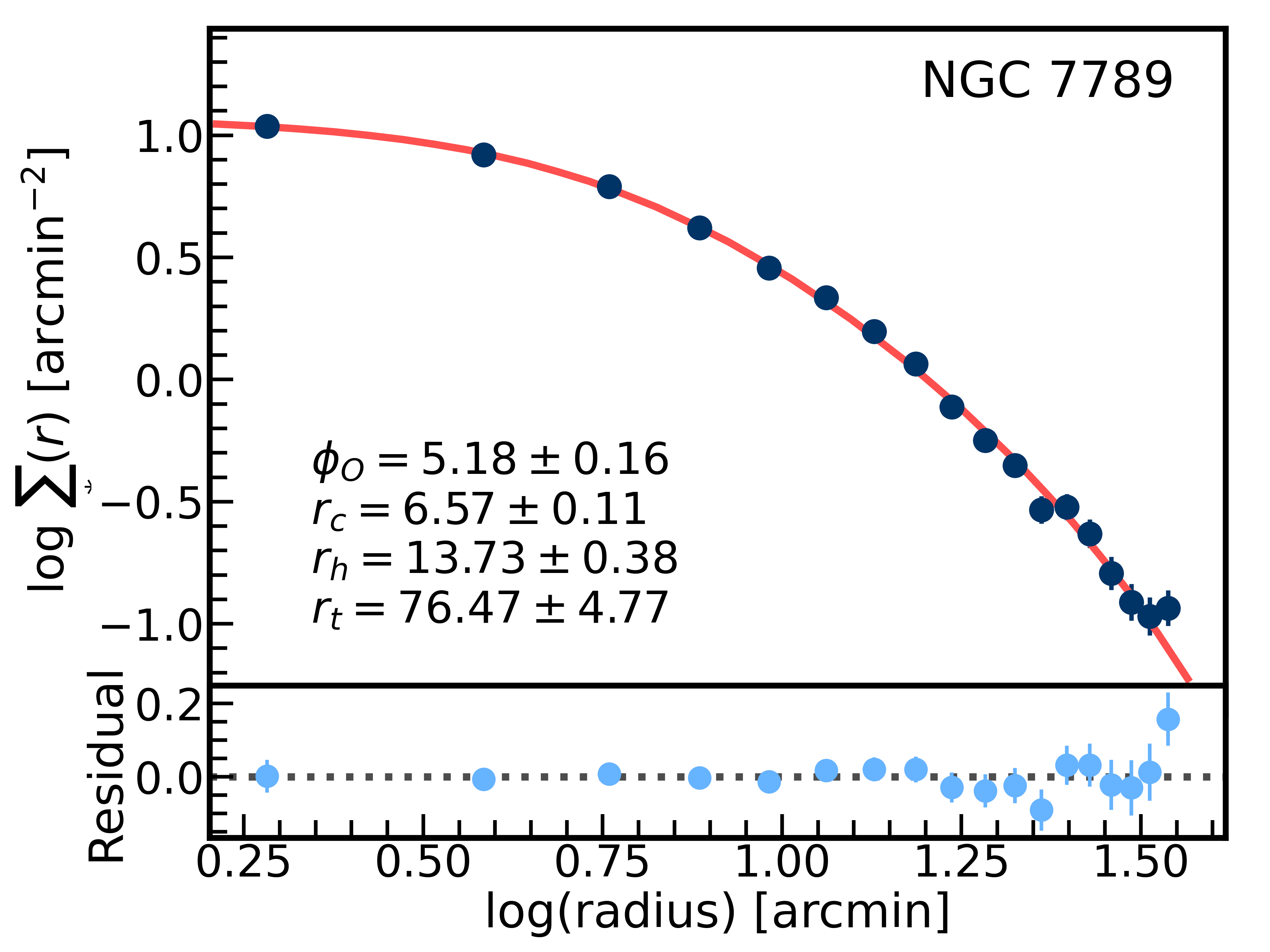}
		\caption*{}
	\end{subfigure}
	\begin{subfigure}[b]{0.32\textwidth}
   		\includegraphics[width=0.99\textwidth]{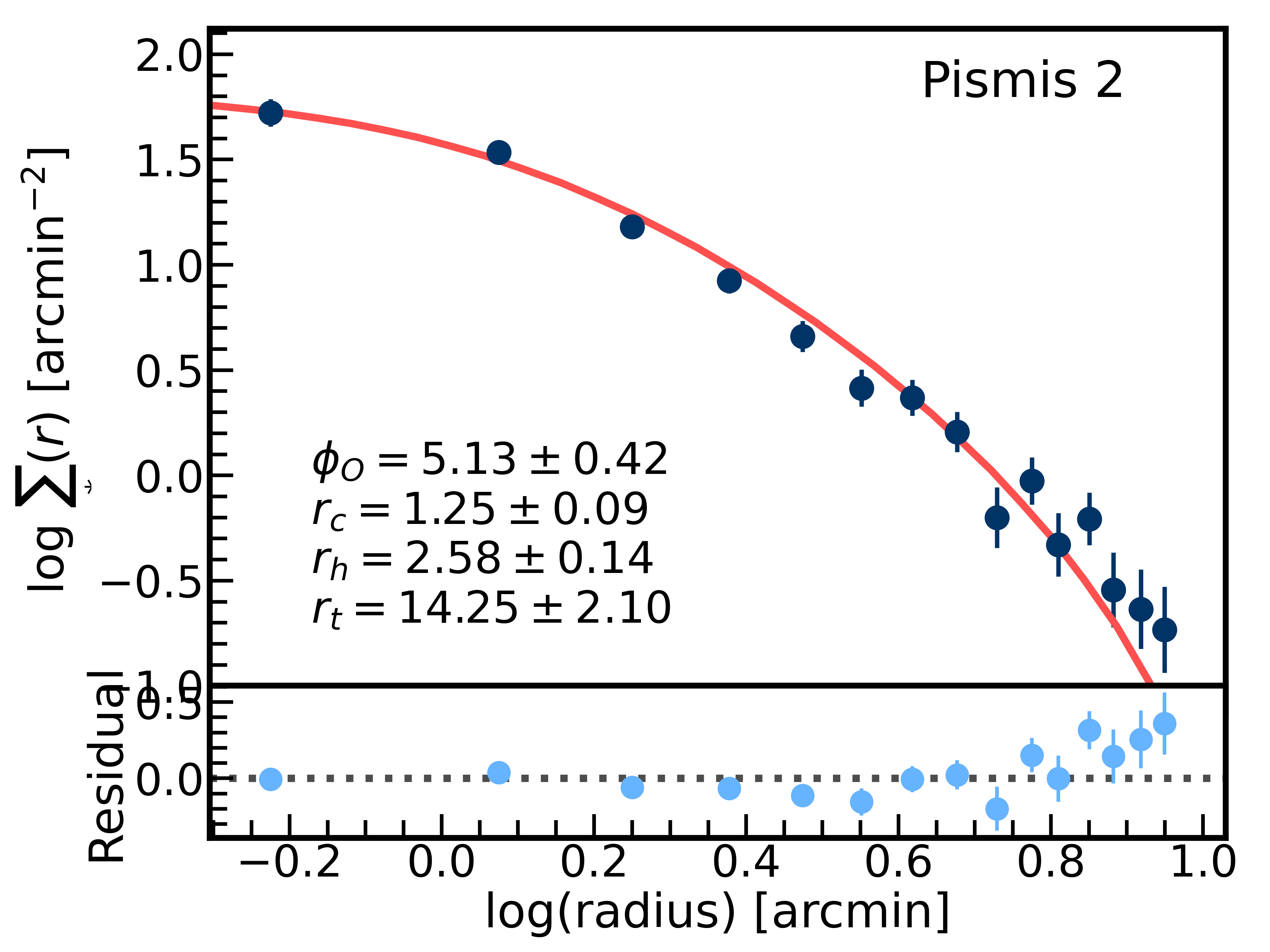}
		\caption*{}
	\end{subfigure}
	\begin{subfigure}[b]{0.32\textwidth}
   		\includegraphics[width=0.99\textwidth]{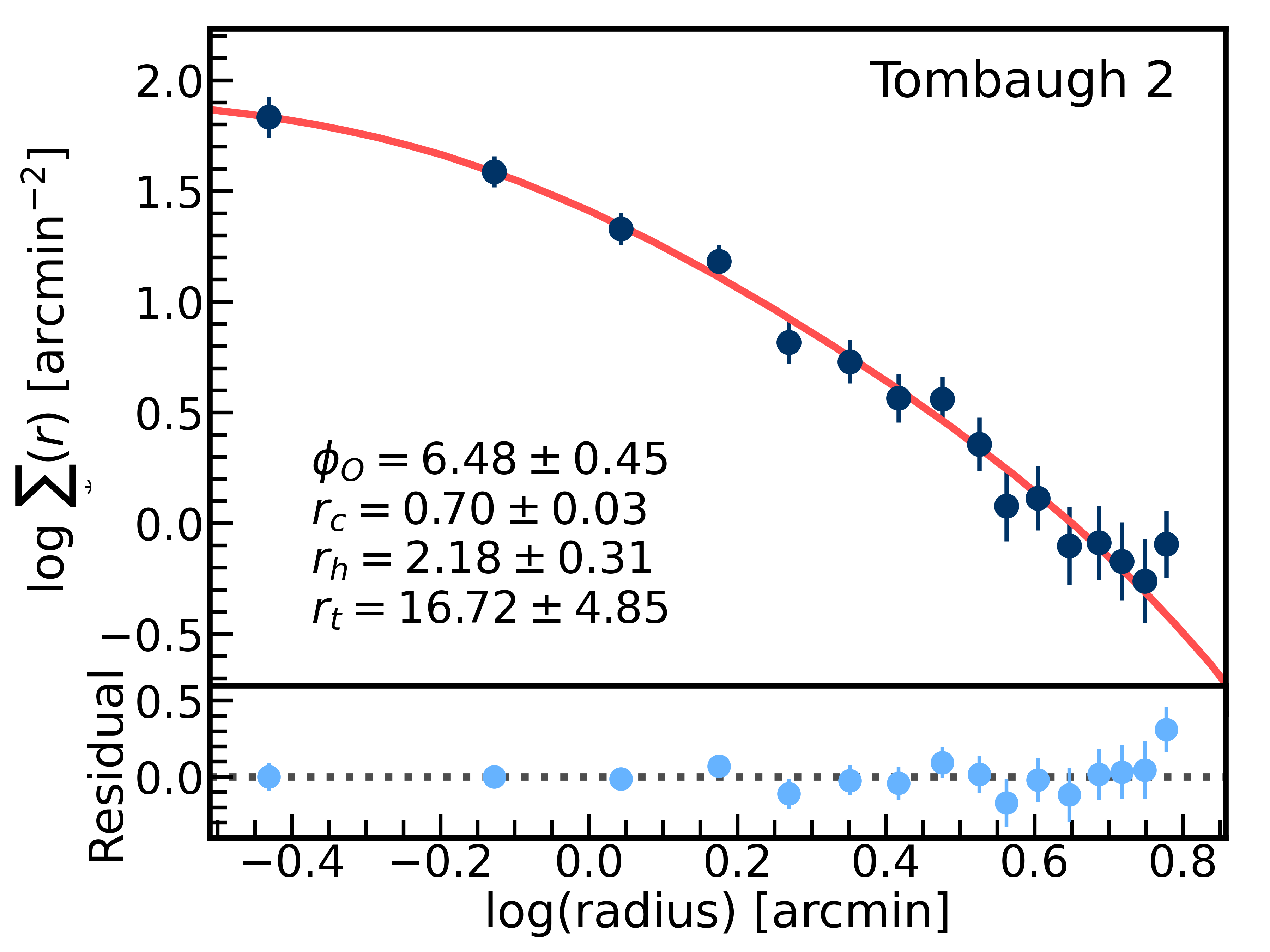}
		\caption*{}
	\end{subfigure}
	\begin{subfigure}[b]{0.32\textwidth}
   		\includegraphics[width=0.99\textwidth]{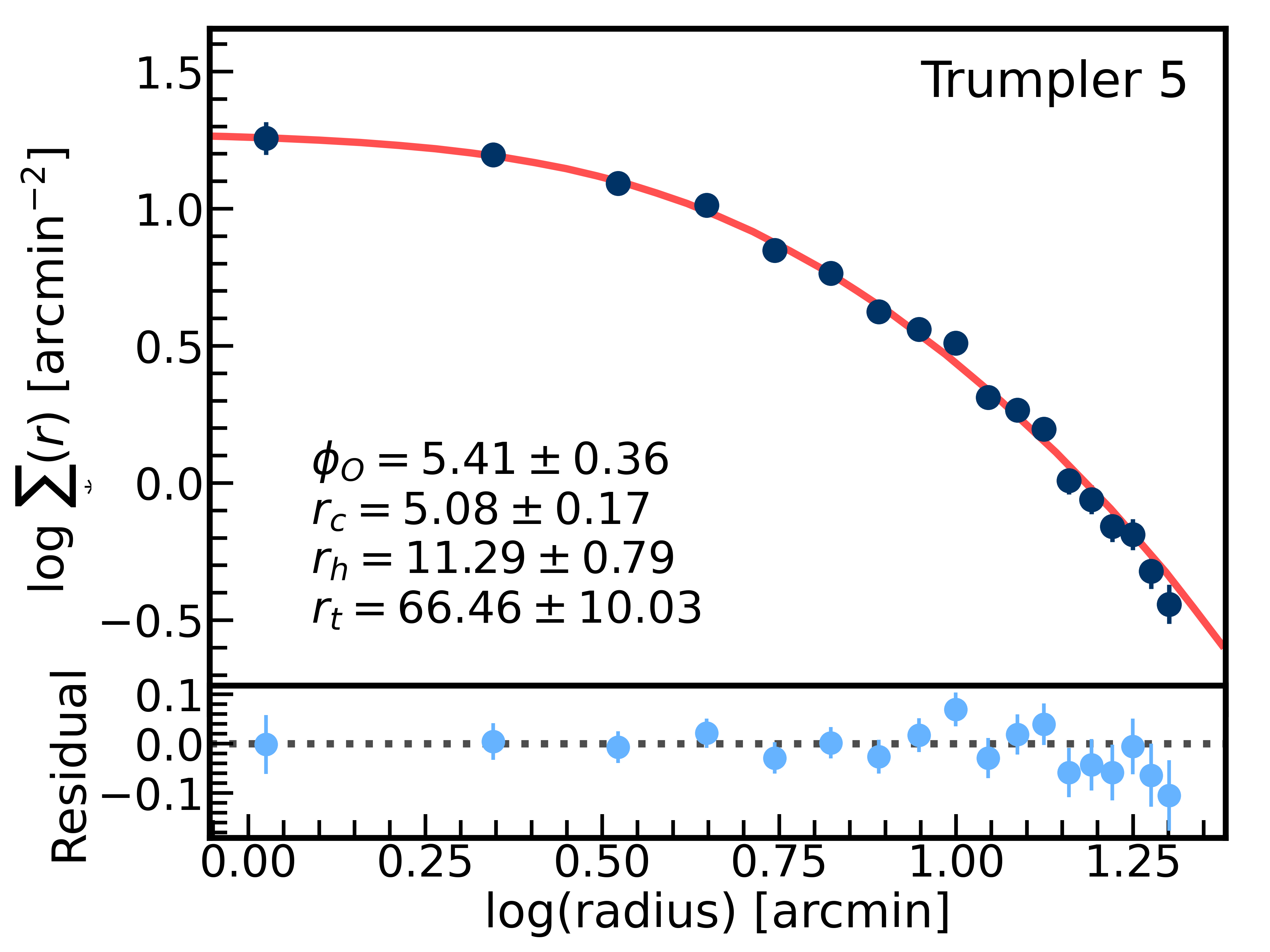}
		\caption*{}
	\end{subfigure}
	\begin{subfigure}[b]{0.32\textwidth}
   		\includegraphics[width=0.99\textwidth]{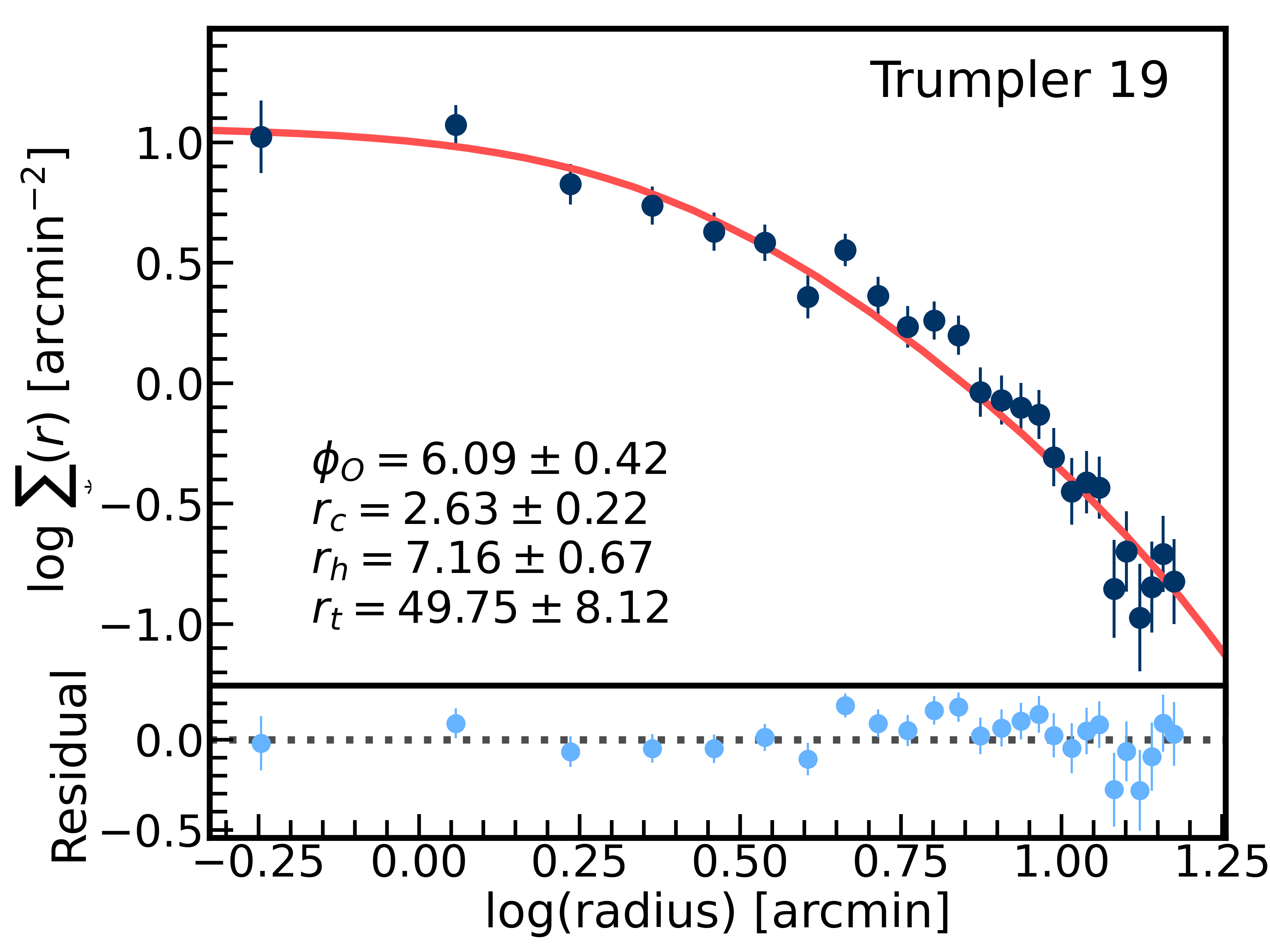}
		\caption*{}
	\end{subfigure}
	\caption{Same as Fig. \ref{fig:kings_profiles} for the other 11 OCs.}
\end{figure*}

\begin{figure*}
    \centering
	\begin{subfigure}[b]{0.24\textwidth}
   		\includegraphics[width=1.0\textwidth]{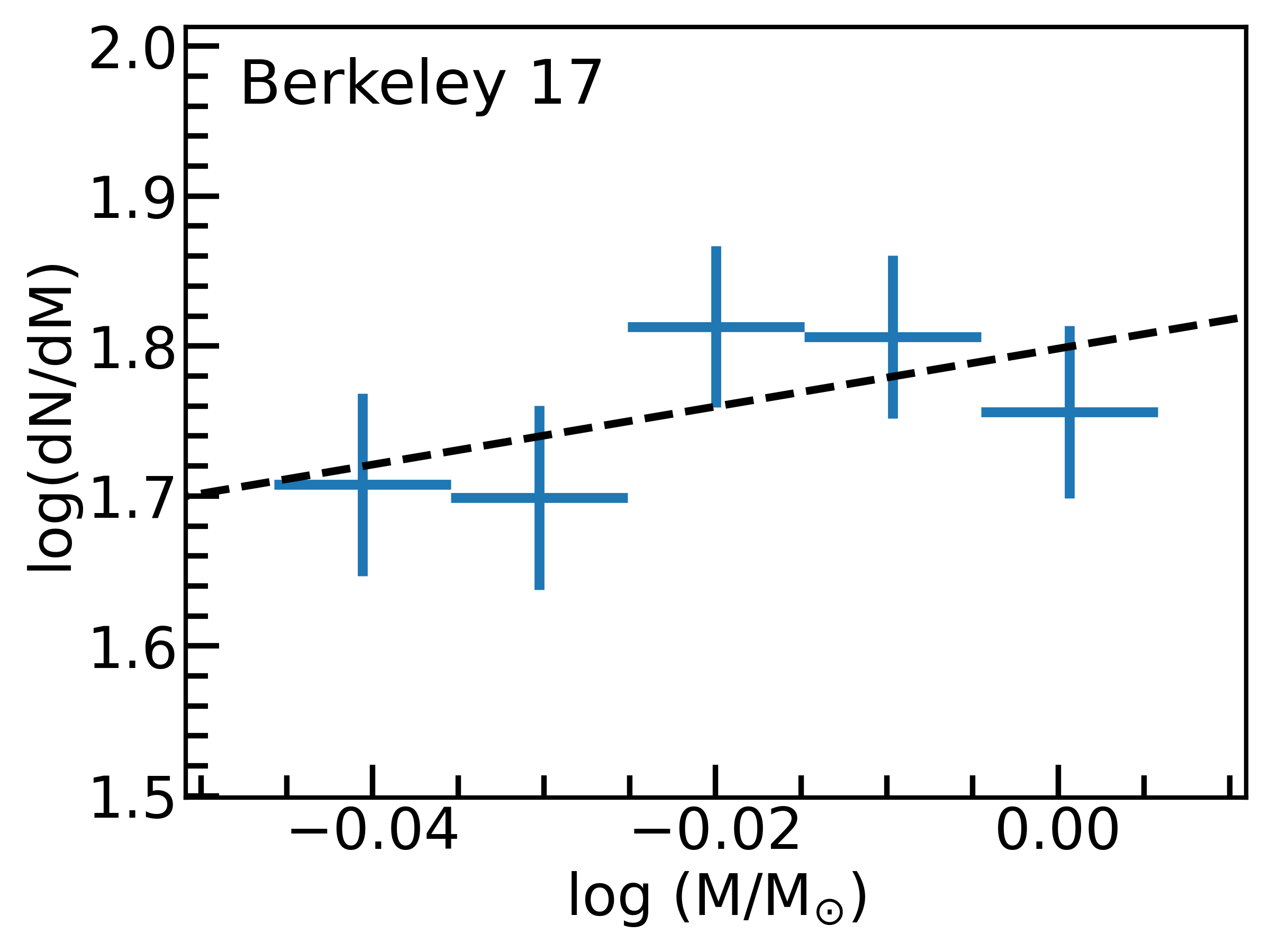}
		\caption*{}
	\end{subfigure}
	\begin{subfigure}[b]{0.24\textwidth}
   		\includegraphics[width=1.0\textwidth]{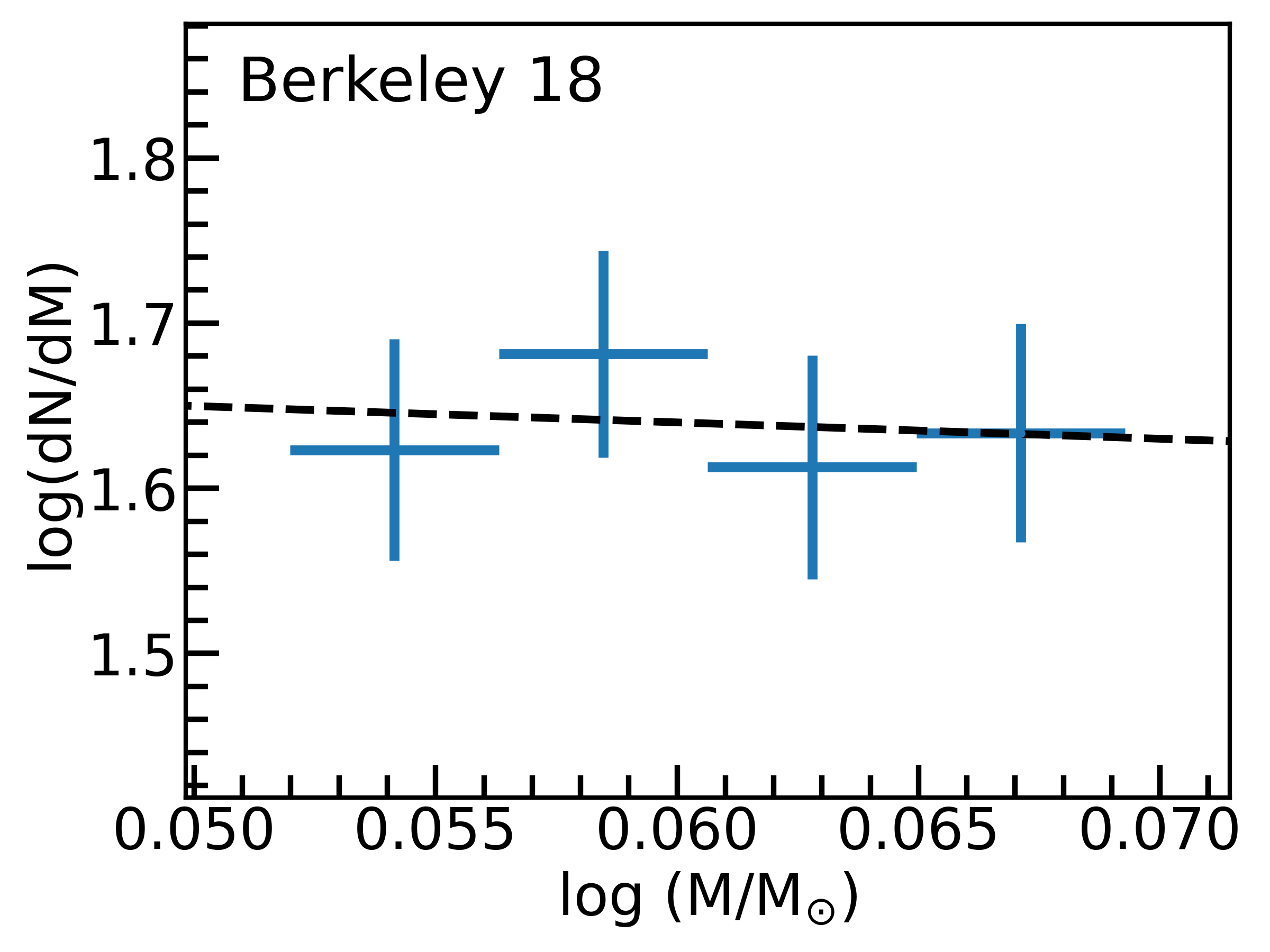}
		\caption*{}
	\end{subfigure}
	\begin{subfigure}[b]{0.24\textwidth}
   		\includegraphics[width=1.0\textwidth]{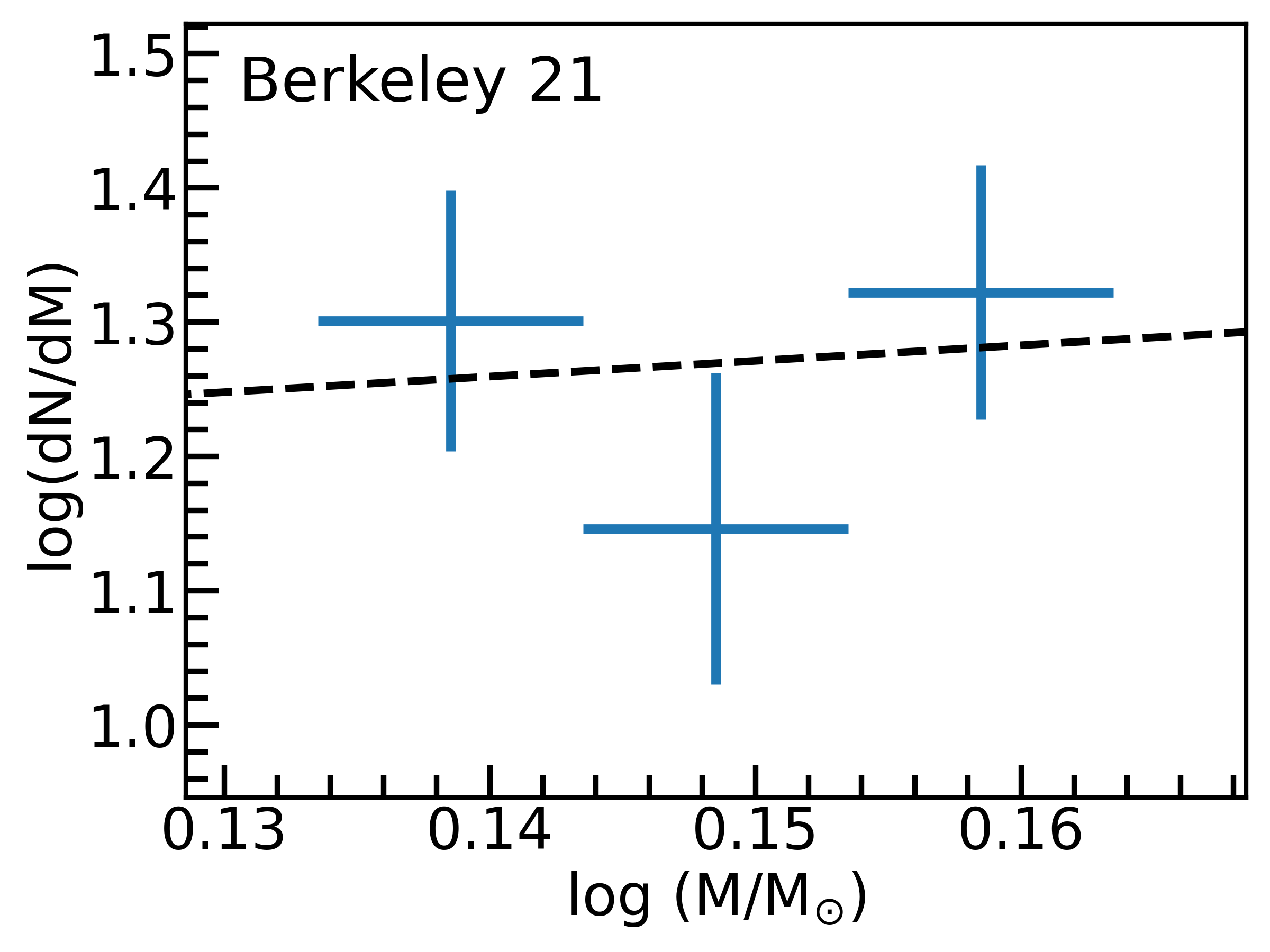}
		\caption*{}
	\end{subfigure}
	\begin{subfigure}[b]{0.24\textwidth}
   		\includegraphics[width=1.0\textwidth]{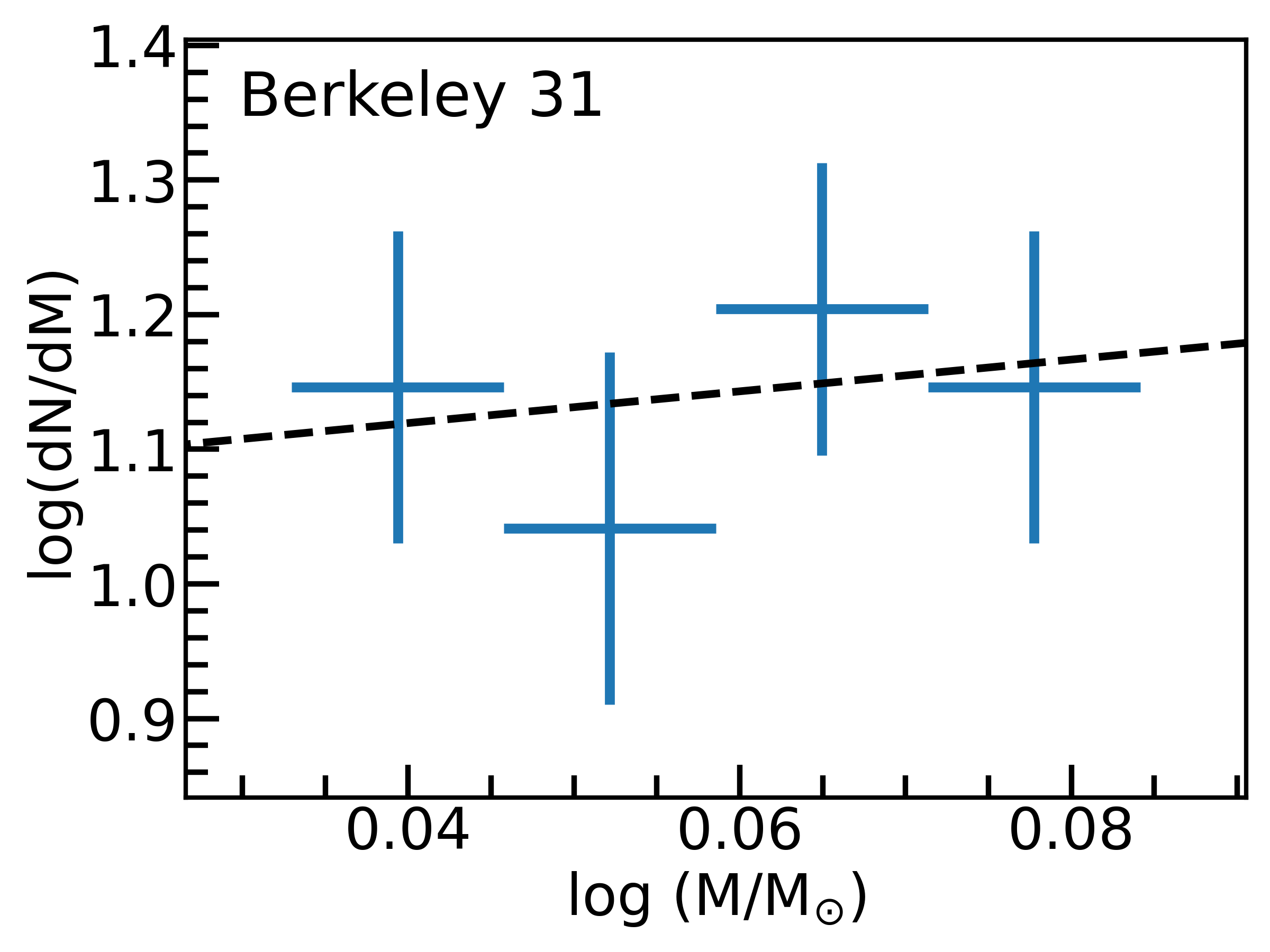}
		\caption*{}
	\end{subfigure}
	\begin{subfigure}[b]{0.24\textwidth}
   		\includegraphics[width=1.0\textwidth]{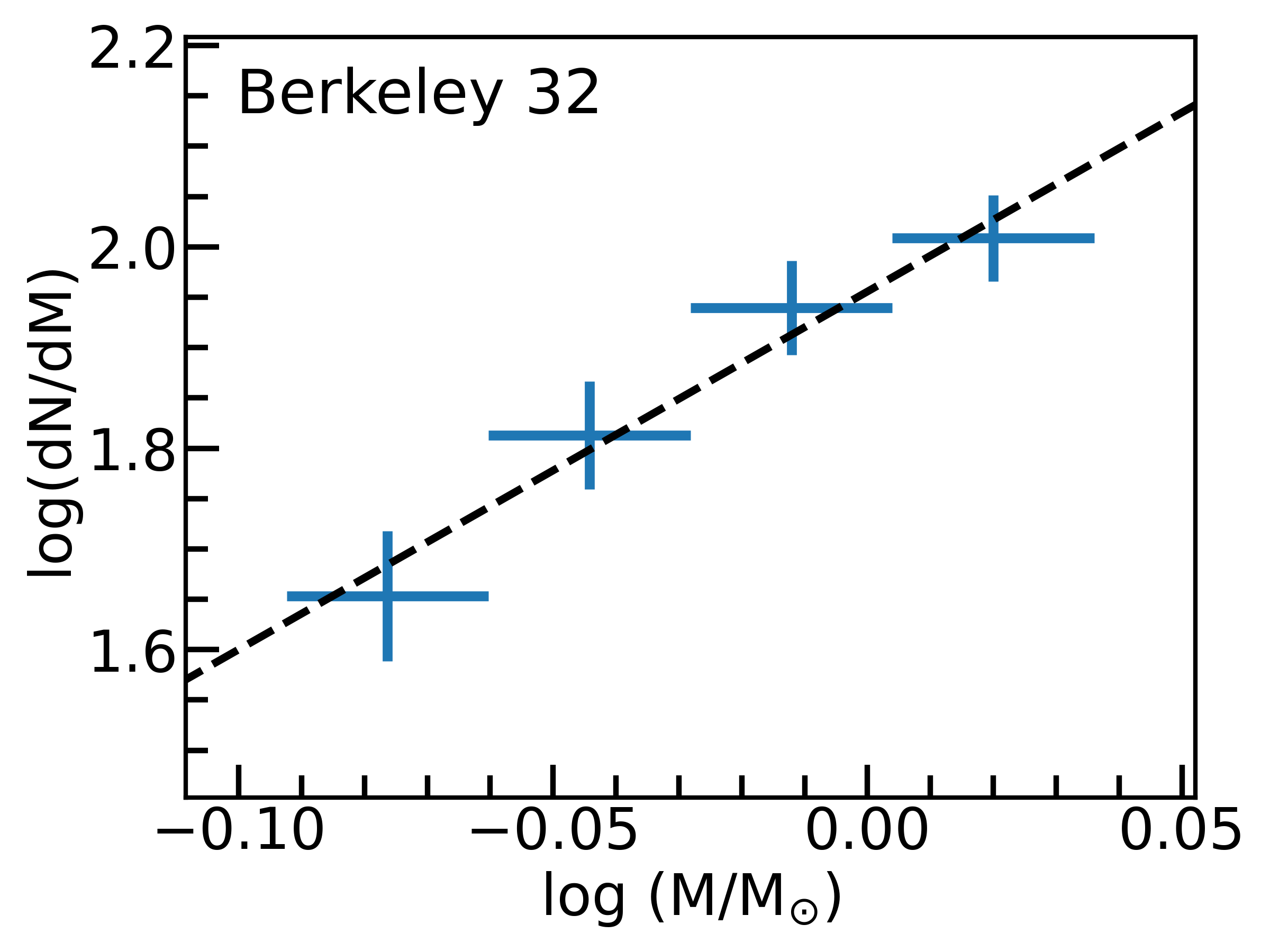}
		\caption*{}
	\end{subfigure}
	\begin{subfigure}[b]{0.24\textwidth}
   		\includegraphics[width=1.0\textwidth]{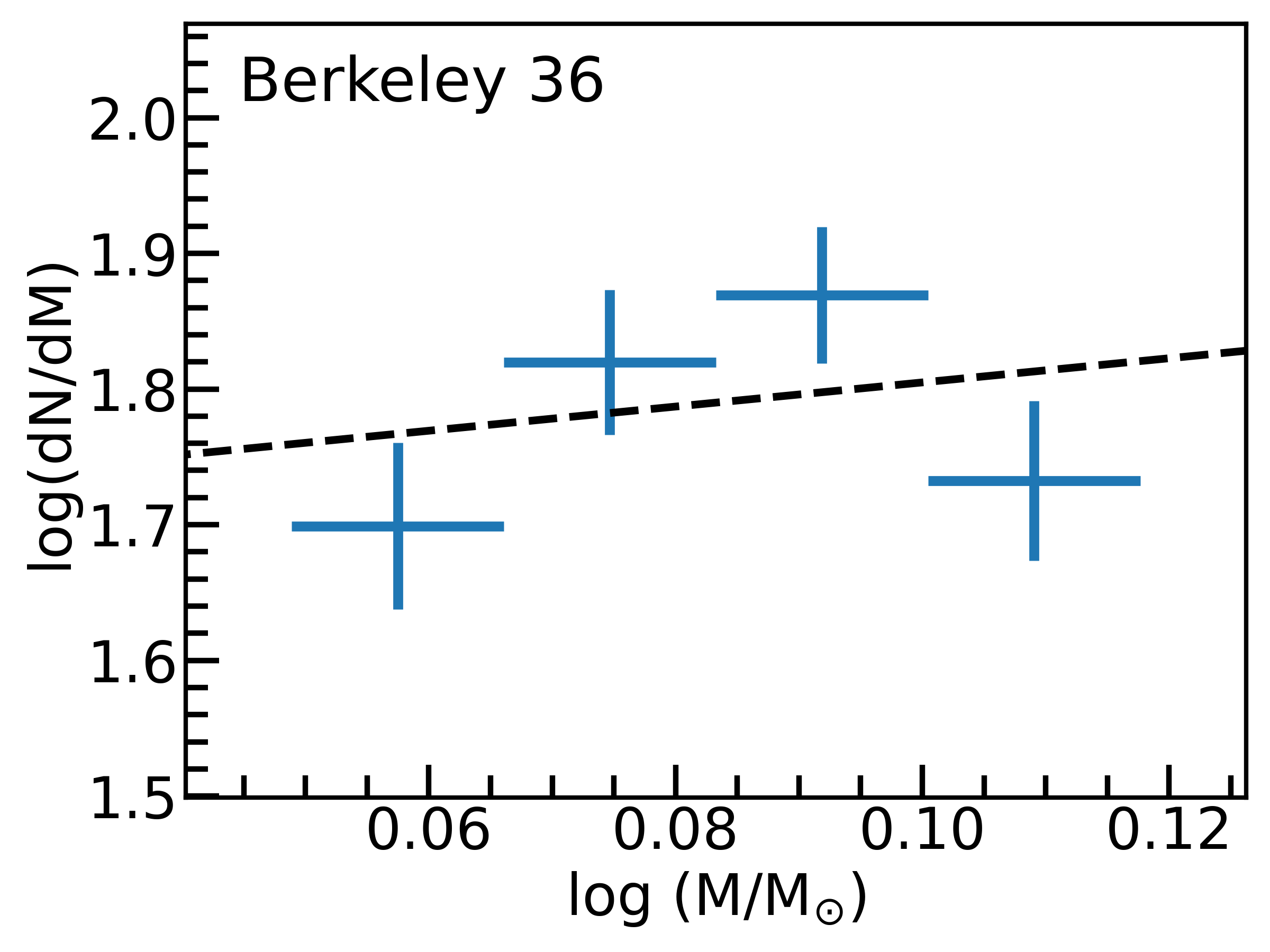}
		\caption*{}
	\end{subfigure}
	\begin{subfigure}[b]{0.24\textwidth}
   		\includegraphics[width=1.0\textwidth]{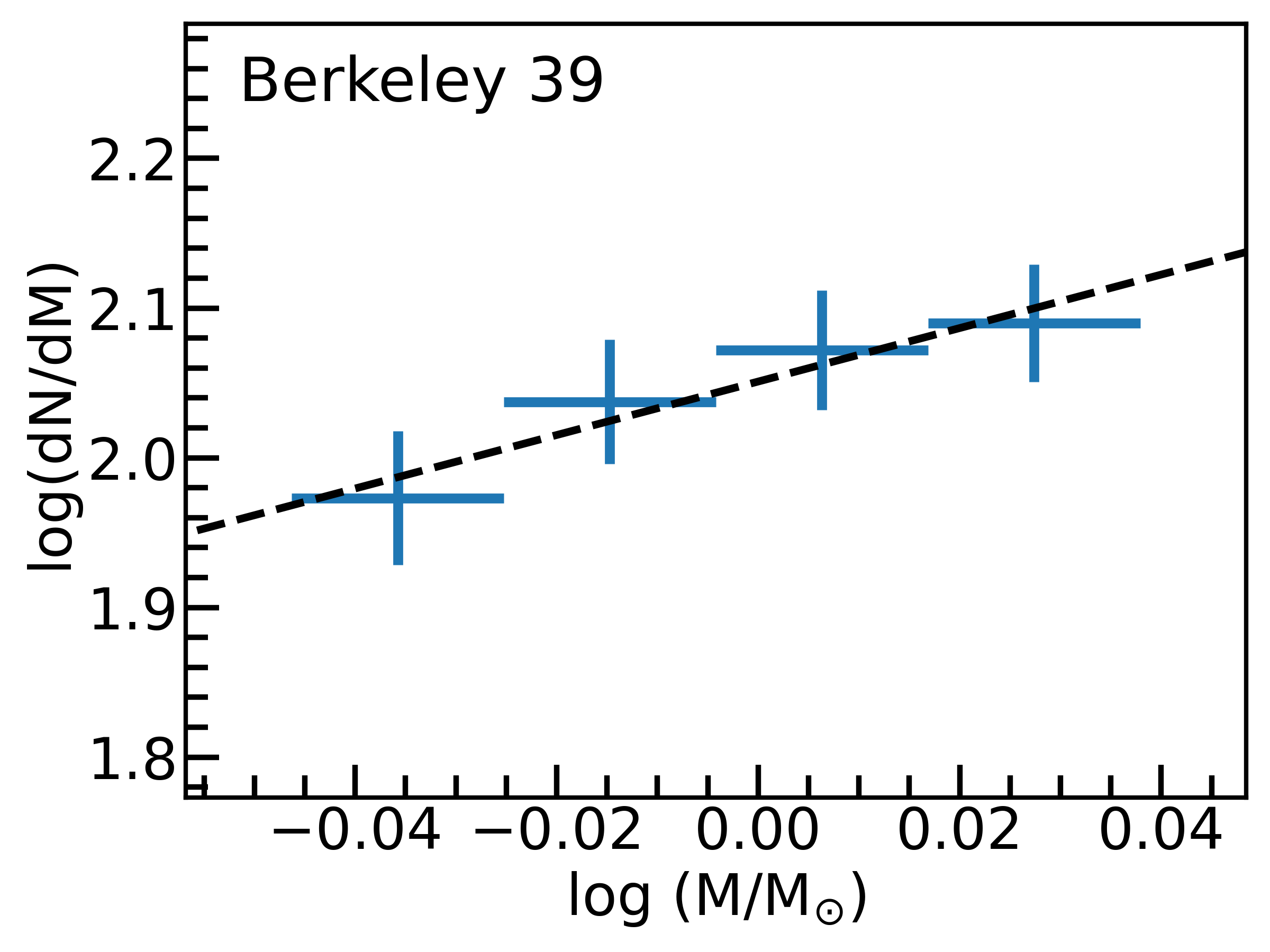}
		\caption*{}
	\end{subfigure}
	\begin{subfigure}[b]{0.24\textwidth}
   		\includegraphics[width=1.0\textwidth]{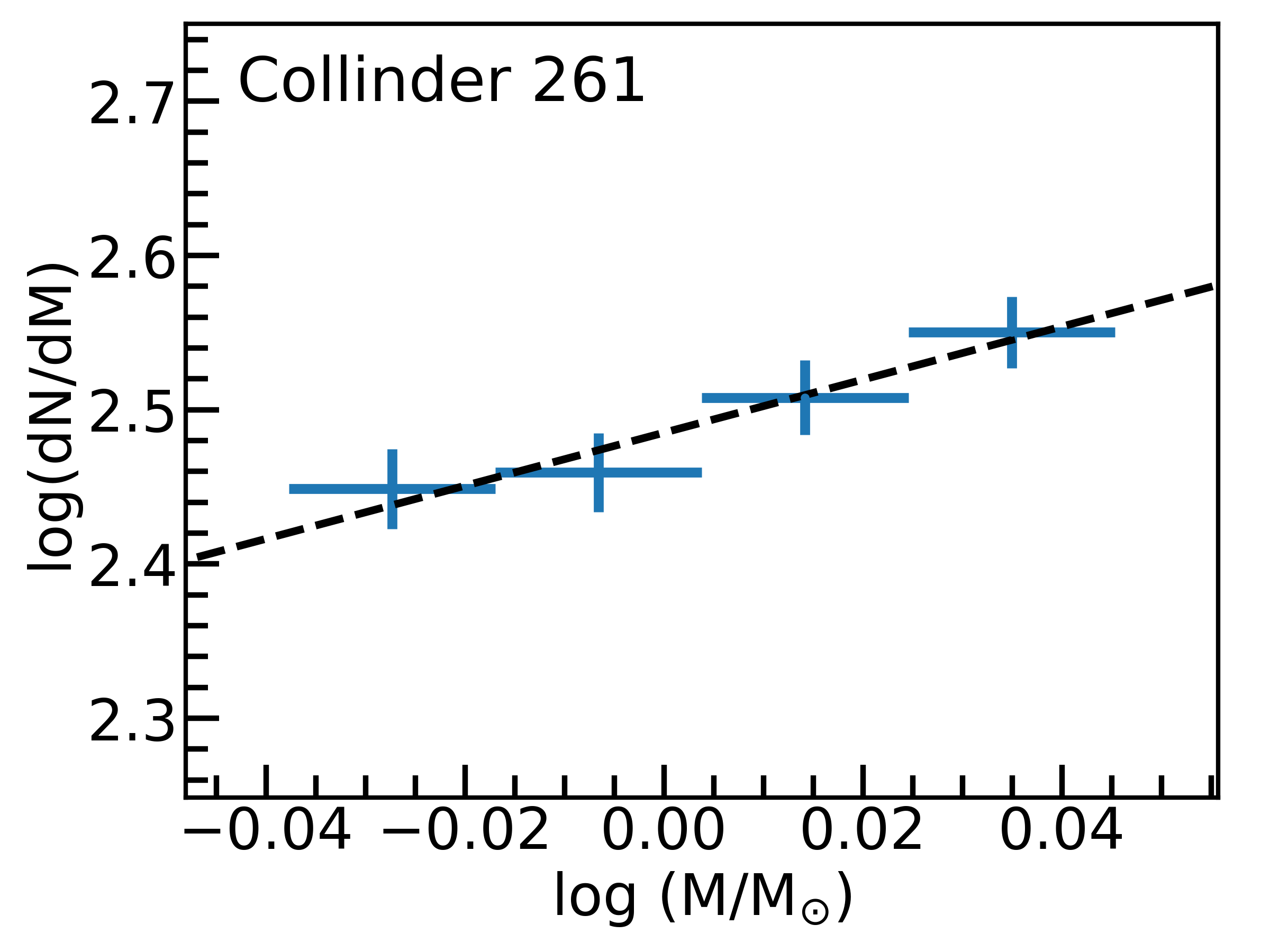}
		\caption*{}
	\end{subfigure}
	\begin{subfigure}[b]{0.24\textwidth}
   		\includegraphics[width=1.0\textwidth]{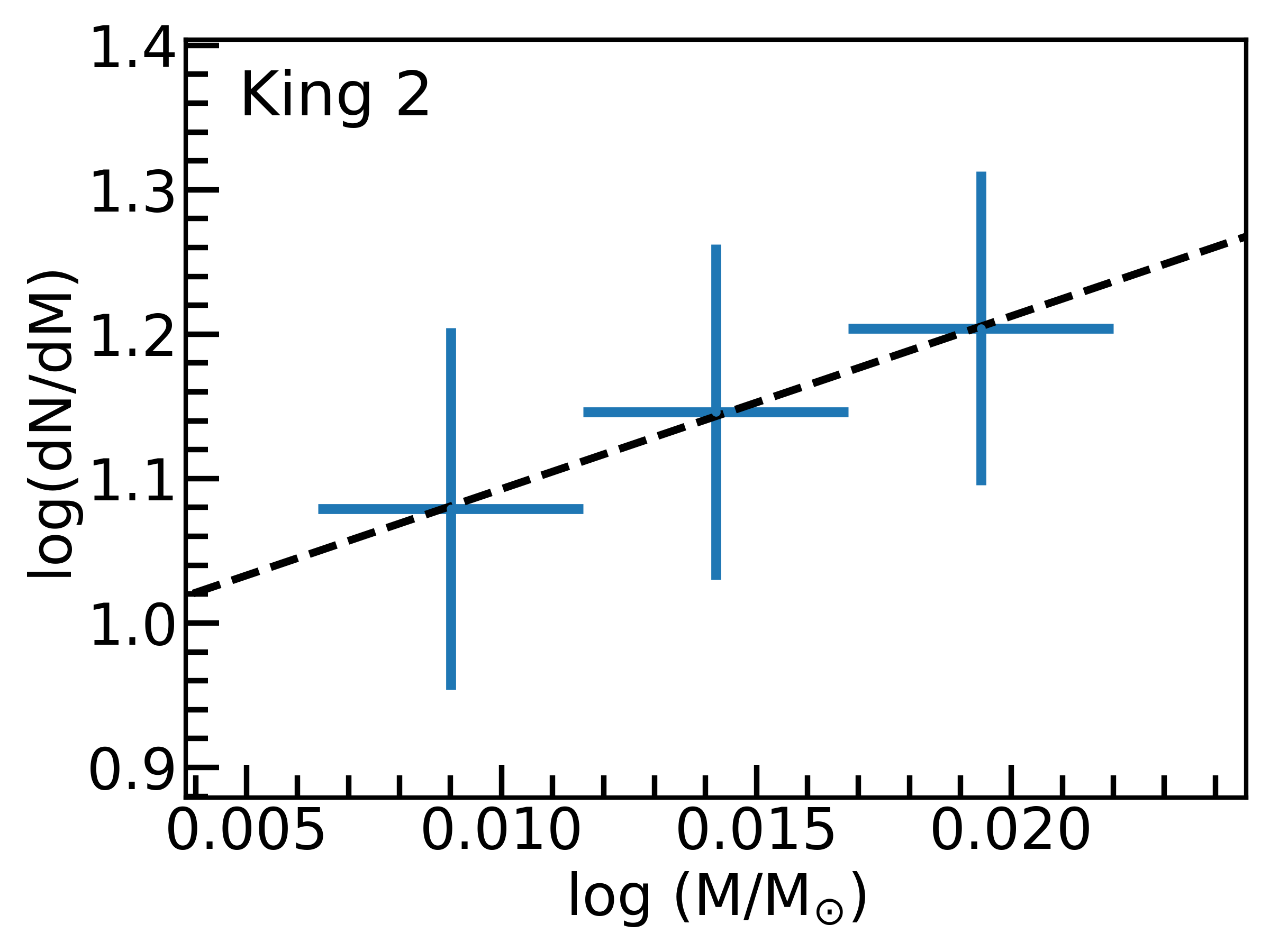}
		\caption*{}
	\end{subfigure}
	\begin{subfigure}[b]{0.24\textwidth}
   		\includegraphics[width=1.0\textwidth]{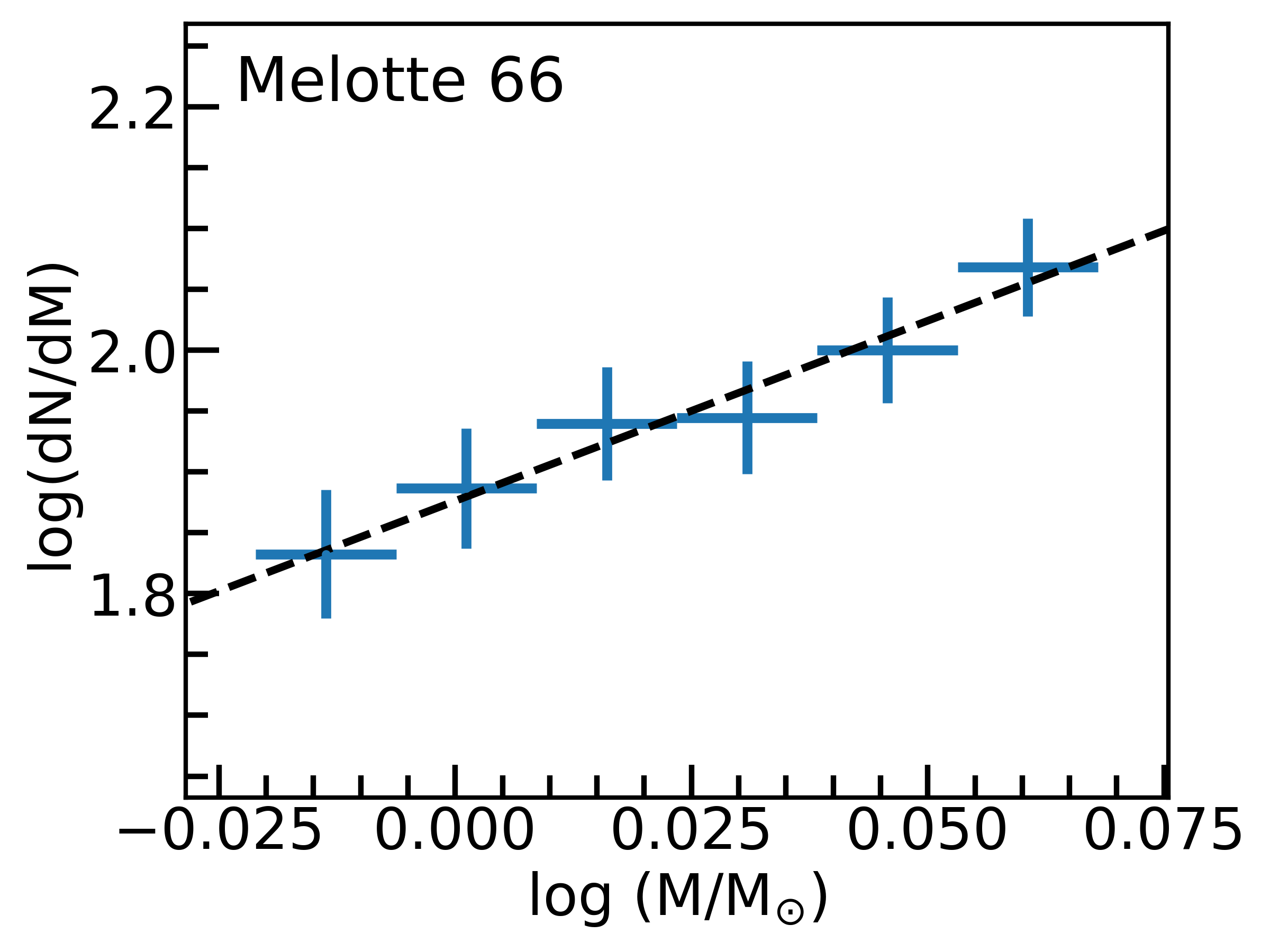}
		\caption*{}
	\end{subfigure}
	\begin{subfigure}[b]{0.24\textwidth}
   		\includegraphics[width=1.0\textwidth]{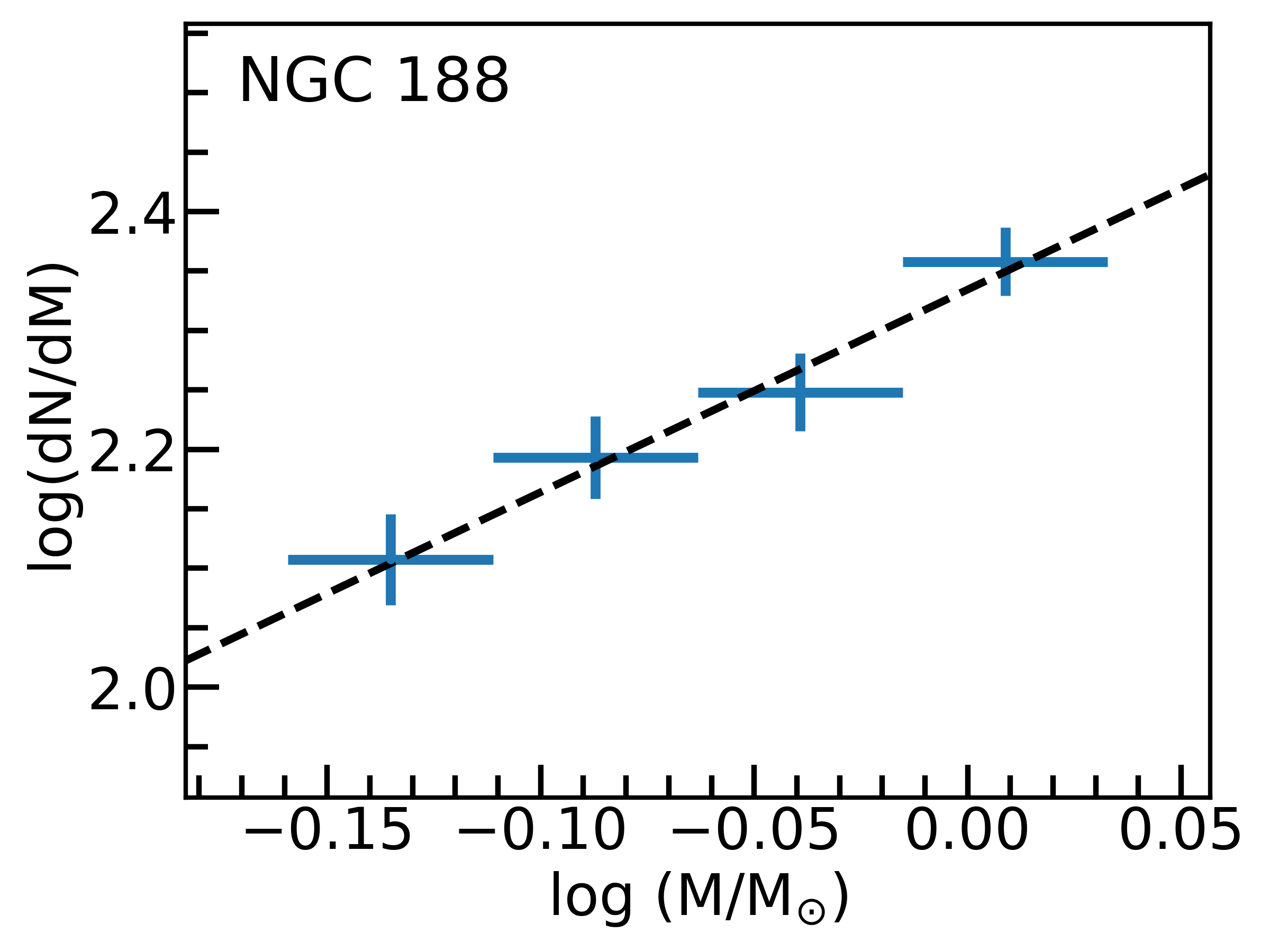}
		\caption*{}
	\end{subfigure}
	\begin{subfigure}[b]{0.24\textwidth}
   		\includegraphics[width=1.0\textwidth]{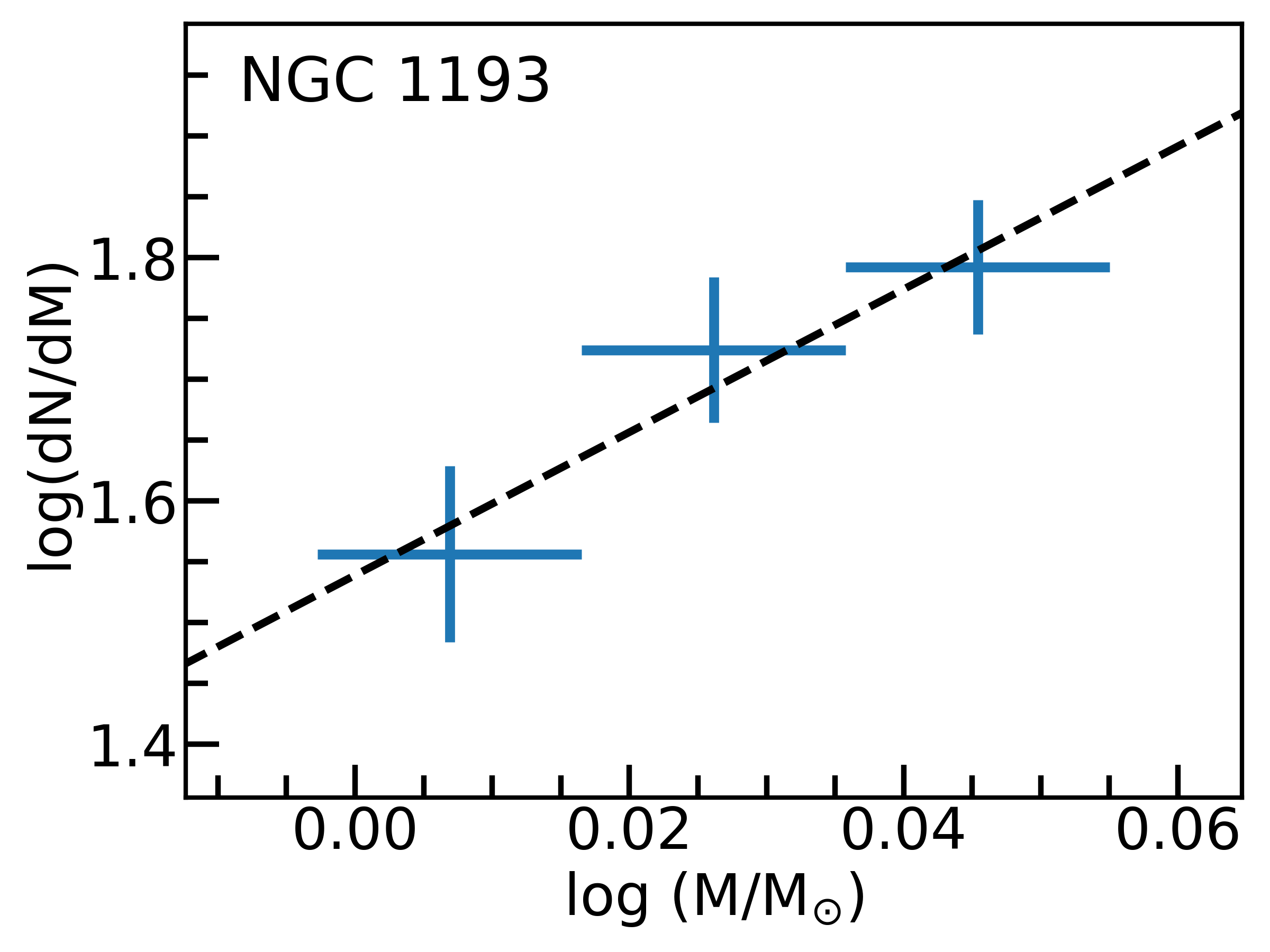}
		\caption*{}
	\end{subfigure}
	\begin{subfigure}[b]{0.24\textwidth}
   		\includegraphics[width=1.0\textwidth]{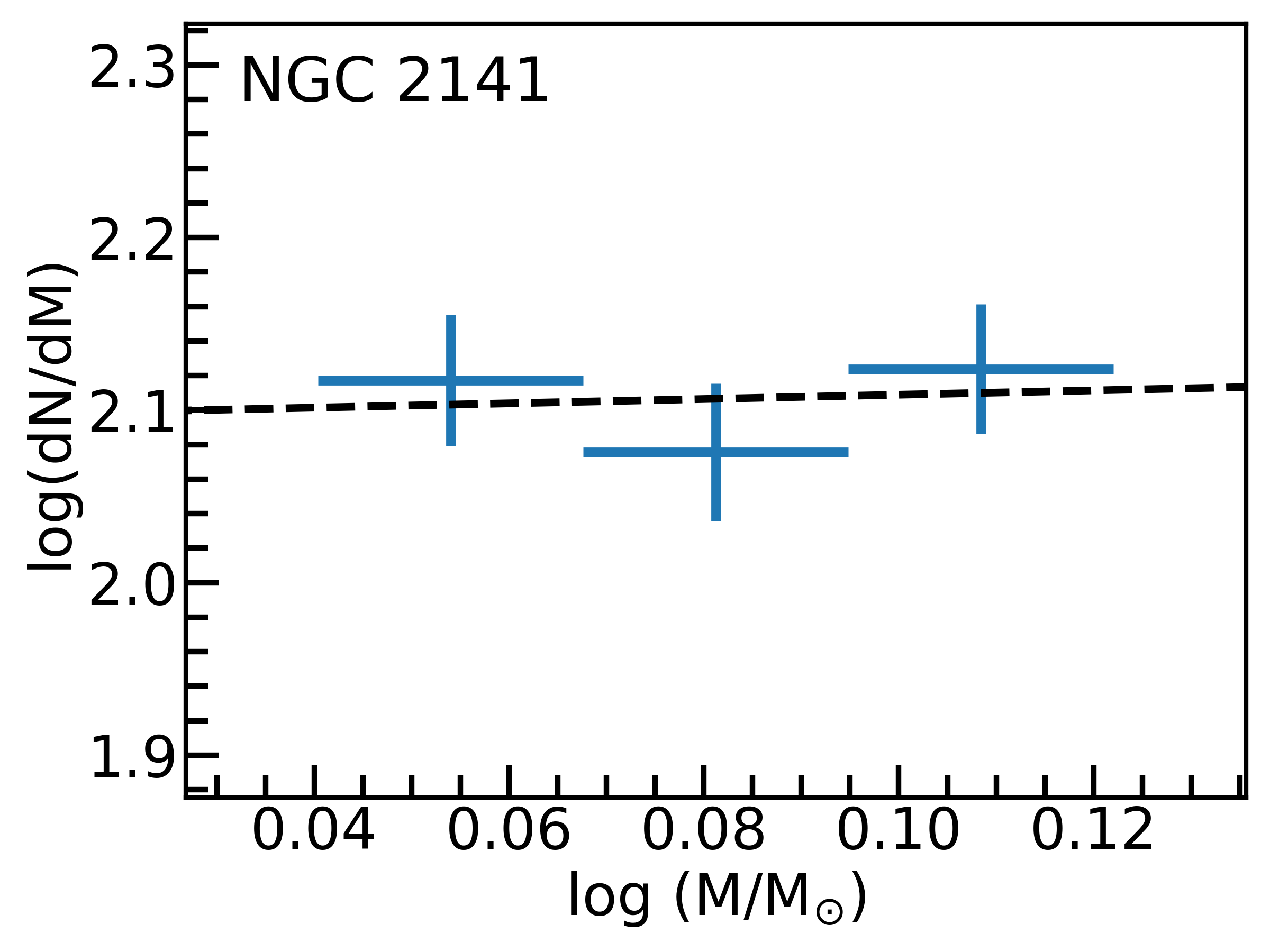}
		\caption*{}
	\end{subfigure}
	\begin{subfigure}[b]{0.24\textwidth}
   		\includegraphics[width=1.0\textwidth]{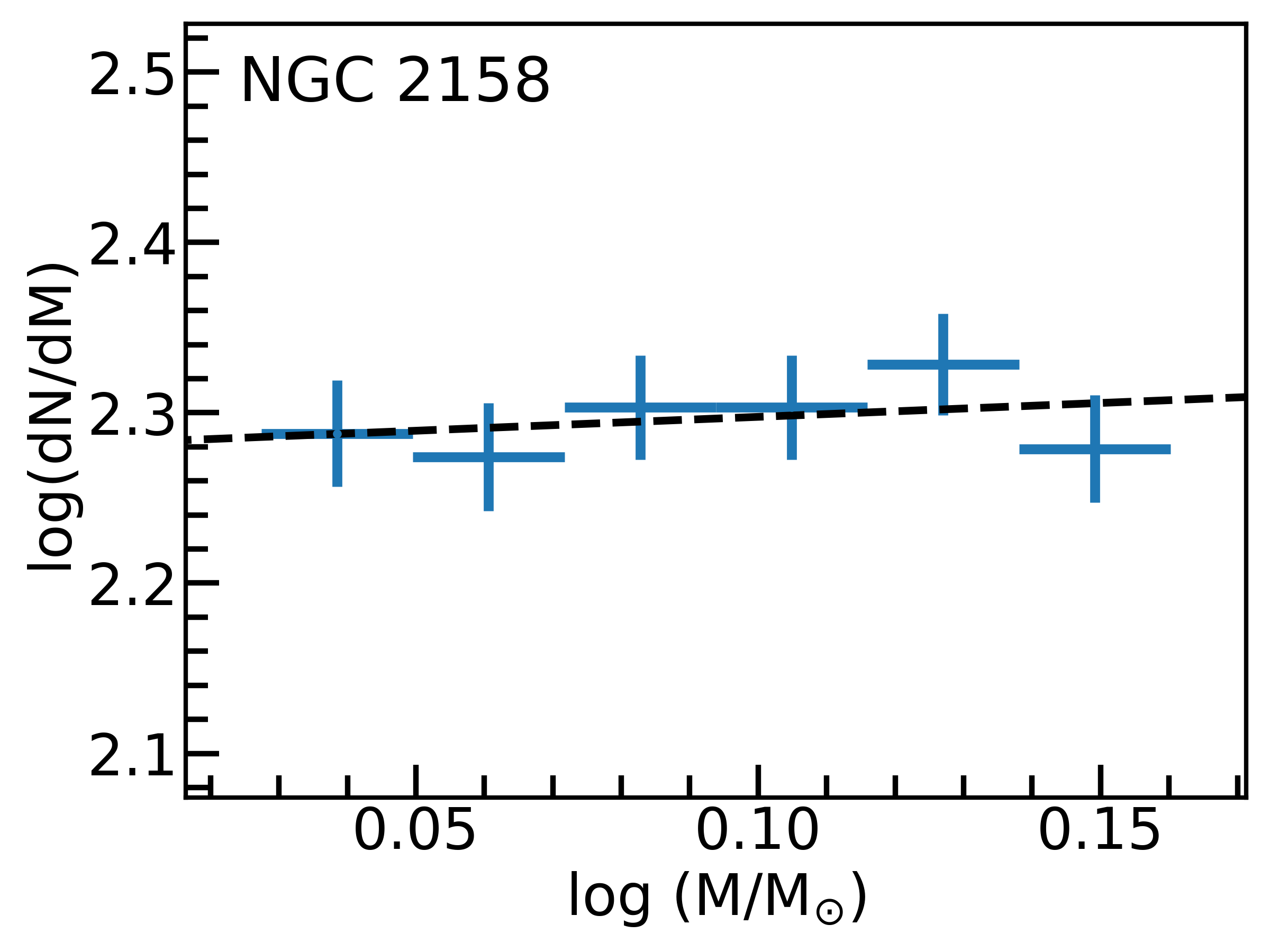}
		\caption*{}
	\end{subfigure}
	\begin{subfigure}[b]{0.24\textwidth}
   		\includegraphics[width=1.0\textwidth]{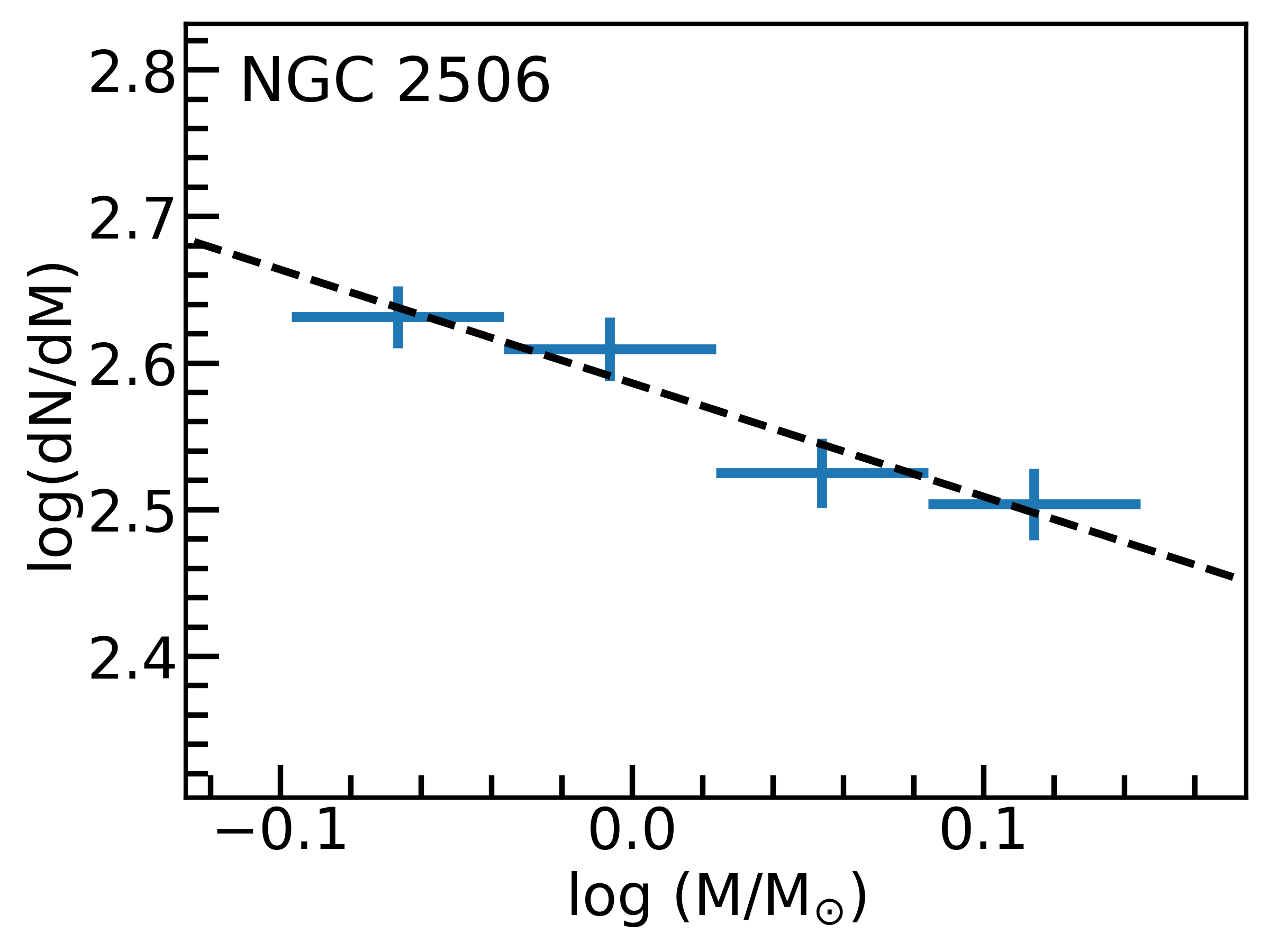}
		\caption*{}
	\end{subfigure}
	\begin{subfigure}[b]{0.24\textwidth}
   		\includegraphics[width=1.0\textwidth]{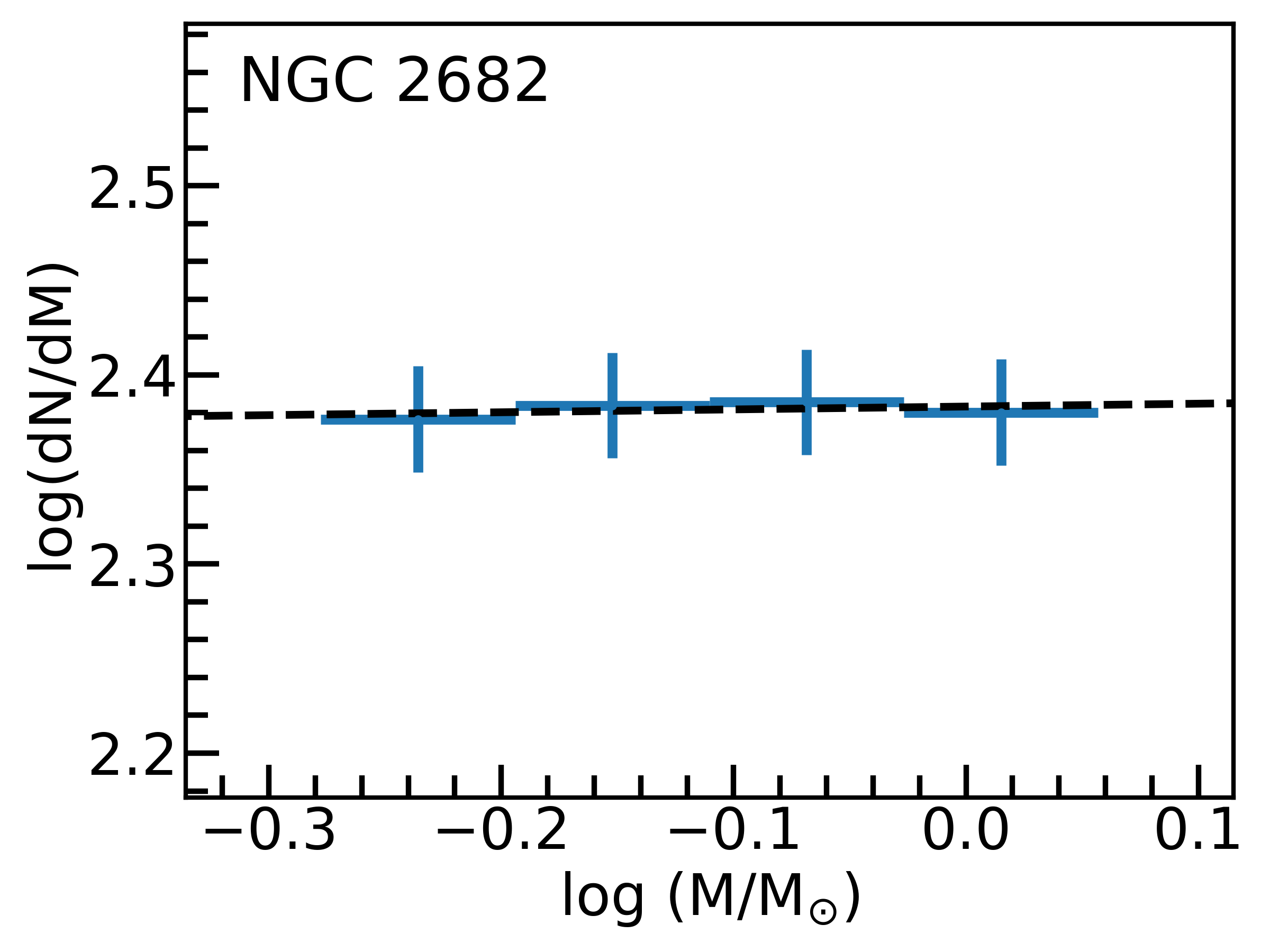}
		\caption*{}
	\end{subfigure}
	\begin{subfigure}[b]{0.24\textwidth}
   		\includegraphics[width=1.0\textwidth]{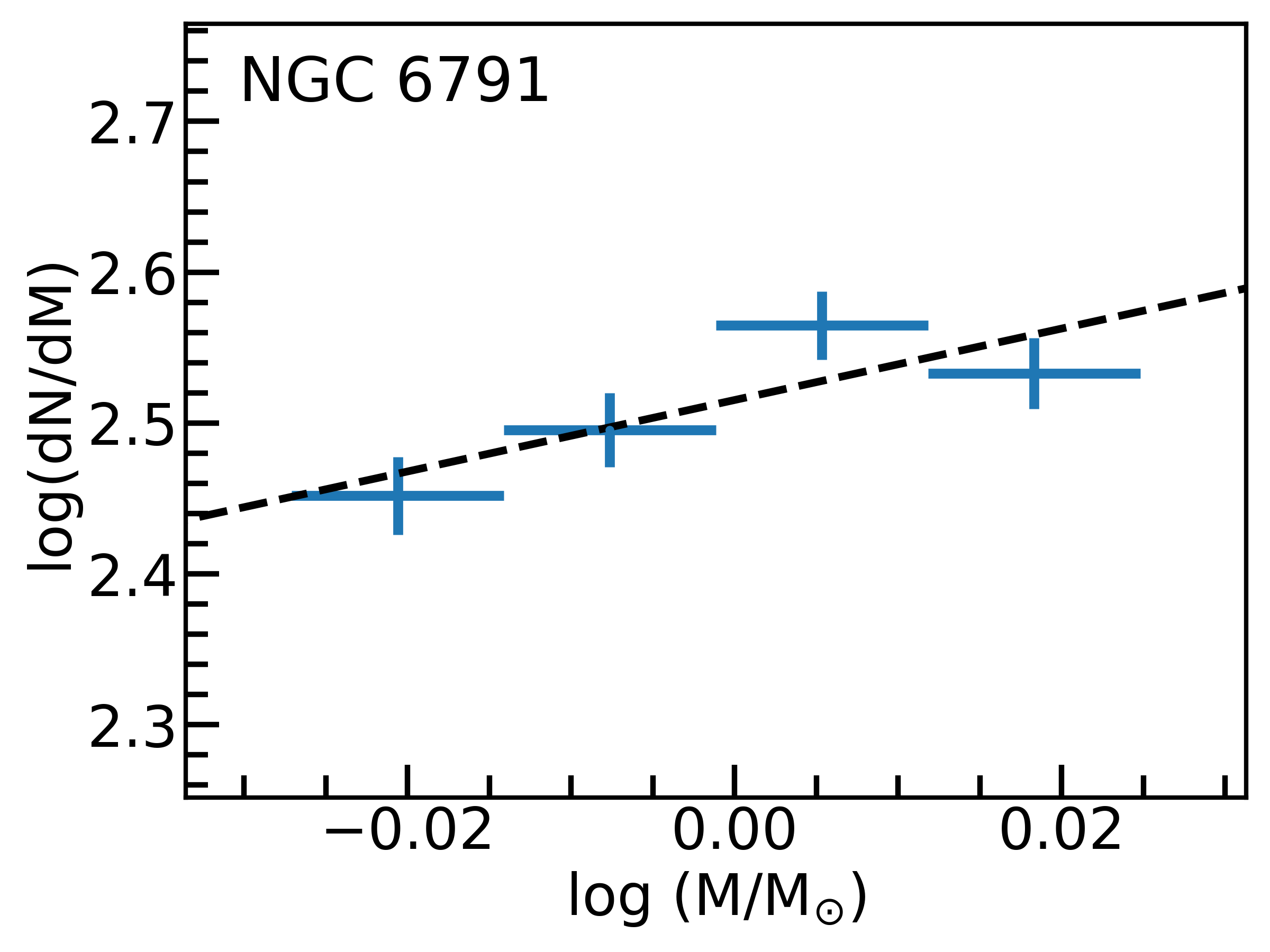}
		\caption*{}
	\end{subfigure}
	\begin{subfigure}[b]{0.24\textwidth}
   		\includegraphics[width=1.0\textwidth]{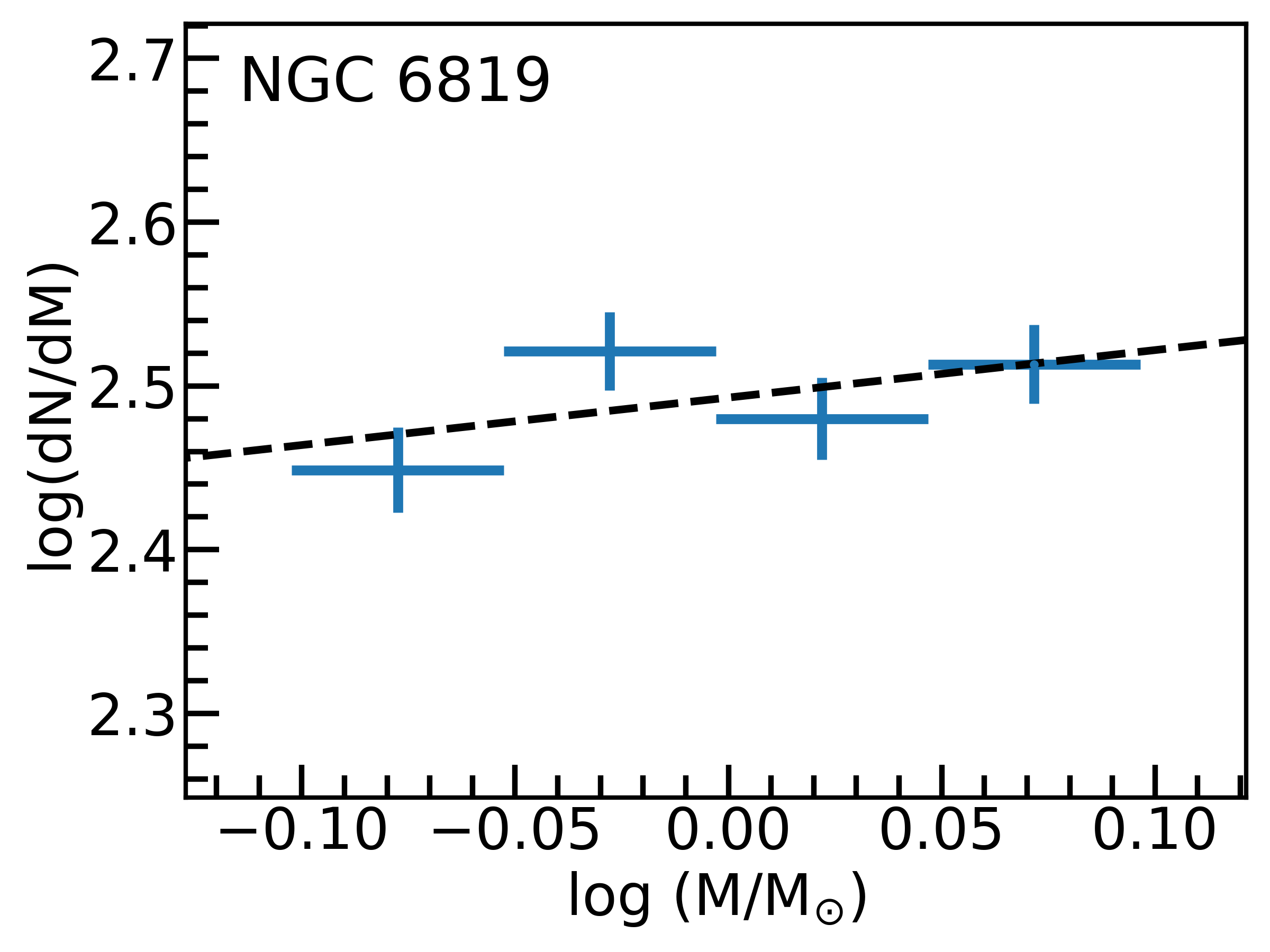}
		\caption*{}
	\end{subfigure}
	\begin{subfigure}[b]{0.24\textwidth}
   		\includegraphics[width=1.0\textwidth]{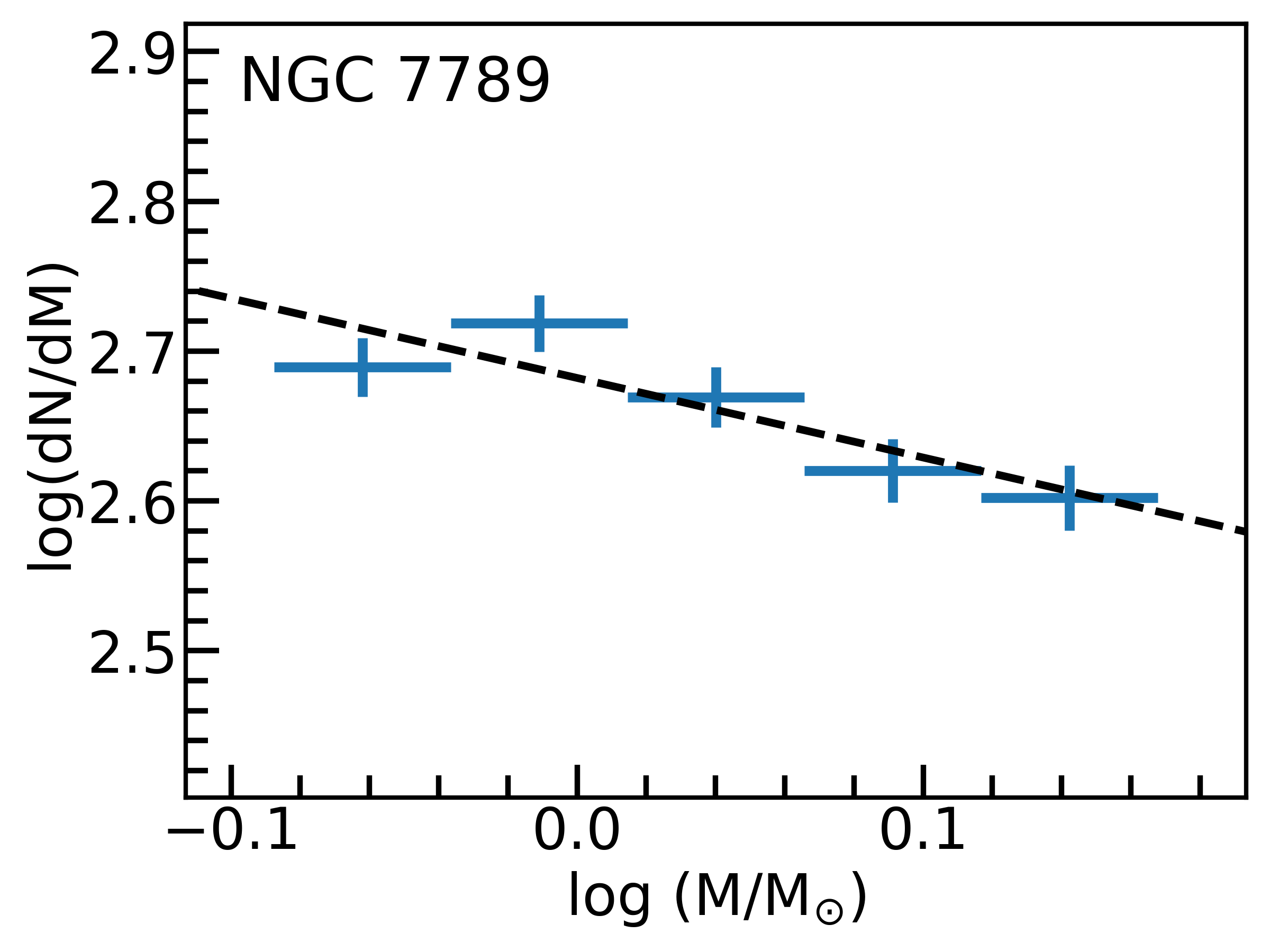}
		\caption*{}
	\end{subfigure}
	\begin{subfigure}[b]{0.24\textwidth}
   		\includegraphics[width=1.0\textwidth]{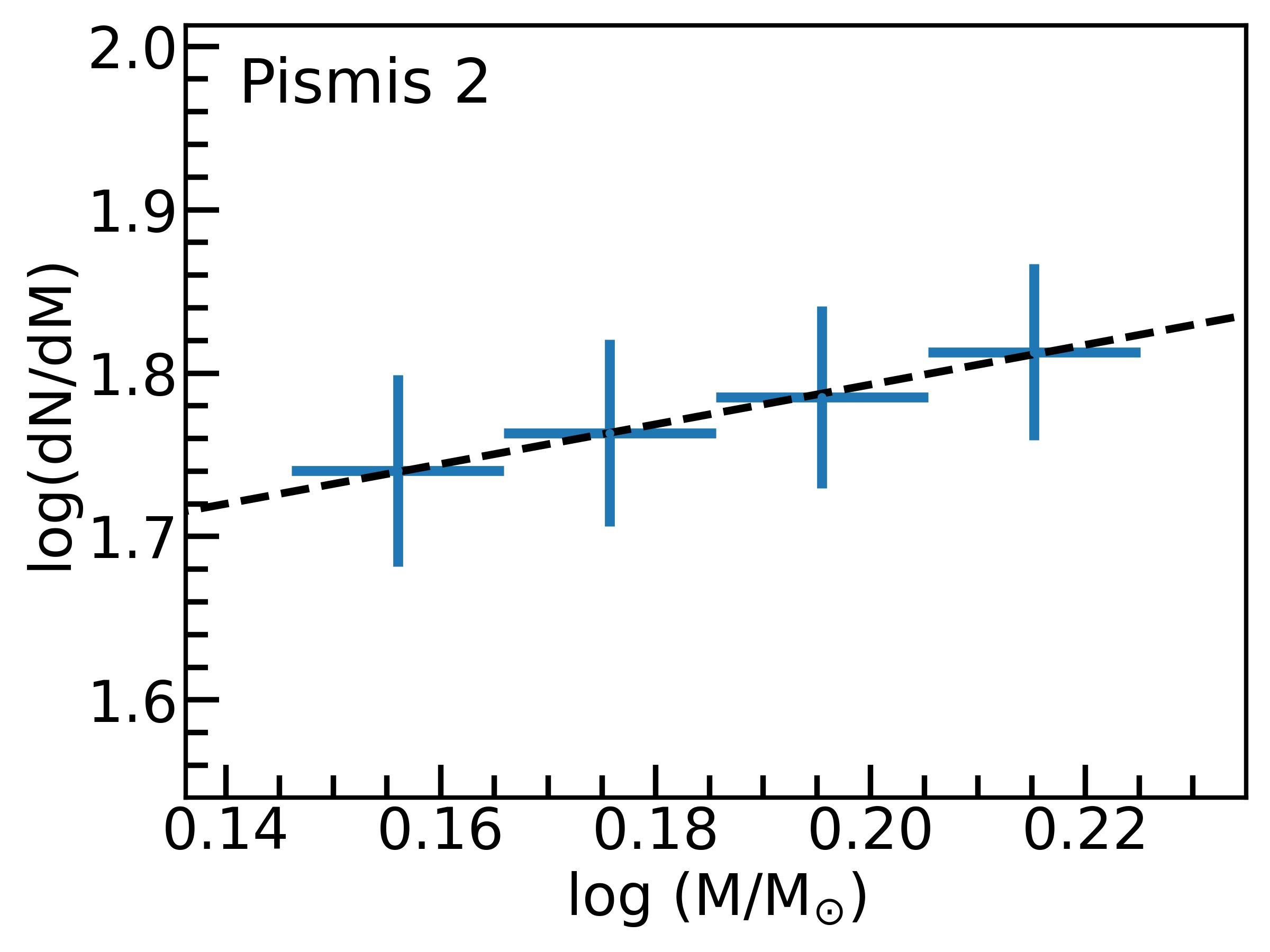}
		\caption*{}
	\end{subfigure}
	\begin{subfigure}[b]{0.24\textwidth}
   		\includegraphics[width=1.0\textwidth]{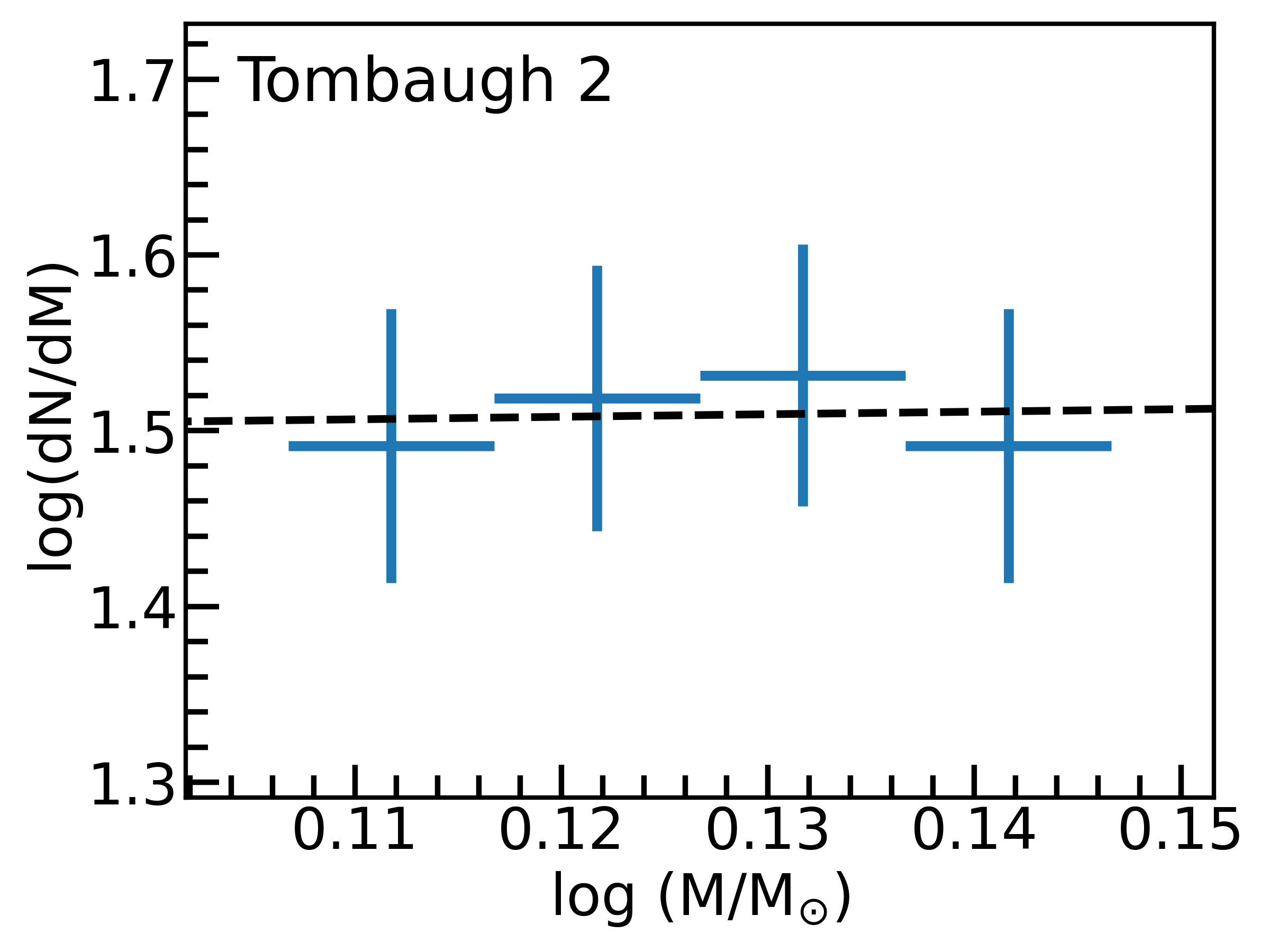}
		\caption*{}
	\end{subfigure}
	\begin{subfigure}[b]{0.24\textwidth}
   		\includegraphics[width=1.0\textwidth]{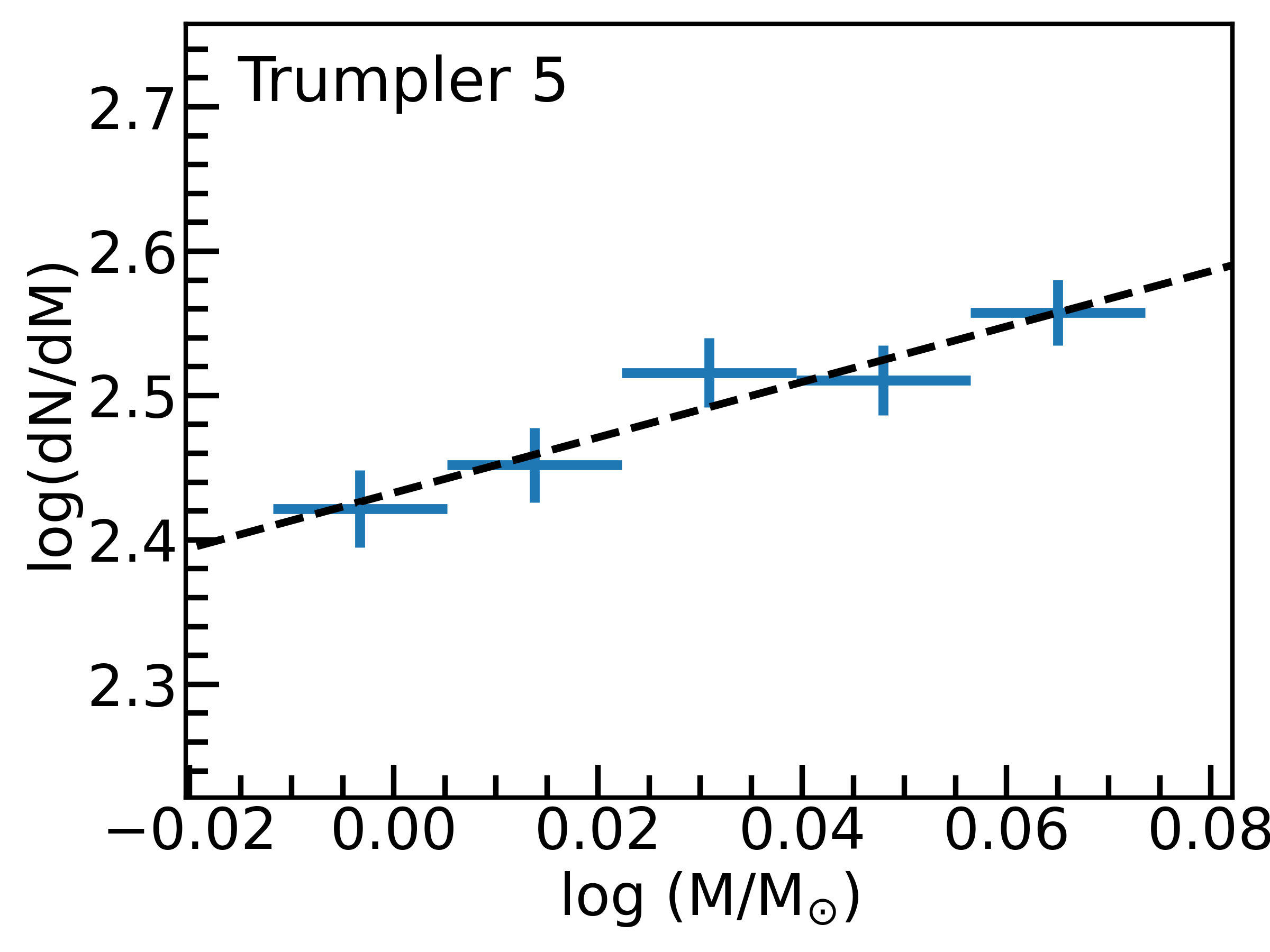}
		\caption*{}
	\end{subfigure}
	\begin{subfigure}[b]{0.24\textwidth}
   		\includegraphics[width=1.0\textwidth]{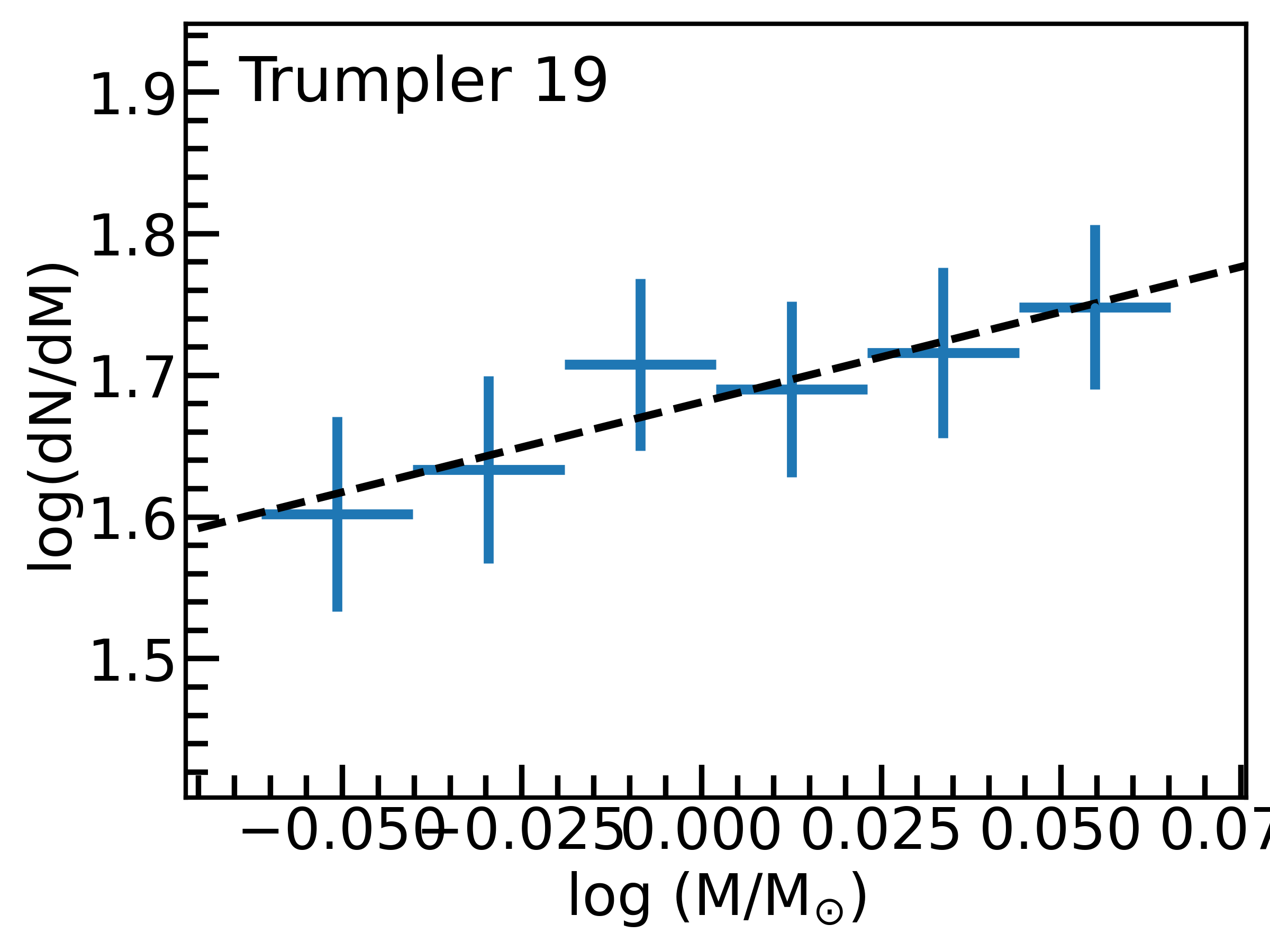}
		\caption*{}
	\end{subfigure}
	\caption{The present day mass functions derived for 23 OCs from the fitted PARSEC isochrones. The error bars are the 1$\sigma$ Poisson errors. The black dashed lines represent the best-fitted mass functions.}
	\label{fig:PDMFs}
\end{figure*}

\section{Relation of \texorpdfstring{$A^+_{\mathrm{rh}}$}{A+} vs \texorpdfstring{N$_{\mathrm{relax}}$}{Nrelax} for all cluster populations as REF}
\begin{figure}
	\includegraphics[width=0.48\textwidth]{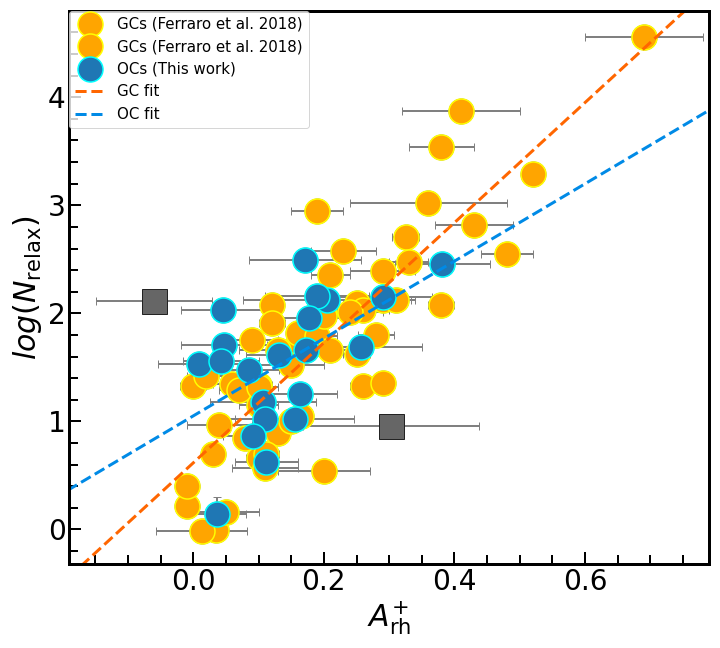}
    \caption{The correlation between the values of $A^+_{\mathrm{rh}}$ and the number of current central relaxation, $N_{\mathrm{relax}}$, for 22 OCs (blue) and 58 GCs (orange) of \citet{Ferraro2018,Ferraro2019,Cadelano2022,Dresbach2022,Beccari2023}. Here, we use SGBs, RGBs, RCs, MSTOs, and MS stars ad REF to calculate $A^+_{\mathrm{rh}}$. The blue dashed line represents the best-fitted line for the 21 OCs excluding Melotte 66 and Berkeley 31 whereas the orange dashed line shows the best-fitted line for the GCs \citep{Ferraro2018}. }
    \label{fig:A_plus_all}
\end{figure}

In addition to calculating $A^+_{\mathrm{rh}}$ using MSTO and MS stars as a REF, we estimated $A^+_{\mathrm{rh}}$ by combining all cluster populations except BSS, such as SGBs, RGBs, RCs, MSTOs, and MS stars, as REF. Fig. \ref{fig:A_plus_all}, shows $N_{\text{relax}}$ vs $A^+_{\mathrm{rh}}$ plot for OCs and GCs, where the $A^+_{\mathrm{rh}}$ is estimated using all the populations. We observe that the relationship between $N_{\text{relax}}$ and $A^+_{\mathrm{rh}}$ for OCs remains the same within errors regardless of which REF population we use.

%% file: Extended_A+.bbl
\begin{thebibliography}{}
\makeatletter
\relax
\def\mn@urlcharsother{\let\do\@makeother \do\$\do\&\do\#\do\^\do\_\do\%\do\~}
\def\mn@doi{\begingroup\mn@urlcharsother \@ifnextchar [ {\mn@doi@}
  {\mn@doi@[]}}
\def\mn@doi@[#1]#2{\def\@tempa{#1}\ifx\@tempa\@empty \href
  {http://dx.doi.org/#2} {doi:#2}\else \href {http://dx.doi.org/#2} {#1}\fi
  \endgroup}
\def\mn@eprint#1#2{\mn@eprint@#1:#2::\@nil}
\def\mn@eprint@arXiv#1{\href {http://arxiv.org/abs/#1} {{\tt arXiv:#1}}}
\def\mn@eprint@dblp#1{\href {http://dblp.uni-trier.de/rec/bibtex/#1.xml}
  {dblp:#1}}
\def\mn@eprint@#1:#2:#3:#4\@nil{\def\@tempa {#1}\def\@tempb {#2}\def\@tempc
  {#3}\ifx \@tempc \@empty \let \@tempc \@tempb \let \@tempb \@tempa \fi \ifx
  \@tempb \@empty \def\@tempb {arXiv}\fi \@ifundefined
  {mn@eprint@\@tempb}{\@tempb:\@tempc}{\expandafter \expandafter \csname
  mn@eprint@\@tempb\endcsname \expandafter{\@tempc}}}

\bibitem[\protect\citeauthoryear{{Agarwal}, {Rao}, {Vaidya}  \&
  {Bhattacharya}}{{Agarwal} et~al.}{2021}]{Agarwal2021}
{Agarwal} M.,  {Rao} K.~K.,  {Vaidya} K.,   {Bhattacharya} S.,  2021, \mn@doi
  [\mnras] {10.1093/mnras/stab118}, \href
  {https://ui.adsabs.harvard.edu/abs/2021MNRAS.502.2582A} {502, 2582}

\bibitem[\protect\citeauthoryear{{Alessandrini}, {Lanzoni}, {Ferraro},
  {Miocchi}  \& {Vesperini}}{{Alessandrini} et~al.}{2016}]{Alessandrini2016}
{Alessandrini} E.,  {Lanzoni} B.,  {Ferraro} F.~R.,  {Miocchi} P.,
  {Vesperini} E.,  2016, \mn@doi [\apj] {10.3847/1538-4357/833/2/252}, \href
  {https://ui.adsabs.harvard.edu/abs/2016ApJ...833..252A} {833, 252}

\bibitem[\protect\citeauthoryear{{Allison}, {Goodwin}, {Parker}, {Portegies
  Zwart}, {de Grijs}  \& {Kouwenhoven}}{{Allison} et~al.}{2009}]{Allison2009}
{Allison} R.~J.,  {Goodwin} S.~P.,  {Parker} R.~J.,  {Portegies Zwart} S.~F.,
  {de Grijs} R.,   {Kouwenhoven} M.~B.~N.,  2009, \mn@doi [\mnras]
  {10.1111/j.1365-2966.2009.14508.x}, \href
  {https://ui.adsabs.harvard.edu/abs/2009MNRAS.395.1449A} {395, 1449}

\bibitem[\protect\citeauthoryear{{Andreuzzi}, {Bragaglia}, {Tosi}  \&
  {Marconi}}{{Andreuzzi} et~al.}{2011}]{Gloria2011}
{Andreuzzi} G.,  {Bragaglia} A.,  {Tosi} M.,   {Marconi} G.,  2011, \mn@doi
  [\mnras] {10.1111/j.1365-2966.2010.17986.x}, \href
  {https://ui.adsabs.harvard.edu/abs/2011MNRAS.412.1265A} {412, 1265}

\bibitem[\protect\citeauthoryear{{Astropy Collaboration}}{{Astropy
  Collaboration}}{2013}]{Astropy2013}
{Astropy Collaboration} 2013, \mn@doi [\aap] {10.1051/0004-6361/201322068},
  \href {https://ui.adsabs.harvard.edu/abs/2013A&A...558A..33A} {558, A33}

\bibitem[\protect\citeauthoryear{{Bailer-Jones}, {Rybizki}, {Fouesneau},
  {Demleitner}  \& {Andrae}}{{Bailer-Jones} et~al.}{2021}]{Bailer2021}
{Bailer-Jones} C.~A.~L.,  {Rybizki} J.,  {Fouesneau} M.,  {Demleitner} M.,
  {Andrae} R.,  2021, \mn@doi [\aj]
  {10.3847/1538-3881/abd80610.48550/arXiv.2012.05220}, \href
  {https://ui.adsabs.harvard.edu/abs/2021AJ....161..147B} {161, 147}

\bibitem[\protect\citeauthoryear{{Bailyn}}{{Bailyn}}{1995}]{Bailyn1995}
{Bailyn} C.~D.,  1995, \mn@doi [\araa] {10.1146/annurev.aa.33.090195.001025},
  \href {https://ui.adsabs.harvard.edu/abs/1995ARA&A..33..133B} {33, 133}

\bibitem[\protect\citeauthoryear{{Baumgardt} \& {Makino}}{{Baumgardt} \&
  {Makino}}{2003}]{Baumgardt2003}
{Baumgardt} H.,  {Makino} J.,  2003, \mn@doi [\mnras]
  {10.1046/j.1365-8711.2003.06286.x}, \href
  {https://ui.adsabs.harvard.edu/abs/2003MNRAS.340..227B} {340, 227}

\bibitem[\protect\citeauthoryear{{Beccari}, {Cadelano}  \&
  {Dalessandro}}{{Beccari} et~al.}{2023}]{Beccari2023}
{Beccari} G.,  {Cadelano} M.,   {Dalessandro} E.,  2023, \mn@doi [\aap]
  {10.1051/0004-6361/202244288}, \href
  {https://ui.adsabs.harvard.edu/abs/2023A&A...670A..11B} {670, A11}

\bibitem[\protect\citeauthoryear{{Bhattacharya}, {Mahulkar}, {Pandaokar}  \&
  {Singh}}{{Bhattacharya} et~al.}{2017a}]{BhattacharyaandMahulkar2017}
{Bhattacharya} S.,  {Mahulkar} V.,  {Pandaokar} S.,   {Singh} P.~K.,  2017a,
  \mn@doi [Astronomy and Computing] {10.1016/j.ascom.2016.10.001}, \href
  {https://ui.adsabs.harvard.edu/abs/2017A&C....18....1B} {18, 1}

\bibitem[\protect\citeauthoryear{{Bhattacharya}, {Mishra}, {Vaidya}  \&
  {Chen}}{{Bhattacharya} et~al.}{2017b}]{Bhattacharya2017}
{Bhattacharya} S.,  {Mishra} I.,  {Vaidya} K.,   {Chen} W.~P.,  2017b, \mn@doi
  [\apj] {10.3847/1538-4357/aa89e2}, \href
  {https://ui.adsabs.harvard.edu/abs/2017ApJ...847..138B} {847, 138}

\bibitem[\protect\citeauthoryear{{Bhattacharya}, {Vaidya}, {Chen}  \&
  {Beccari}}{{Bhattacharya} et~al.}{2019}]{Bhattacharya2019}
{Bhattacharya} S.,  {Vaidya} K.,  {Chen} W.~P.,   {Beccari} G.,  2019, \mn@doi
  [\aap] {10.1051/0004-6361/201834449}, \href
  {https://ui.adsabs.harvard.edu/abs/2019A&A...624A..26B} {624, A26}

\bibitem[\protect\citeauthoryear{{Bhattacharya}, {Agarwal}, {Rao}  \&
  {Vaidya}}{{Bhattacharya} et~al.}{2021}]{Bhattacharya2021}
{Bhattacharya} S.,  {Agarwal} M.,  {Rao} K.~K.,   {Vaidya} K.,  2021, \mn@doi
  [\mnras] {10.1093/mnras/stab1404}, \href
  {https://ui.adsabs.harvard.edu/abs/2021MNRAS.505.1607B} {505, 1607}

\bibitem[\protect\citeauthoryear{{Bhattacharya}, {Rao}, {Agarwal}, {Balan}  \&
  {Vaidya}}{{Bhattacharya} et~al.}{2022}]{Bhattacharya2022}
{Bhattacharya} S.,  {Rao} K.~K.,  {Agarwal} M.,  {Balan} S.,   {Vaidya} K.,
  2022, \mn@doi [\mnras] {10.1093/mnras/stac2906}, \href
  {https://ui.adsabs.harvard.edu/abs/2022MNRAS.517.3525B} {517, 3525}

\bibitem[\protect\citeauthoryear{{Bonatto} \& {Bica}}{{Bonatto} \&
  {Bica}}{2003}]{Bonatto2003}
{Bonatto} C.,  {Bica} E.,  2003, \mn@doi [\aap] {10.1051/0004-6361:20030205},
  \href {https://ui.adsabs.harvard.edu/abs/2003A&A...405..525B} {405, 525}

\bibitem[\protect\citeauthoryear{{Bragaglia}, {Tosi}, {Andreuzzi}  \&
  {Marconi}}{{Bragaglia} et~al.}{2006}]{Bragaglia2006}
{Bragaglia} A.,  {Tosi} M.,  {Andreuzzi} G.,   {Marconi} G.,  2006, \mn@doi
  [\mnras] {10.1111/j.1365-2966.2006.10266.x}, \href
  {https://ui.adsabs.harvard.edu/abs/2006MNRAS.368.1971B} {368, 1971}

\bibitem[\protect\citeauthoryear{{Bragaglia} et~al.,}{{Bragaglia}
  et~al.}{2022}]{Bragaglia2022}
{Bragaglia} A.,  et~al., 2022, \mn@doi [\aap] {10.1051/0004-6361/202142674},
  \href {https://ui.adsabs.harvard.edu/abs/2022A&A...659A.200B} {659, A200}

\bibitem[\protect\citeauthoryear{{Cadelano}, {Ferraro}, {Dalessandro},
  {Lanzoni}, {Pallanca}  \& {Saracino}}{{Cadelano} et~al.}{2022}]{Cadelano2022}
{Cadelano} M.,  {Ferraro} F.~R.,  {Dalessandro} E.,  {Lanzoni} B.,  {Pallanca}
  C.,   {Saracino} S.,  2022, \mn@doi [\apj] {10.3847/1538-4357/aca016}, \href
  {https://ui.adsabs.harvard.edu/abs/2022ApJ...941...69C} {941, 69}

\bibitem[\protect\citeauthoryear{{Cantat-Gaudin} et~al.,}{{Cantat-Gaudin}
  et~al.}{2018}]{Cantat2018}
{Cantat-Gaudin} T.,  et~al., 2018, \mn@doi [\aap]
  {10.1051/0004-6361/201833476}, \href
  {https://ui.adsabs.harvard.edu/abs/2018A&A...618A..93C} {618, A93}

\bibitem[\protect\citeauthoryear{{Carraro}, {de Silva}, {Monaco}, {Milone}  \&
  {Mateluna}}{{Carraro} et~al.}{2014}]{Carraro2014}
{Carraro} G.,  {de Silva} G.,  {Monaco} L.,  {Milone} A.~P.,   {Mateluna} R.,
  2014, \mn@doi [\aap] {10.1051/0004-6361/20142371410.48550/arXiv.1404.6748},
  \href {https://ui.adsabs.harvard.edu/abs/2014A&A...566A..39C} {566, A39}

\bibitem[\protect\citeauthoryear{{Carrera} \& {Pancino}}{{Carrera} \&
  {Pancino}}{2011}]{Carrera2011}
{Carrera} R.,  {Pancino} E.,  2011, \mn@doi [\aap]
  {10.1051/0004-6361/201117473}, \href
  {https://ui.adsabs.harvard.edu/abs/2011A&A...535A..30C} {535, A30}

\bibitem[\protect\citeauthoryear{{Carrera} et~al.,}{{Carrera}
  et~al.}{2019}]{Carrera2019}
{Carrera} R.,  et~al., 2019, \mn@doi [\aap] {10.1051/0004-6361/201935599},
  \href {https://ui.adsabs.harvard.edu/abs/2019A&A...627A.119C} {627, A119}

\bibitem[\protect\citeauthoryear{{Chandrasekhar}}{{Chandrasekhar}}{1943}]{Chandrasekhar1943}
{Chandrasekhar} S.,  1943, \mn@doi [\apj] {10.1086/144517}, \href
  {https://ui.adsabs.harvard.edu/abs/1943ApJ....97..255C} {97, 255}

\bibitem[\protect\citeauthoryear{Cover \& Hart}{Cover \&
  Hart}{1967}]{Cover_kNN}
Cover T.,  Hart P.,  1967, \mn@doi [IEEE Transactions on Information Theory]
  {10.1109/TIT.1967.1053964}, 13, 21

\bibitem[\protect\citeauthoryear{{Dias}, {Alessi}, {Moitinho}  \&
  {L{\'e}pine}}{{Dias} et~al.}{2002}]{Dias2002}
{Dias} W.~S.,  {Alessi} B.~S.,  {Moitinho} A.,   {L{\'e}pine} J.~R.~D.,  2002,
  \mn@doi [\aap] {10.1051/0004-6361:20020668}, \href
  {https://ui.adsabs.harvard.edu/abs/2002A&A...389..871D} {389, 871}

\bibitem[\protect\citeauthoryear{{Djorgovski}}{{Djorgovski}}{1993}]{Djorgovski1993}
{Djorgovski} S.,  1993, in {Djorgovski} S.~G.,  {Meylan} G.,  eds,
  Astronomical Society of the Pacific Conference Series Vol. 50, Structure and
  Dynamics of Globular Clusters. p.~373

\bibitem[\protect\citeauthoryear{{Dresbach}, {Massari}, {Lanzoni}, {Ferraro},
  {Dalessandro}, {Raso}, {Bellini}  \& {Libralato}}{{Dresbach}
  et~al.}{2022}]{Dresbach2022}
{Dresbach} F.,  {Massari} D.,  {Lanzoni} B.,  {Ferraro} F.~R.,  {Dalessandro}
  E.,  {Raso} S.,  {Bellini} A.,   {Libralato} M.,  2022, \mn@doi [\apj]
  {10.3847/1538-4357/ac5406}, \href
  {https://ui.adsabs.harvard.edu/abs/2022ApJ...928...47D} {928, 47}

\bibitem[\protect\citeauthoryear{{Fan} et~al.,}{{Fan} et~al.}{1996}]{Fan19968}
{Fan} X.,  et~al., 1996, \mn@doi [\aj] {10.1086/118039}, \href
  {https://ui.adsabs.harvard.edu/abs/1996AJ....112..628F} {112, 628}

\bibitem[\protect\citeauthoryear{{Ferraro} et~al.,}{{Ferraro}
  et~al.}{2012}]{Ferraro2012}
{Ferraro} F.~R.,  et~al., 2012, \mn@doi [\nat] {10.1038/nature11686}, \href
  {https://ui.adsabs.harvard.edu/abs/2012Natur.492..393F} {492, 393}

\bibitem[\protect\citeauthoryear{{Ferraro} et~al.,}{{Ferraro}
  et~al.}{2018}]{Ferraro2018}
{Ferraro} F.~R.,  et~al., 2018, \mn@doi [\apj] {10.3847/1538-4357/aac01c},
  \href {https://ui.adsabs.harvard.edu/abs/2018ApJ...860...36F} {860, 36}

\bibitem[\protect\citeauthoryear{{Ferraro}, {Lanzoni}, {Dalessandro},
  {Cadelano}, {Raso}, {Mucciarelli}, {Beccari}  \& {Pallanca}}{{Ferraro}
  et~al.}{2019}]{Ferraro2019}
{Ferraro} F.~R.,  {Lanzoni} B.,  {Dalessandro} E.,  {Cadelano} M.,  {Raso} S.,
  {Mucciarelli} A.,  {Beccari} G.,   {Pallanca} C.,  2019, \mn@doi [Nature
  Astronomy] {10.1038/s41550-019-0865-1}, \href
  {https://ui.adsabs.harvard.edu/abs/2019NatAs...3.1149F} {3, 1149}

\bibitem[\protect\citeauthoryear{{Ferraro}, {Lanzoni}  \&
  {Dalessandro}}{{Ferraro} et~al.}{2020}]{Ferraro2020}
{Ferraro} F.~R.,  {Lanzoni} B.,   {Dalessandro} E.,  2020, arXiv e-prints,
  \href {https://ui.adsabs.harvard.edu/abs/2020arXiv200107435F} {p.
  arXiv:2001.07435}

\bibitem[\protect\citeauthoryear{{Fiorentino}, {Lanzoni}, {Dalessandro},
  {Ferraro}, {Bono}  \& {Marconi}}{{Fiorentino} et~al.}{2014}]{Fiorentino2014}
{Fiorentino} G.,  {Lanzoni} B.,  {Dalessandro} E.,  {Ferraro} F.~R.,  {Bono}
  G.,   {Marconi} M.,  2014, \mn@doi [\apj] {10.1088/0004-637X/783/1/34}, \href
  {https://ui.adsabs.harvard.edu/abs/2014ApJ...783...34F} {783, 34}

\bibitem[\protect\citeauthoryear{Fisher}{Fisher}{1992}]{fisher1992}
Fisher R.~A.,  1992, in , Breakthroughs in statistics.
Springer, pp 66--70

\bibitem[\protect\citeauthoryear{{Foreman-Mackey}, {Hogg}, {Lang}  \&
  {Goodman}}{{Foreman-Mackey} et~al.}{2013}]{Foreman2013}
{Foreman-Mackey} D.,  {Hogg} D.~W.,  {Lang} D.,   {Goodman} J.,  2013, \mn@doi
  [\pasp] {10.1086/670067}, \href
  {https://ui.adsabs.harvard.edu/abs/2013PASP..125..306F} {125, 306}

\bibitem[\protect\citeauthoryear{{Friel}, {Janes}, {Tavarez}, {Scott},
  {Katsanis}, {Lotz}, {Hong}  \& {Miller}}{{Friel} et~al.}{2002}]{Friel2002}
{Friel} E.~D.,  {Janes} K.~A.,  {Tavarez} M.,  {Scott} J.,  {Katsanis} R.,
  {Lotz} J.,  {Hong} L.,   {Miller} N.,  2002, \mn@doi [\aj] {10.1086/344161},
  \href {https://ui.adsabs.harvard.edu/abs/2002AJ....124.2693F} {124, 2693}

\bibitem[\protect\citeauthoryear{{Friel}, {Jacobson}  \& {Pilachowski}}{{Friel}
  et~al.}{2005}]{Friel2005}
{Friel} E.~D.,  {Jacobson} H.~R.,   {Pilachowski} C.~A.,  2005, \mn@doi [\aj]
  {10.1086/430146}, \href
  {https://ui.adsabs.harvard.edu/abs/2005AJ....129.2725F} {129, 2725}

\bibitem[\protect\citeauthoryear{{Friel}, {Jacobson}  \& {Pilachowski}}{{Friel}
  et~al.}{2010}]{Friel2010}
{Friel} E.~D.,  {Jacobson} H.~R.,   {Pilachowski} C.~A.,  2010, \mn@doi [\aj]
  {10.1088/0004-6256/139/5/1942}, \href
  {https://ui.adsabs.harvard.edu/abs/2010AJ....139.1942F} {139, 1942}

\bibitem[\protect\citeauthoryear{{Gaia Collaboration}, {Brown}, {Vallenari},
  {Prusti}, {de Bruijne}, {Babusiaux}  \& {Biermann}}{{Gaia Collaboration}
  et~al.}{2021}]{Gaiaedr32021}
{Gaia Collaboration} {Brown} A.~G.~A.,  {Vallenari} A.,  {Prusti} T.,  {de
  Bruijne} J.~H.~J.,  {Babusiaux} C.,   {Biermann} M.,  2021, \mn@doi [\aap]
  {10.1051/0004-6361/202039657}, \href
  {https://ui.adsabs.harvard.edu/abs/2021A&A...649A...1G} {649, A1}

\bibitem[\protect\citeauthoryear{{Gieles} \& {Zocchi}}{{Gieles} \&
  {Zocchi}}{2015}]{Gieles2015}
{Gieles} M.,  {Zocchi} A.,  2015, \mn@doi [\mnras] {10.1093/mnras/stv1848},
  \href {https://ui.adsabs.harvard.edu/abs/2015MNRAS.454..576G} {454, 576}

\bibitem[\protect\citeauthoryear{{Gieles} \& {Zocchi}}{{Gieles} \&
  {Zocchi}}{2017}]{Gieles2017}
{Gieles} M.,  {Zocchi} A.,  2017, {LIMEPY: Lowered Isothermal Model Explorer in
  PYthon}, Astrophysics Source Code Library, record ascl:1710.023 (\mn@eprint
  {ascl} {1710.023})

\bibitem[\protect\citeauthoryear{Harris et~al.,}{Harris
  et~al.}{2020}]{Harris2020}
Harris C.~R.,  et~al., 2020, \mn@doi [Nature] {10.1038/s41586-020-2649-2}, 585,
  357–362

\bibitem[\protect\citeauthoryear{{Hunter}}{{Hunter}}{2007}]{Hunter4160265}
{Hunter} J.~D.,  2007, Computing in Science Engineering, 9, 90

\bibitem[\protect\citeauthoryear{{Jadhav}, {Pandey}, {Subramaniam}  \&
  {Sagar}}{{Jadhav} et~al.}{2021}]{Jadhav2021}
{Jadhav} V.~V.,  {Pandey} S.,  {Subramaniam} A.,   {Sagar} R.,  2021, \mn@doi
  [Journal of Astrophysics and Astronomy] {10.1007/s12036-021-09746-y}, \href
  {https://ui.adsabs.harvard.edu/abs/2021JApA...42...89J} {42, 89}

\bibitem[\protect\citeauthoryear{{Janes} \& {Phelps}}{{Janes} \&
  {Phelps}}{1994}]{Janes1994}
{Janes} K.~A.,  {Phelps} R.~L.,  1994, \mn@doi [\aj] {10.1086/117192}, \href
  {https://ui.adsabs.harvard.edu/abs/1994AJ....108.1773J} {108, 1773}

\bibitem[\protect\citeauthoryear{{Kaluzny}}{{Kaluzny}}{1989}]{Kaluzny1989}
{Kaluzny} J.,  1989, \actaa, \href
  {https://ui.adsabs.harvard.edu/abs/1989AcA....39...13K} {39, 13}

\bibitem[\protect\citeauthoryear{{King}}{{King}}{1966}]{King1966}
{King} I.~R.,  1966, \mn@doi [\aj] {10.1086/109857}, \href
  {https://ui.adsabs.harvard.edu/abs/1966AJ.....71...64K} {71, 64}

\bibitem[\protect\citeauthoryear{{Krause} et~al.,}{{Krause}
  et~al.}{2020}]{Krause2020}
{Krause} M. G.~H.,  et~al., 2020, \mn@doi [\ssr] {10.1007/s11214-020-00689-4},
  \href {https://ui.adsabs.harvard.edu/abs/2020SSRv..216...64K} {216, 64}

\bibitem[\protect\citeauthoryear{{Krumholz}, {McKee}  \&
  {Bland-Hawthorn}}{{Krumholz} et~al.}{2019}]{Krumholz2019}
{Krumholz} M.~R.,  {McKee} C.~F.,   {Bland-Hawthorn} J.,  2019, \mn@doi [\araa]
  {10.1146/annurev-astro-091918-104430}, \href
  {https://ui.adsabs.harvard.edu/abs/2019ARA&A..57..227K} {57, 227}

\bibitem[\protect\citeauthoryear{{Lanzoni}, {Ferraro}, {Alessandrini},
  {Dalessand ro}, {Vesperini}  \& {Raso}}{{Lanzoni} et~al.}{2016}]{Lanzoni2016}
{Lanzoni} B.,  {Ferraro} F.~R.,  {Alessandrini} E.,  {Dalessand ro} E.,
  {Vesperini} E.,   {Raso} S.,  2016, \mn@doi [\apjl]
  {10.3847/2041-8213/833/2/L29}, \href
  {https://ui.adsabs.harvard.edu/abs/2016ApJ...833L..29L} {833, L29}

\bibitem[\protect\citeauthoryear{{Magrini} et~al.,}{{Magrini}
  et~al.}{2021}]{Magrini2021}
{Magrini} L.,  et~al., 2021, \mn@doi [\aap] {10.1051/0004-6361/202141275},
  \href {https://ui.adsabs.harvard.edu/abs/2021A&A...655A..23M} {655, A23}

\bibitem[\protect\citeauthoryear{Massari et~al.,}{Massari
  et~al.}{2012}]{Massari2012}
Massari D.,  et~al., 2012, \mn@doi [The Astrophysical Journal Letters]
  {10.1088/2041-8205/755/2/L32}, 755, L32

\bibitem[\protect\citeauthoryear{McLachlan \& Peel}{McLachlan \&
  Peel}{2000}]{mclachlan200001}
McLachlan G.~J.,  Peel D.,  2000, Finite mixture models.
 Probability and Statistics -- Applied Probability and Statistics Section Vol.
  299, Wiley, New York

\bibitem[\protect\citeauthoryear{{Netopil}, {Paunzen}, {Heiter}  \&
  {Soubiran}}{{Netopil} et~al.}{2016}]{Netopil2016}
{Netopil} M.,  {Paunzen} E.,  {Heiter} U.,   {Soubiran} C.,  2016, \mn@doi
  [\aap] {10.1051/0004-6361/201526370}, \href
  {https://ui.adsabs.harvard.edu/abs/2016A&A...585A.150N} {585, A150}

\bibitem[\protect\citeauthoryear{Newville, Stensitzki, Allen  \&
  Ingargiola}{Newville et~al.}{2014}]{Newville2014}
Newville M.,  Stensitzki T.,  Allen D.~B.,   Ingargiola A.,  2014, {LMFIT:
  Non-Linear Least-Square Minimization and Curve-Fitting for Python},
  \mn@doi{10.5281/zenodo.11813}, \url {https://doi.org/10.5281/zenodo.11813}

\bibitem[\protect\citeauthoryear{{Ortolani}, {Bica}, {Barbuy}  \&
  {Zoccali}}{{Ortolani} et~al.}{2005}]{Ortolani2005}
{Ortolani} S.,  {Bica} E.,  {Barbuy} B.,   {Zoccali} M.,  2005, \mn@doi [\aap]
  {10.1051/0004-6361:20041458}, \href
  {https://ui.adsabs.harvard.edu/abs/2005A&A...429..607O} {429, 607}

\bibitem[\protect\citeauthoryear{{Overbeek}, {Friel}  \& {Jacobson}}{{Overbeek}
  et~al.}{2016}]{Overbeek2016}
{Overbeek} J.~C.,  {Friel} E.~D.,   {Jacobson} H.~R.,  2016, \mn@doi [\apj]
  {10.3847/0004-637X/824/2/75}, \href
  {https://ui.adsabs.harvard.edu/abs/2016ApJ...824...75O} {824, 75}

\bibitem[\protect\citeauthoryear{{Panthi}, {Vaidya}, {Jadhav}, {Rao},
  {Subramaniam}, {Agarwal}  \& {Pandey}}{{Panthi} et~al.}{2022}]{Panthi2022}
{Panthi} A.,  {Vaidya} K.,  {Jadhav} V.,  {Rao} K.~K.,  {Subramaniam} A.,
  {Agarwal} M.,   {Pandey} S.,  2022, \mn@doi [\mnras]
  {10.1093/mnras/stac2421}, \href
  {https://ui.adsabs.harvard.edu/abs/2022MNRAS.516.5318P} {516, 5318}

\bibitem[\protect\citeauthoryear{{Piskunov}, {Kharchenko}, {Schilbach},
  {R{\"o}ser}, {Scholz}  \& {Zinnecker}}{{Piskunov}
  et~al.}{2008}]{Piskunov2008}
{Piskunov} A.~E.,  {Kharchenko} N.~V.,  {Schilbach} E.,  {R{\"o}ser} S.,
  {Scholz} R.~D.,   {Zinnecker} H.,  2008, \mn@doi [\aap]
  {10.1051/0004-6361:200809505}, \href
  {https://ui.adsabs.harvard.edu/abs/2008A&A...487..557P} {487, 557}

\bibitem[\protect\citeauthoryear{{Piskunov}, {Kharchenko}, {Schilbach},
  {R{\"o}ser}, {Scholz}  \& {Zinnecker}}{{Piskunov}
  et~al.}{2011}]{Piskunov2011}
{Piskunov} A.~E.,  {Kharchenko} N.~V.,  {Schilbach} E.,  {R{\"o}ser} S.,
  {Scholz} R.~D.,   {Zinnecker} H.,  2011, \mn@doi [\aap]
  {10.1051/0004-6361/201015376}, \href
  {https://ui.adsabs.harvard.edu/abs/2011A&A...525A.122P} {525, A122}

\bibitem[\protect\citeauthoryear{{Portegies Zwart}, {McMillan}  \&
  {Gieles}}{{Portegies Zwart} et~al.}{2010}]{Portegies2010}
{Portegies Zwart} S.~F.,  {McMillan} S. L.~W.,   {Gieles} M.,  2010, \mn@doi
  [\araa] {10.1146/annurev-astro-081309-130834}, \href
  {https://ui.adsabs.harvard.edu/abs/2010ARA&A..48..431P} {48, 431}

\bibitem[\protect\citeauthoryear{{Rain}, {Carraro}, {Ahumada}, {Villanova},
  {Boffin}, {Monaco}  \& {Beccari}}{{Rain} et~al.}{2020}]{Rain2020a}
{Rain} M.~J.,  {Carraro} G.,  {Ahumada} J.~A.,  {Villanova} S.,  {Boffin} H.,
  {Monaco} L.,   {Beccari} G.,  2020, \mn@doi [\aj] {10.3847/1538-3881/ab5f0b},
  \href {https://ui.adsabs.harvard.edu/abs/2020AJ....159...59R} {159, 59}

\bibitem[\protect\citeauthoryear{{Rain}, {Carraro}, {Ahumada}, {Villanova},
  {Boffin}  \& {Monaco}}{{Rain} et~al.}{2021}]{Rain2020b}
{Rain} M.~J.,  {Carraro} G.,  {Ahumada} J.~A.,  {Villanova} S.,  {Boffin} H.,
  {Monaco} L.,  2021, \mn@doi [\aj] {10.3847/1538-3881/abc1ee}, \href
  {https://ui.adsabs.harvard.edu/abs/2021AJ....161...37R} {161, 37}

\bibitem[\protect\citeauthoryear{{Rao}, {Vaidya}, {Agarwal}  \&
  {Bhattacharya}}{{Rao} et~al.}{2021}]{Rao2021}
{Rao} K.~K.,  {Vaidya} K.,  {Agarwal} M.,   {Bhattacharya} S.,  2021, \mn@doi
  [\mnras] {10.1093/mnras/stab2894}, \href
  {https://ui.adsabs.harvard.edu/abs/2021MNRAS.508.4919R} {508, 4919}

\bibitem[\protect\citeauthoryear{{Rao}, {Vaidya}, {Agarwal}, {Panthi}, {Jadhav}
   \& {Subramaniam}}{{Rao} et~al.}{2022}]{Rao2022}
{Rao} K.~K.,  {Vaidya} K.,  {Agarwal} M.,  {Panthi} A.,  {Jadhav} V.,
  {Subramaniam} A.,  2022, \mn@doi [\mnras] {10.1093/mnras/stac2241}, \href
  {https://ui.adsabs.harvard.edu/abs/2022MNRAS.516.2444R} {516, 2444}

\bibitem[\protect\citeauthoryear{{Rao}, {Bhattacharya}, {Vaidya}  \&
  {Agarwal}}{{Rao} et~al.}{2023}]{Rao2023}
{Rao} K.~K.,  {Bhattacharya} S.,  {Vaidya} K.,   {Agarwal} M.,  2023, \mn@doi
  [\mnras] {10.1093/mnrasl/slac122}, \href
  {https://ui.adsabs.harvard.edu/abs/2023MNRAS.518L...7R} {518, L7}

\bibitem[\protect\citeauthoryear{{Renaud}}{{Renaud}}{2018}]{Renaud2018}
{Renaud} F.,  2018, \mn@doi [\nar] {10.1016/j.newar.2018.03.001}, \href
  {https://ui.adsabs.harvard.edu/abs/2018NewAR..81....1R} {81, 1}

\bibitem[\protect\citeauthoryear{{Ricker} et~al.,}{{Ricker}
  et~al.}{2015}]{Ricker2015}
{Ricker} G.~R.,  et~al., 2015, \mn@doi [Journal of Astronomical Telescopes,
  Instruments, and Systems] {10.1117/1.JATIS.1.1.014003}, \href
  {https://ui.adsabs.harvard.edu/abs/2015JATIS...1a4003R} {1, 014003}

\bibitem[\protect\citeauthoryear{{Ritter}}{{Ritter}}{2010}]{Ritter2010}
{Ritter} H.,  2010, \memsai, \href
  {https://ui.adsabs.harvard.edu/abs/2010MmSAI..81..849R} {81, 849}

\bibitem[\protect\citeauthoryear{{Roming} et~al.,}{{Roming}
  et~al.}{2005}]{Roming2005}
{Roming} P. W.~A.,  et~al., 2005, \mn@doi [\ssr] {10.1007/s11214-005-5095-4},
  \href {https://ui.adsabs.harvard.edu/abs/2005SSRv..120...95R} {120, 95}

\bibitem[\protect\citeauthoryear{{Rosvick}}{{Rosvick}}{1995}]{Rosvick1995}
{Rosvick} J.~M.,  1995, \mn@doi [\mnras] {10.1093/mnras/277.4.1379}, \href
  {https://ui.adsabs.harvard.edu/abs/1995MNRAS.277.1379R} {277, 1379}

\bibitem[\protect\citeauthoryear{{Salpeter}}{{Salpeter}}{1955}]{Salpeter1955}
{Salpeter} E.~E.,  1955, \mn@doi [\apj] {10.1086/145971}, \href
  {https://ui.adsabs.harvard.edu/abs/1955ApJ...121..161S} {121, 161}

\bibitem[\protect\citeauthoryear{{Sandage}}{{Sandage}}{1953}]{Sandage1953}
{Sandage} A.~R.,  1953, \mn@doi [\aj] {10.1086/106822}, \href
  {https://ui.adsabs.harvard.edu/abs/1953AJ.....58...61S} {58, 61}

\bibitem[\protect\citeauthoryear{{Sariya}, {Jiang}, {Bisht}, {Yadav}  \&
  {Rangwal}}{{Sariya} et~al.}{2021}]{Sariya2021}
{Sariya} D.~P.,  {Jiang} I.-G.,  {Bisht} D.,  {Yadav} R.~K.~S.,   {Rangwal} G.,
   2021, \mn@doi [\aj] {10.3847/1538-3881/abd31f}, \href
  {https://ui.adsabs.harvard.edu/abs/2021AJ....161..102S} {161, 102}

\bibitem[\protect\citeauthoryear{{Shara}, {Saffer}  \& {Livio}}{{Shara}
  et~al.}{1997}]{Shara1997}
{Shara} M.~M.,  {Saffer} R.~A.,   {Livio} M.,  1997, \mn@doi [\apjl]
  {10.1086/310952}, \href
  {https://ui.adsabs.harvard.edu/abs/1997ApJ...489L..59S} {489, L59}

\bibitem[\protect\citeauthoryear{Storn \& Price}{Storn \&
  Price}{1997}]{Storn1997}
Storn R.,  Price K.,  1997, Journal of global optimization, 11, 341

\bibitem[\protect\citeauthoryear{{Stryker}}{{Stryker}}{1993}]{Stryker1993}
{Stryker} L.~L.,  1993, \mn@doi [\pasp] {10.1086/133286}, \href
  {https://ui.adsabs.harvard.edu/abs/1993PASP..105.1081S} {105, 1081}

\bibitem[\protect\citeauthoryear{{Tosi}, {Pulone}, {Marconi}  \&
  {Bragaglia}}{{Tosi} et~al.}{1998}]{Tosi1998}
{Tosi} M.,  {Pulone} L.,  {Marconi} G.,   {Bragaglia} A.,  1998, \mn@doi
  [\mnras] {10.1046/j.1365-8711.1998.01812.x}, \href
  {https://ui.adsabs.harvard.edu/abs/1998MNRAS.299..834T} {299, 834}

\bibitem[\protect\citeauthoryear{{Vaidya}, {Rao}, {Agarwal}  \&
  {Bhattacharya}}{{Vaidya} et~al.}{2020}]{Vaidya2020}
{Vaidya} K.,  {Rao} K.~K.,  {Agarwal} M.,   {Bhattacharya} S.,  2020, \mn@doi
  [\mnras] {10.1093/mnras/staa1667}, \href
  {https://ui.adsabs.harvard.edu/abs/2020MNRAS.496.2402V} {496, 2402}

\bibitem[\protect\citeauthoryear{{Vaidya}, {Panthi}, {Agarwal}, {Pandey},
  {Rao}, {Jadhav}  \& {Subramaniam}}{{Vaidya} et~al.}{2022}]{Vaidya2022}
{Vaidya} K.,  {Panthi} A.,  {Agarwal} M.,  {Pandey} S.,  {Rao} K.~K.,  {Jadhav}
  V.,   {Subramaniam} A.,  2022, \mn@doi [\mnras] {10.1093/mnras/stac207},
  \href {https://ui.adsabs.harvard.edu/abs/2022MNRAS.511.2274V} {511, 2274}

\bibitem[\protect\citeauthoryear{{Vande Putte}, {Garnier}, {Ferreras},
  {Mignani}  \& {Cropper}}{{Vande Putte} et~al.}{2010}]{Vande2010}
{Vande Putte} D.,  {Garnier} T.~P.,  {Ferreras} I.,  {Mignani} R.~P.,
  {Cropper} M.,  2010, \mn@doi [\mnras] {10.1111/j.1365-2966.2010.17025.x},
  \href {https://ui.adsabs.harvard.edu/abs/2010MNRAS.407.2109V} {407, 2109}

\bibitem[\protect\citeauthoryear{{Vesperini}}{{Vesperini}}{2010}]{Vesperini2010}
{Vesperini} E.,  2010, \mn@doi [Philosophical Transactions of the Royal Society
  of London Series A] {10.1098/rsta.2009.0260}, \href
  {https://ui.adsabs.harvard.edu/abs/2010RSPTA.368..829V} {368, 829}

\bibitem[\protect\citeauthoryear{{Villanova}, {Randich}, {Geisler}, {Carraro}
  \& {Costa}}{{Villanova} et~al.}{2010}]{Villanova2010}
{Villanova} S.,  {Randich} S.,  {Geisler} D.,  {Carraro} G.,   {Costa} E.,
  2010, \mn@doi [\aap] {10.1051/0004-6361/200913258}, \href
  {https://ui.adsabs.harvard.edu/abs/2010A&A...509A.102V} {509, A102}

\bibitem[\protect\citeauthoryear{{Virtanen} et~al.,}{{Virtanen}
  et~al.}{2020}]{Virtanen2020}
{Virtanen} P.,  et~al., 2020, \mn@doi [Nature Methods]
  {10.1038/s41592-019-0686-2}, \href
  {https://ui.adsabs.harvard.edu/abs/2020NatMe..17..261V} {17, 261}

\bibitem[\protect\citeauthoryear{{Viscasillas V{\'a}zquez}
  et~al.,}{{Viscasillas V{\'a}zquez} et~al.}{2022}]{Viscasillas2022}
{Viscasillas V{\'a}zquez} C.,  et~al., 2022, \mn@doi [\aap]
  {10.1051/0004-6361/202142937}, \href
  {https://ui.adsabs.harvard.edu/abs/2022A&A...660A.135V} {660, A135}

\bibitem[\protect\citeauthoryear{{Wang} \& {Chen}}{{Wang} \&
  {Chen}}{2019}]{Wang2019}
{Wang} S.,  {Chen} X.,  2019, \mn@doi [\apj] {10.3847/1538-4357/ab1c61}, \href
  {https://ui.adsabs.harvard.edu/abs/2019ApJ...877..116W} {877, 116}

\bibitem[\protect\citeauthoryear{{Yong}, {Carney}  \& {Teixera de
  Almeida}}{{Yong} et~al.}{2005}]{Yong2005}
{Yong} D.,  {Carney} B.~W.,   {Teixera de Almeida} M.~L.,  2005, \mn@doi [\aj]
  {10.1086/430934}, \href
  {https://ui.adsabs.harvard.edu/abs/2005AJ....130..597Y} {130, 597}

\bibitem[\protect\citeauthoryear{Zou}{Zou}{2007}]{zou2007}
Zou G.~Y.,  2007, Psychological methods, 12, 399

\makeatother
\end{thebibliography}
